AI Ethics and Governance in Practice Programme

# AI Sustainability in Practice

## Part Two: Sustainability Throughout the AI Workflow

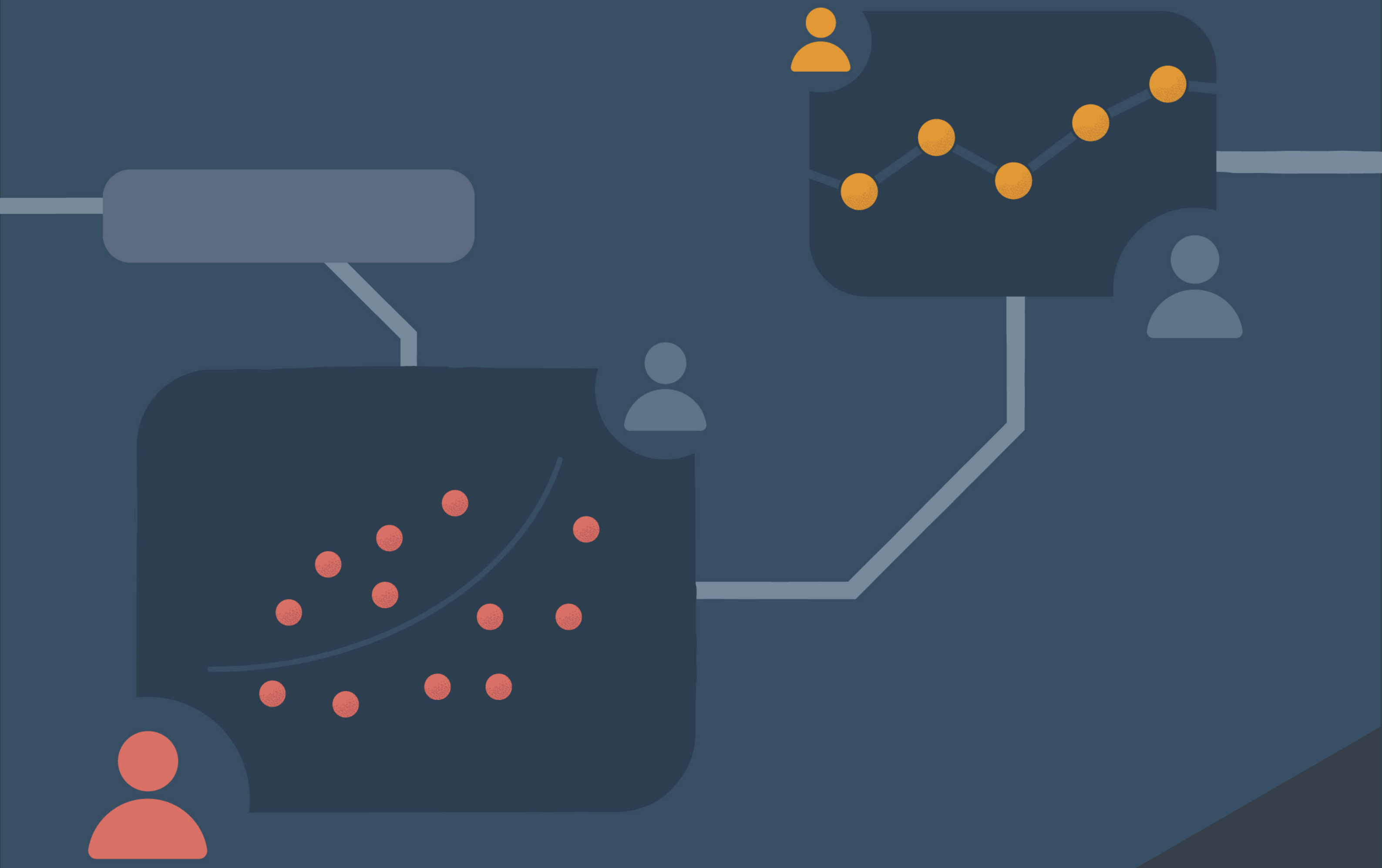

**For Facilitators**

This workbook is annotated to support facilitators in delivering the accompanying activities.

# Acknowledgements

This workbook was written by David Leslie, Cami Rincón, Morgan Briggs, Antonella Perini, Smera Jayadeva, Ann Borda, SJ Bennett, Christopher Burr, Mhairi Aitken, Michael Katell, Claudia Fischer, Janis Wong, and Ismael Kherroubi Garcia.

The creation of this workbook would not have been possible without the support and efforts of various partners and collaborators. As ever, all members of our brilliant team of researchers in the Ethics Theme of the Public Policy Programme at The Alan Turing Institute have been crucial and inimitable supports of this project from its inception several years ago, as have our Public Policy Programme Co-Directors, Helen Margetts and Cosmina Dorobantu. We are deeply thankful to Conor Rigby, who led the design of this workbook and provided extraordinary feedback across its iterations. We also want to acknowledge Johnny Lighthands, who created various illustrations for this document, and Alex Krook and John Gilbert, whose input and insights helped get the workbook over the finish line. Special thanks must be given to the Ministry of Justice for helping us test the activities and review the content included in this workbook. Lastly, we want to thank Youmna Hashem (The Alan Turing Institute) and Sabeehah Mahomed (The Alan Turing Institute) for their meticulous peer review and timely feedback, which greatly enriched this document.

This work was supported by Wave 1 of The UKRI Strategic Priorities Fund under the EPSRC Grant EP/W006022/1, particularly the Public Policy Programme theme within that grant & The Alan Turing Institute; Towards Turing 2.0 under the EPSRC Grant EP/W037211/1 & The Alan Turing Institute; and the Ecosystem Leadership Award under the EPSRC Grant EP/X03870X/1 & The Alan Turing Institute.

Cite this work as: Leslie, D., Rincón, C., Briggs, M., Perini, A., Jayadeva, S., Borda, A., Bennett, SJ. Burr, C., Aitken, M., Katell, M., Fischer, C., Wong, J., and Kherroubi Garcia, I. (2023). *AI Sustainability in Practice Part Two: Sustainability Throughout the AI Workflow.* The Alan Turing Institute.

# Contents

# About the AI Ethics and Governance in Practice Workbook Series

## Who We Are

The Public Policy Programme at The Alan Turing Institute was set up in May 2018 with the aim of developing research, tools, and techniques that help governments innovate with data-intensive technologies and improve the quality of people's lives. We work alongside policymakers to explore how data science and artificial intelligence can inform public policy and improve the provision of public services. We believe that governments can reap the benefits of these technologies only if they make considerations of ethics and safety a first priority.

## Origins of the Workbook Series

In 2019, The Alan Turing Institute's Public Policy Programme, in collaboration with the UK's Office for Artificial Intelligence and the Government Digital Service, published the UK Government's official Public Sector Guidance on AI Ethics and Safety. This document provides end-to-end guidance on how to apply principles of AI ethics and safety to the design, development, and implementation of algorithmic systems in the public sector. It provides a governance framework designed to assist AI project teams in ensuring that the AI technologies they build, procure, or use are ethical, safe, and responsible.

In 2021, the UK's National AI Strategy recommended as a 'key action' the update and expansion of this original guidance. From 2021 to 2023, with the support of funding from the Office for AI and the Engineering and Physical Sciences Research Council as well as with the assistance of several public sector bodies, we undertook this updating and expansion. The result is the AI Ethics and Governance in Practice Programme, a bespoke series of eight workbooks and a forthcoming digital platform designed to equip the public sector with tools, training, and support for adopting what we call a Process-Based Governance (PBG) Framework to carry out projects in line with state-of-the-art practices in responsible and trustworthy AI innovation.

# About the Workbooks

The AI Ethics and Governance in Practice Programme curriculum is composed of a series of eight workbooks. Each of the workbooks in the series covers how to implement a key component of the PBG Framework. These include Sustainability, Technical Safety, Accountability, Fairness, Explainability, and Data Stewardship. Each of the workbooks also focuses on a specific domain, so that case studies can be used to promote ethical reflection and animate the Key Concepts.

## Programme Curriculum: AI Ethics and Governance in Practice Workbook Series

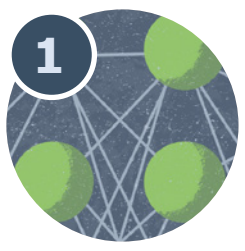

**AI Ethics and Governance in Practice: An Introduction**
*Multiple Domains*

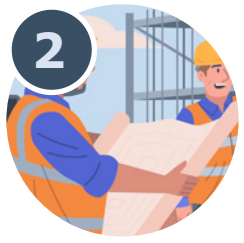

**AI Sustainability in Practice Part One**
*AI in Urban Planning*

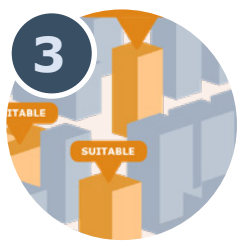

**AI Sustainability in Practice Part Two**
*AI in Urban Planning*

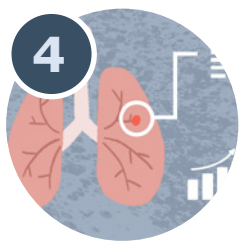

**AI Fairness in Practice**
*AI in Healthcare*

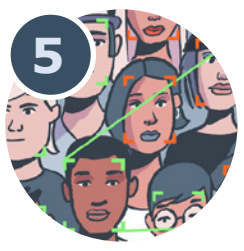

**Responsible Data Stewardship in Practice**
*AI in Policing and Criminal Justice*

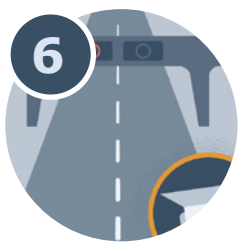

**AI Safety in Practice**
*AI in Transport*

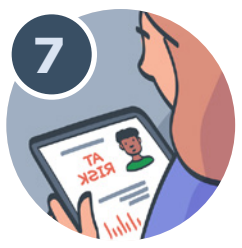

**AI Transparency and Explainability in Practice**
*AI in Social Care*

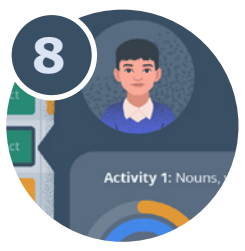

**AI Accountability in Practice**
*AI in Education*

Taken together, the workbooks are intended to provide public sector bodies with the skills required for putting AI ethics and governance principles into practice through the full implementation of the guidance. To this end, they contain activities with instructions for either facilitating or participating in capacity-building workshops.

Please note, these workbooks are living documents that will evolve and improve with input from users, affected stakeholders, and interested parties. We need your participation. Please share feedback with us at policy@turing.ac.uk.

## Programme Roadmap

The graphic below visualises this workbook in context alongside key frameworks, values and principles discussed within this programme. For more information on how these elements build upon one another, refer to [AI Ethics and Governance in Practice: An Introduction].

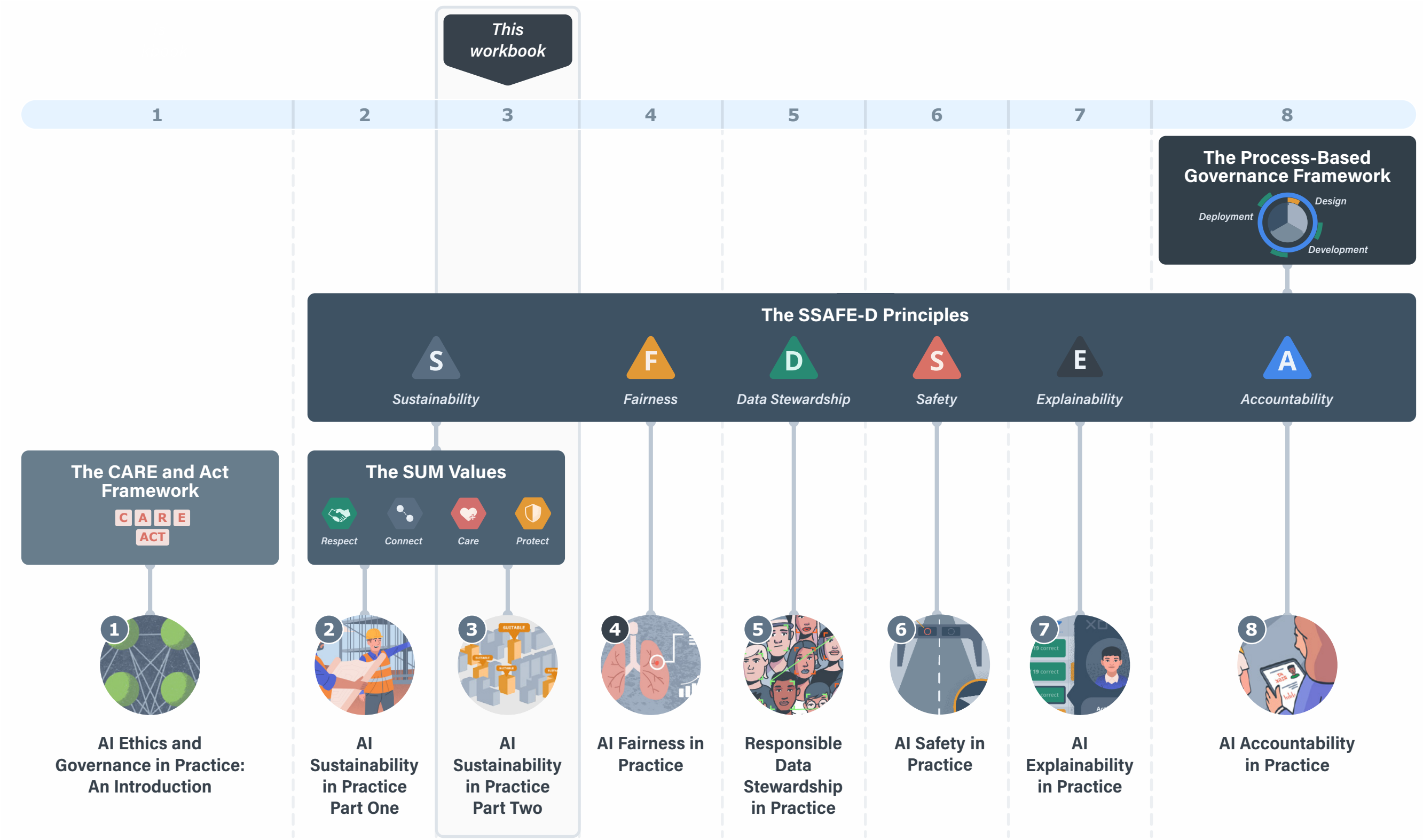

# Intended Audience

This workbook series is primarily aimed at civil servants engaging in the AI Ethics and Governance in Practice Programme - either AI Ethics Champions delivering the curriculum within their organisations by facilitating peer-learning workshops, or participants completing the programme by attending workshops. Anyone interested in learning about AI ethics, however, can make use of the programme curriculum, the workbooks, and resources provided. These have been designed to serve as stand-alone, open access resources. Find out more at turing.ac.uk/ai-ethics-governance.

There are two versions of each workbook:

- **Annotated workbooks** (such as this document) are intended for facilitators. These contain guidance and resources for preparing and facilitating training workshops.

- **Non-annotated workbooks** are intended for workshop participants to engage with in preparation for, and during, workshops.

# Introduction to This Workbook

This workbook is part two of two workbooks: Foundations for Sustainable AI Projects and Sustainability Throughout the AI Workflow. Both workbooks are intended to help faciliate the delivery of a two-part workshop on the concepts of SUM Values and Sustainability.

## AI Sustainability in Practice Part Two: Sustainability Throughout the AI Workflow

This workbook explores how to put the SUM Values and the principle of Sustainability into practice throughout the Design, Development, and Deployment Phases of the AI lifecycle. It discusses Stakeholder Impact Assessments in depth, providing tools and training resources to help AI project teams to conduct these. This workbook is divided into two sections, Key Concepts and Activities:

**Key Concepts Section**

This section discusses frameworks for establishing the foundations for sustainable AI projects:

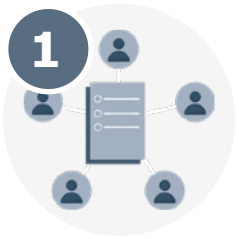

Introduction to Sustainability: Stakeholder Impact Assessments

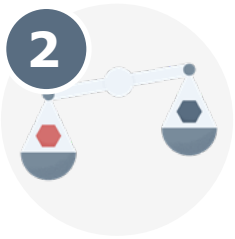

A Closer Look at Stakeholder Impact Assessments

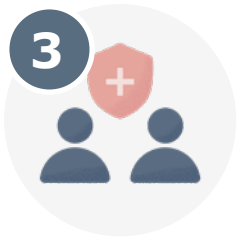

Skills for Conducting Stakeholder Impact Assessments: Consequences-Based and Values-Based Approaches to Balancing Values

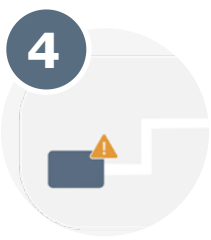

Sustainability Throughout the AI Lifecycle

## Activities Section

This section contains instructions for group-based activities (each corresponding to a section in the Key Concepts). These activities are intended to increase understanding of Key Concepts by using them.

*Case studies within the AI Ethics and Governance in Practice workbook series are grounded in public sector use cases, but do not reference specific AI projects.*

### Stakeholder Impact Assessment (Design Phase)

Practise answering key questions within Stakeholder Impact Assessments (SIAs).

### Balancing Values

Practise weighing tensions between values when assessing the ethical permissibility of AI projects by considering consequence-based and values-based approaches and engaging in deliberation.

### Revisiting Engagement Method

Practise undertaking practical considerations of resources, capacities, timeframes, and logistics as well as stakeholder needs to establish an engagement method for the following SIA.

### Stakeholder Impact Assessment (Deployment Phase)

Practise using SIAs to formulate proportional monitoring activities for the development and deployment of AI models.

**Note for Facilitators**

Additionally, you will find facilitator instructions (and where appropriate, considerations) required for facilitating activities and delivering capacity-building workshops.

AI Sustainability in Practice Part Two: Sustainability Throughout the AI Workflow

# Key Concepts

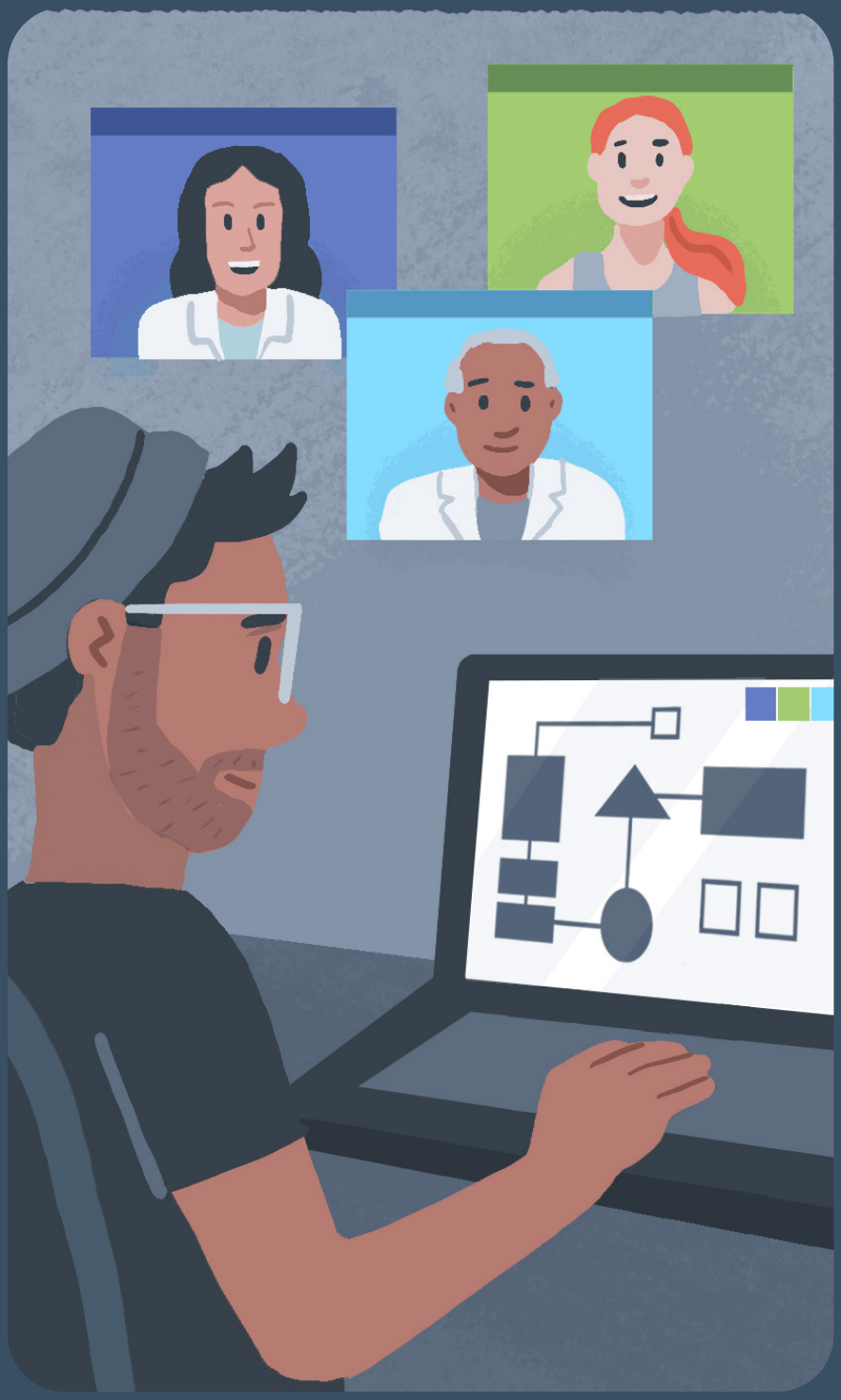

# Introduction to Sustainability: Stakeholder Impact Assessments

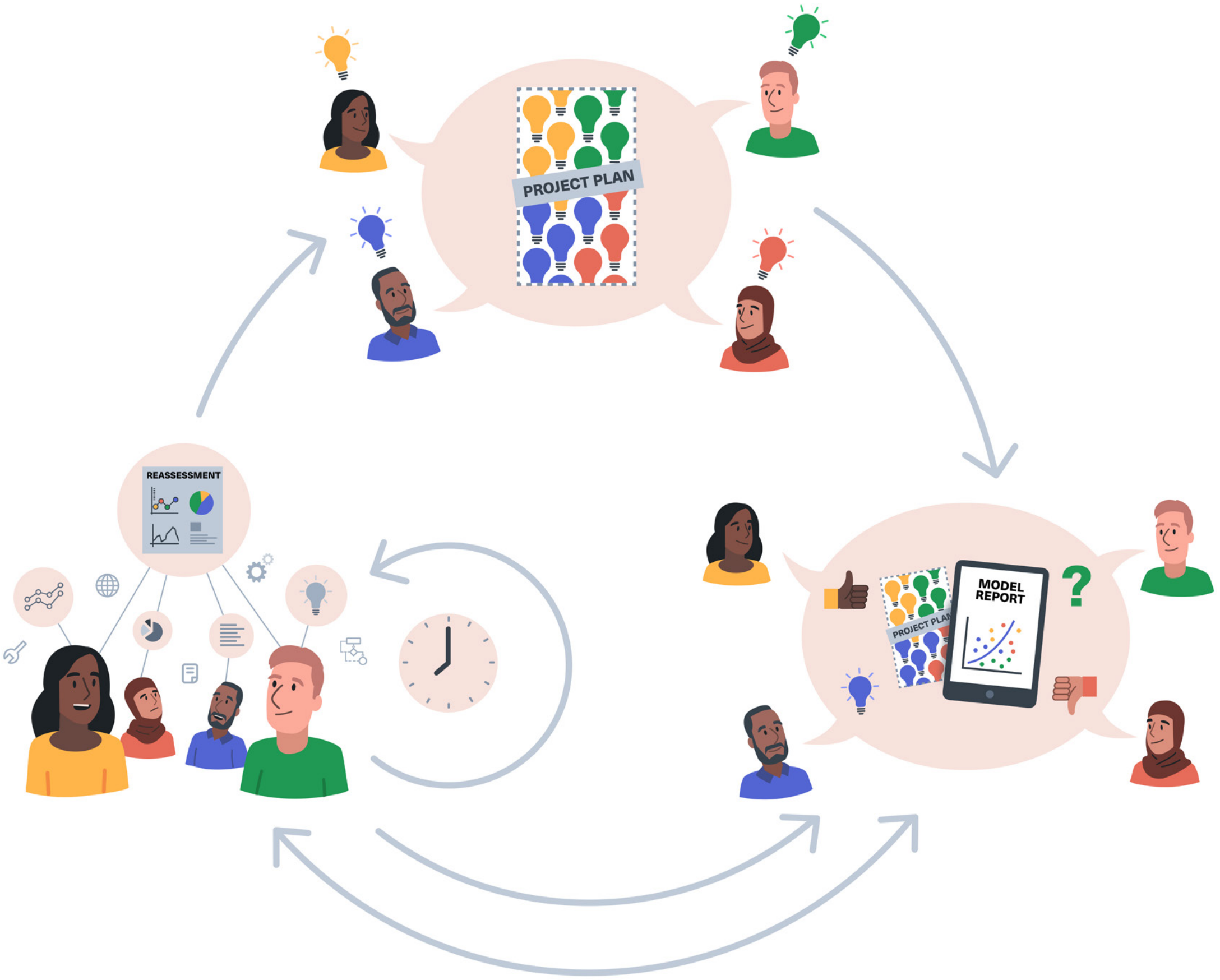

AI systems may have transformative and long-term effects on individuals and society. Designers and users of AI systems should remain aware of this. To ensure that the deployment of your AI system remains sustainable and supports the sustainability of the communities it will affect, you and your team should proceed with a continuous sensitivity to its real-world effects. You and your project team should come together to evaluate the social impact and sustainability of your AI project through a Stakeholder Impact Assessment (SIA).

The SUM Values introduced in the AI Sustainability in Practice Part One workbook form the basis of the SIA. They are not intended to provide a comprehensive inventory of moral concerns and solutions. Instead, they are a launching point for open and inclusive conversations about the individual and societal impacts of data science research and AI innovation projects. When starting a project, the SUM Values should provide the normative point of departure for collaborative and anticipatory reflection. They should also allow for the respectful and interculturally sensitive inclusion of other points of view.

KEY CONCEPT

## Stakeholder Impact Assessment (SIA)

Over the past few years, several different types of "impact assessment" have become relevant for public sector AI innovation projects. Data Protection Law requires **Data Protection Impact Assessments (DPIAs)** to be carried out in cases where the processing of personal data is likely to result in a high risk to individuals.[1] DPIAs assess the necessity and proportionality of the processing of personal data, identify risks that may emerge in that processing, and present measures taken to mitigate those risks. **Equality Impact Assessments (EIAs)** assist public authorities in fulfilling the requirements of the equality duties, specifically regarding race, gender, and disability equality. They identify the ways government can proactively promote equality.

DPIAs and EIAs provide relevant insights into the ethical stakes of AI innovation projects. However, they go only part of the way in identifying and assessing the full range of potential individual and societal impacts of the design, development, and deployment of AI and data-intensive technologies. Reaching a comprehensive assessment of these impacts is the purpose of SIAs. SIAs are tools that create a procedure for, and a means of, documenting the collaborative evaluation and reflective anticipation of the possible harms and benefits of AI innovation projects. SIAs are not intended to replace DPIAs or EIAs, which are obligatory. Rather, SIAs are meant to be integrated into the wider impact assessment regime. This demonstrates that sufficient attention has been paid to the ethical permissibility, transparency, accountability, and equity of AI innovation projects.

The purpose of carrying out an SIA is multidimensional. SIAs can serve several purposes, some of which include:

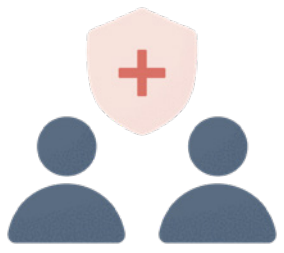

Help to build public confidence that the design and deployment of your AI system has been done responsibly.

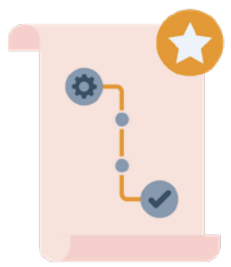

Underwrite well-informed decision-making and transparent innovation practices.

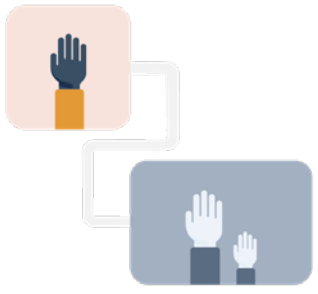

Facilitate and strengthen your accountability framework.

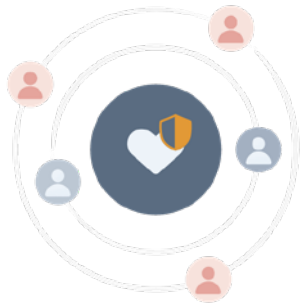

Demonstrate forethought and due diligence not only within your organisation, but also to the wider public.

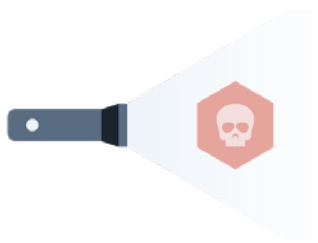

Shed light on unseen risks that threaten to affect individuals and the public good.

Your team should convene to evaluate the social impact and sustainability of your AI project through the SIA at three critical points in the project delivery lifecycle:

## SECTION 1: Design Phase
### Problem Formulation

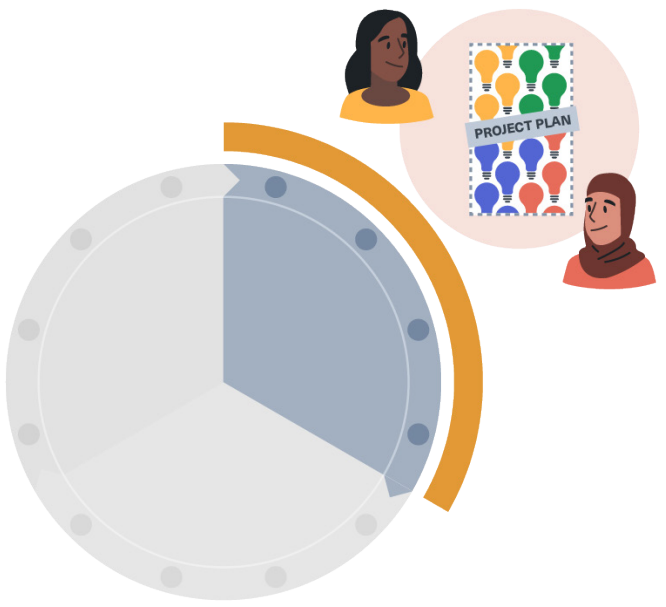

Carry out an initial SIA to determine the ethical permissibility of the project. Refer to the SUM Values as a starting point for the considerations of the possible effects of your project on individual wellbeing and public welfare. You should include stakeholder engagement objectives and methods for the Development Phase SIA established in your initial Project Summary Report (PS Report), described in the previous workbook. This will help to consider public views in ways that are proportional to potential project impacts, and appropriate to team positionality. The participation of a more representative range of stakeholders will bolster the inclusion of a diversity of voices and opinions into the design and development processes.[2] [3] [4] [5] The Design Phase SIA includes a revisitation of the PS Report. Revisions of the engagement objectives and methods, as well as other relevant revisions, should be reflected in an update to the PS Report.

## SECTION 2: Development Phase
### Model Reporting

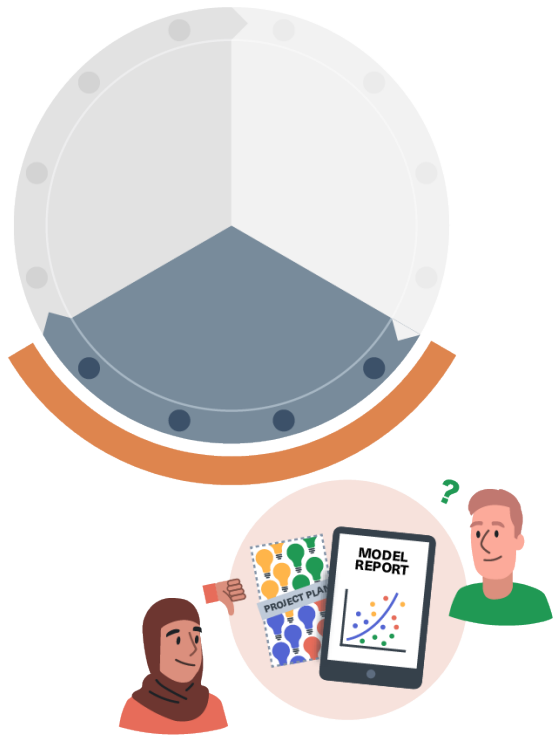

Once your model has been trained, tested, and validated, you and your team should revisit your initial SIA to confirm that the AI system to be implemented is still in line with the evaluations and conclusions of your original assessment. This check-in should be logged in the Development Phase section of the SIA with any applicable changes added and discussed. The method of stakeholder engagement that accompanies the SIA process will have been initially established in the PS Report and revisited in the Design Phase SIA. This report should be revisited again during the Development Phase SIA and updated where needed. At this point you must also set a timeframe for re-assessment once the system is in operation. The timeframes for these re-assessments should be decided by your team on a case-by-case basis, but should be proportional to the scale of potential impact that the system might have on individuals and communities it will affect.

## SECTION 3: Deployment Phase

System Use and Monitoring

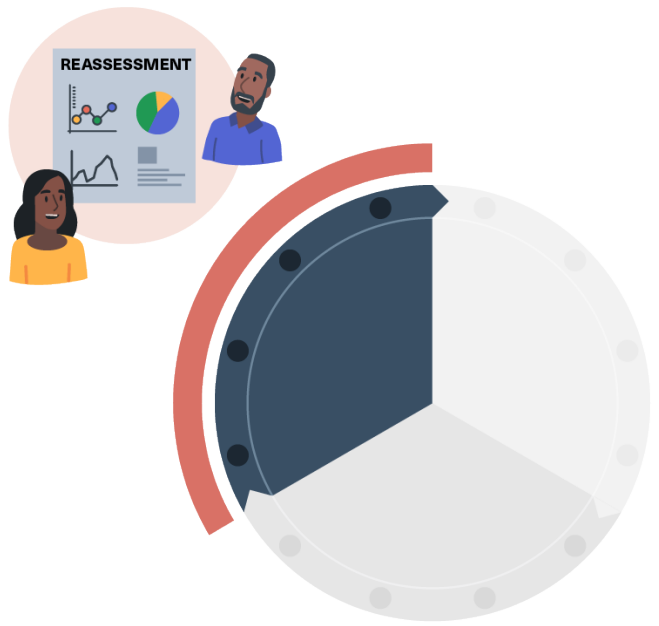

After your AI system has gone live, your team should iteratively revisit and re-evaluate your SIA. These check-ins should be logged in the Deployment Phase section of the SIA with any applicable changes added and discussed. Deployment Phase SIAs should focus on evaluating the existing SIA against real-world impacts. They should also focus on considering how to mitigate the unforeseen or unintended consequences that may have ensued in the wake of the deployment of the system. As with each SIA iteration, the PS Report should be revisited at this point, when objectives, methods, and timeframes for the next Deployment Phase SIA are established.

# A Closer Look at Stakeholder Impact Assessments

Stakeholder Impact Assessments (SIAs) provide you with the opportunity to draw on the learning and insights you have gained in your Stakeholder Engagement Processes (SEPs), and on the lived experience of engaged people, in order to delve more deeply into the potential impacts of your project. Your SIAs should enable you:

- To re-examine and re-evaluate the potential impacts you have already identified in your PS Report.
- To contextualise and corroborate these potential impacts in dialogue with stakeholders, when appropriate.
- To identify and analyse further potential impacts. By engaging in extended reflection and giving stakeholders the opportunity (where appropriate) to uncover previously unexplored harms, gaps in the completeness and comprehensiveness of the previously enumerated harms can be identified.

To illustrate how to implement an SIA, we have provided a we have provided a three-part template, with each part corresponding to a stage of the project development - from Design through to Deployment. Section 1 guides the Design Phase, addressing Project Planning, Problem Formulation, as well as revisitation of Stakeholder Analysis, Positionally Reflection, and Engagement Objectives and Methods. Section 2 provides a touchpoint for evaluation and reflection during Development Phase of models and outputs, and facilitates ongoing model reporting. Section 3 supports ongoing ethical deliberation and reflection during the Deployment Phase of resultant project outputs, recording relevant changes from earlier iterations of the SIA.

PS Report

You might find it helpful to refer back to the Project Summary Report found in Sustainability In Practice Part One, while answering these questions.

# Stakeholder Impact Assessment for: Project Name

SIA

## SECTION 1A: Design Phase (Project Planning)

General questions

Date completed: ........................

Team members involved: ..................................................

External stakeholders consulted: ..................................................

### 1. Horizon-Scanning and the Decision to Design

Have you assessed whether building an AI model or tool is the right solution to help you deliver the desired services given:

a. the existing technologies and processes already in place to solve the problem;

b. current user needs;

c. the current state of available data;

d. the resources (material and human) available to your project;

e. the nature of the policy problem you are trying to solve; and

f. whether an AI-based solution is appropriate for the complexity of its potential use contexts?

Do these initial assessments support the justifiability and reasonableness of choosing to build an AI system or tool to help you deliver the desired services?

For more details on "Assessing if artificial intelligence is the right solution" see guidance by the Office for AI and Central Digital and Data Office. For further details about understanding user needs, see Section 1 of the Data Ethics Framework and the user research section of the Gov.UK Service Manual.

- Has a thorough assessment of the human rights compliant business practices of all businesses, parties, and entities involved in the value chain of the AI product or service been undertaken? This would include all businesses, parties, and entities directly linked to your business lifecycle through supply chains, operations, contracting, sales, consulting, and partnering. If not, do you have plans to do this?

## 2. Goal-Setting and Objective-Mapping

a. How are you defining the outcome (the target variable) that the system is optimising for? Is this a fair, reasonable, and widely acceptable definition?

b. Does the target variable (or its measurable proxy) reflect a reasonable and justifiable translation of the project's objective into the statistical frame?

c. Is this translation justifiable given the general purpose of the project and the potential impacts that the outcomes of its implementation will have on the communities involved?

d. Where appropriate, have you engaged relevant stakeholders to gather input on their views about reasonableness and justifiability of the outcome definition and target variable determination?

## 3. Possible Impacts on the Individual

a. How, if at all, might the use of your AI system impact the abilities of affected stakeholders to make free, independent, and well-informed decisions about their lives? How might it enhance or diminish their autonomy?

b. How, if at all, might the use of your system affect their capacities to flourish and to fully develop themselves?

c. How, if at all, might the use of your system do harm to their physical, mental, or moral integrity? Have risks to individual health and safety been adequately considered and addressed?

d. How, if at all, might the use of your system impact freedoms of thought, conscience, and religion or freedoms of expression and opinion?

e. How, if at all, might the use of your system infringe on the privacy rights of affected stakeholders, both on the data processing end of designing the system and on the implementation end of deploying it? When appropriate, this question should supplement the completion of a Data Protection Impact Assessment.

## 4. Possible Impacts on Interpersonal Relationships, Society, and the Biosphere

a. How, if at all, might the use of your system adversely affect each stakeholder's fair and equal treatment under the law? Are there any aspects of the project that expose historically marginalised, vulnerable, or protected groups to possible discriminatory harm? These questions should supplement the completion of an Equality Impact Assessment.

b. Does the project aim to advance the interests and wellbeingof as many affected individuals as possible? Might any disparate socioeconomic impacts result from its deployment?

c. How, if at all, might the use of your system affect the integrity of interpersonal dialogue, meaningful human connection, and social cohesion?

d. How, if at all, might the use of your system affect freedom of assembly and association?

e. How, if at all, might the use of your system affect the integrity of the information ecosystem, the right to diverse and reliable information, and access to a plurality of ideas and perspectives?

f. How, if at all, might the use of your system affect the right of individuals and communities to participate in the conduct of public affairs?

g. How, if at all, might the use of your system affect the right to effective remedy for violation of rights and freedoms, the right to a fair trial and due process, the right to judicial independence and impartiality, and equality of arms?

h. Have the values of civic participation, inclusion, and diversity been adequately considered in articulating the purpose and setting the goals of the project? If not, how might these values be incorporated into your project design?

i. Have you sufficiently considered the wider impacts of the system on future generations and on the planet and biosphere as a whole?

j. How could the use of the AI system you are planning to build or acquire—or the policies, decisions, and processes behind its design, development, and deployment—lead to the discriminatory harassment of impacted individuals?

k. How could the use of the AI system you are planning to build or acquire—or the policies, decisions, and processes behind its design, development, and deployment—lead to the disproportionate adverse treatment of impacted individuals from protected groups on the basis of their protected characteristics?

l. How could the use of the AI system you are planning to build or acquire—or the policies, decisions, and processes behind its design, development, and deployment—lead to the discriminatory harassment of impacted individuals?

## SECTION 1B: Design Phase (Problem Formulation)

Sector-Specific and Use Case-Specific Questions

Date completed: ........................ Team members involved: ........................ External stakeholders consulted: ........................

In this section, you should consider the sector-specific and use case-specific issues surrounding the social and ethical impacts of your AI project on affected stakeholders. Compile a list of the questions and concerns you anticipate. State how your team is attempting to address these questions and concerns. Where appropriate, engage with relevant stakeholders to gather input about their sector-specific and use case-specific concerns.

## SECTION 1C: Design Phase

Revisiting Project Summary Report

Date completed: ........................ Team members involved: ........................ External stakeholders consulted: ........................

### 1. Revisiting Stakeholder Analysis and Positionality

a. Do the stakeholder groups outlined in the report accurately reflect current stakeholders of this project? Are there other stakeholder groups that should be considered.

b. Do the potential impacts outlined in the report accurately reflect current SIA results?

c. Do the stakeholder groups currently identified as salient represent those groups that are currently likely to be most differentially impacted, vulnerable, or marginalised?

d. Does the team positionality reflection accurately represent the relationship between team members and stakeholders at this stage in the project?

### 2. Revisiting Engagement Objectives and Methods

a. Considering the results of the SIA, are there any new potential project impacts that may lead you to reconsider your engagement objectives and methods? If so, how?

b. Do your chosen engagement objectives and methods seem proportional to the current identified impacts?

c. Do any adjustments need to be made to your chosen engagement objectives and methods given the SIA results? If so, are there any additional practical considerations that need to be addressed to ensure that your engagement objectives and methods are realised?

## 3. Revisiting the Process-Based Governance (PBG) Framework

**a.** Considering SIA results, does the PBG Framework for this project still accurately reflect the human chain of responsibility and create the baseline conditions for the project team to be actively accountable for system impacts? (For further details on the PBG Framework, see Workbook 8, AI Accountability in Practice.)

## SECTION 2: Development Phase

Model Reporting

Date completed: ........................

Team members involved: ....................................................

External stakeholders consulted: ....................................................

After reviewing the results of your initial SIA, answer the following questions:

**a.** Are the trained model's actual objective, design, and testing results still in line with the evaluations and conclusions contained in your original assessment? If not, how does your assessment now differ?

**b.** Have any other areas of concern arisen with regard to possibly harmful social or ethical impacts as you have moved from the Design to the Development Phase?

Re-Assess Questions in the Project Summary Report.

You must also set a reasonable timeframe for Public Consultation and Development Phase re-assessment:

**Dates of Public Consultation on Development Phase Impact Revisitation:**

....................................................

....................................................

**Date of Planned Development Phase Re-Assessment:**

....................................................

....................................................

## SECTION 3: Deployment Phase
System Use and Monitoring

Date completed: ........................ Team members involved: ........................ External stakeholders consulted: ........................

Once you have reviewed the most recent version of your SIA and the results of the public consultation, answer the following questions:

**a.** What steps can be taken to rectify any problems or issues that have emerged?

**b.** Have any unintended harmful consequences ensued in the wake of the deployment of the system? If so, how might these negative impacts be mitigated and redressed?

**c.** Have the maintenance processes for your AI model adequately taken into account the possibility of distributional shifts in the underlying population? Has the model been properly retuned and retrained to accommodate changes in the environment?

Re-Assess Questions in the Project Summary Report.

You must also set a reasonable timeframe for Public Consultation and Deployment Phase re-assessment:

**Dates of Public Consultation on Deployment Phase Impacts:**

........................

........................

**Date of Next Planned Deployment Phase Re-Assessment:**

........................

........................

# Skills for Conducting Stakeholder Impact Assessments

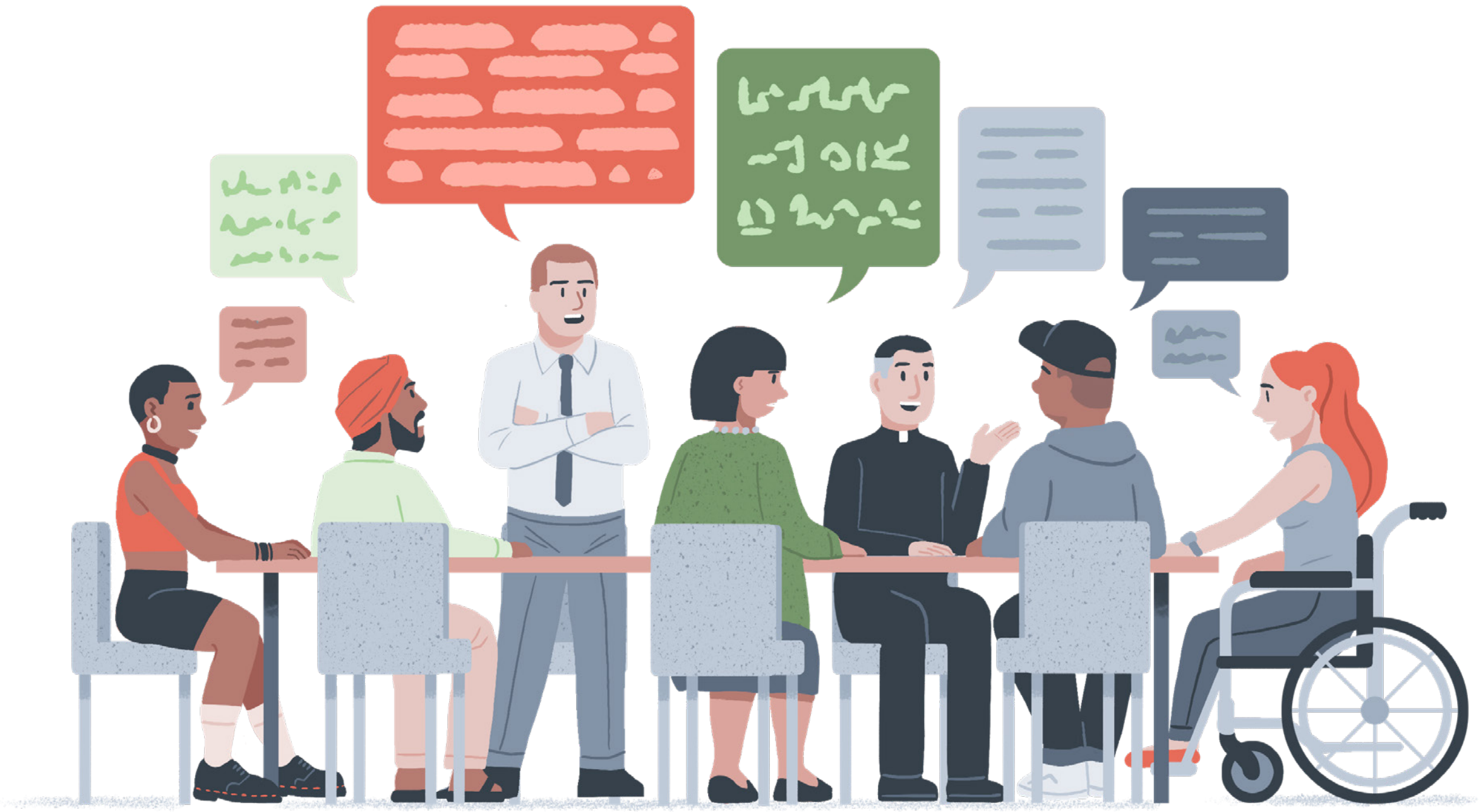

## Weighing The Values and Considering Trade-Offs

Different AI projects may give rise to circumstances when the SUM Values come into tension with each other. As part of conducting SIAs, there should be discussion around how to weigh values against one another when they do not align with one another or conflict. Discussions should include considerations about potential trade-offs between values. For instance, there may be circumstances where the use of an AI system could optimally advance the public interest only at the cost of safeguarding the wellbeing or the autonomy of a given individual. In other cases, the use of an AI system could preserve the wellbeing of a particular individual only at the cost of the autonomy of another, or of the public welfare more generally.

The issue of adjudicating between conflicting values has long been a crucial and thorny dimension of collective life. The problem of discovering reasonable ways to overcome the disagreements that arise as a result of the plurality of human values has occupied thinkers for just as long. Nonetheless, over the course of the development of modern democratic and plural societies, several useful approaches to managing the tension between conflicting values have emerged.

# Consequences-Based and Principles-Based Approaches to Balancing Values

We can find a concrete and agent-centred approach to managing the tension between conflicting values in two of the standard schools of modern ethics:

- consequences-based moral thinking or consequentialism; and
- principles-based moral thinking or deontology.

These positions offer tools for thinking through a given dilemma in weighing values.*

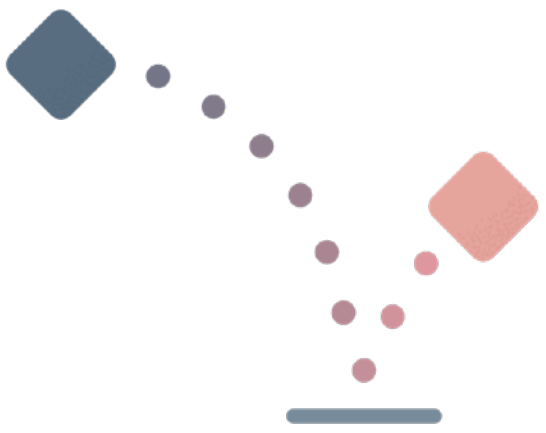

## A Consequences-Based Approach

A consequences-based approach asks that, in judging the moral correctness of an action, you prioritise considerations of the goodness produced by an outcome. In other words, the consequences of your actions and the achievement of your goals matter most. The goodness of these consequences should be maximised. In this view, standards of right and wrong (indicators of what one ought to do) are determined by the goal served as a result of an action taken, rather than by the principles or standards one applies when acting.

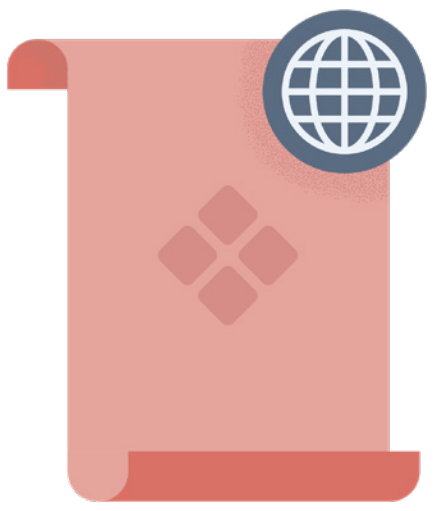

## A Principles-Based Approach

A principles-based approach takes the opposite track. From this standpoint, the rightness of an action is determined by the intentional application of a universally applicable standard, maxim, or principle. This approach does not base the morality of conduct on the ends served by it. Instead, it anchors rightness in the duty or obligation of the individual agent to follow a rationally determined (and therefore "universalisable") principle. Deontological or principles-based ethics holds that the integrity of the principled action and intention matters most, and such constraints must be put on the pursuit of the achievement of one's goals when the actions taken as means to achieve these ends come into conflict with moral standards.

* Learn more about ethics and governance in Leslie, D., & Fischer, C. (2023). Introduction to Normative Ethical Theories. In *AI Ethics and Governance (Turing Commons Skills Track)*. The Alan Turing Institute. https://alan-turing-institute.github.io/turing-commons/skills-tracks/aeg/chapter1/normative/

Knowing when to prioritise consequences and when to prioritise principles in moral deliberations is a tricky matter. This may make sense depending upon the context.

To take a familiar example, lying to a murderer who appears at your front door would save an innocent victim whom you are concealing in your cellar. The prioritisation of consequences makes more sense than the prioritisation of the principle of not lying. However, in another situation, the principle matters. For instance, where you would be constrained, on principle, from deceiving others by taking credit for someone else's work in order to advance in your job.

## PRIORITISING CONSEQUENCES

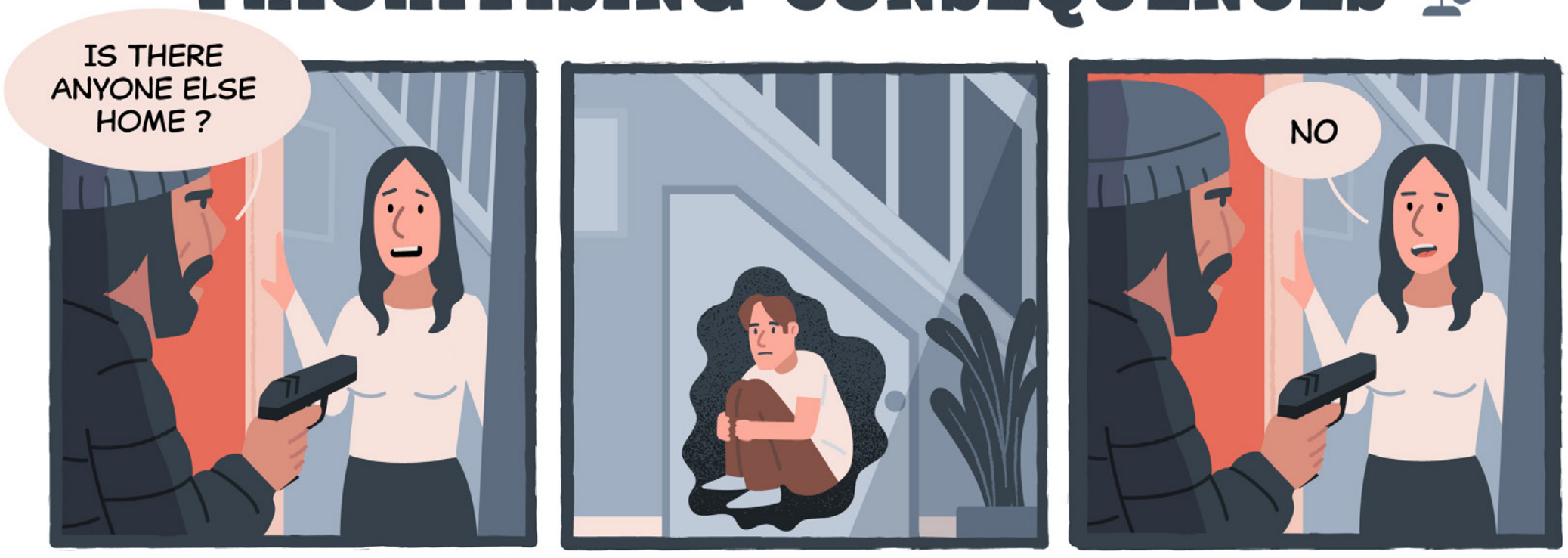

## PRIORITISING PRINCIPLES

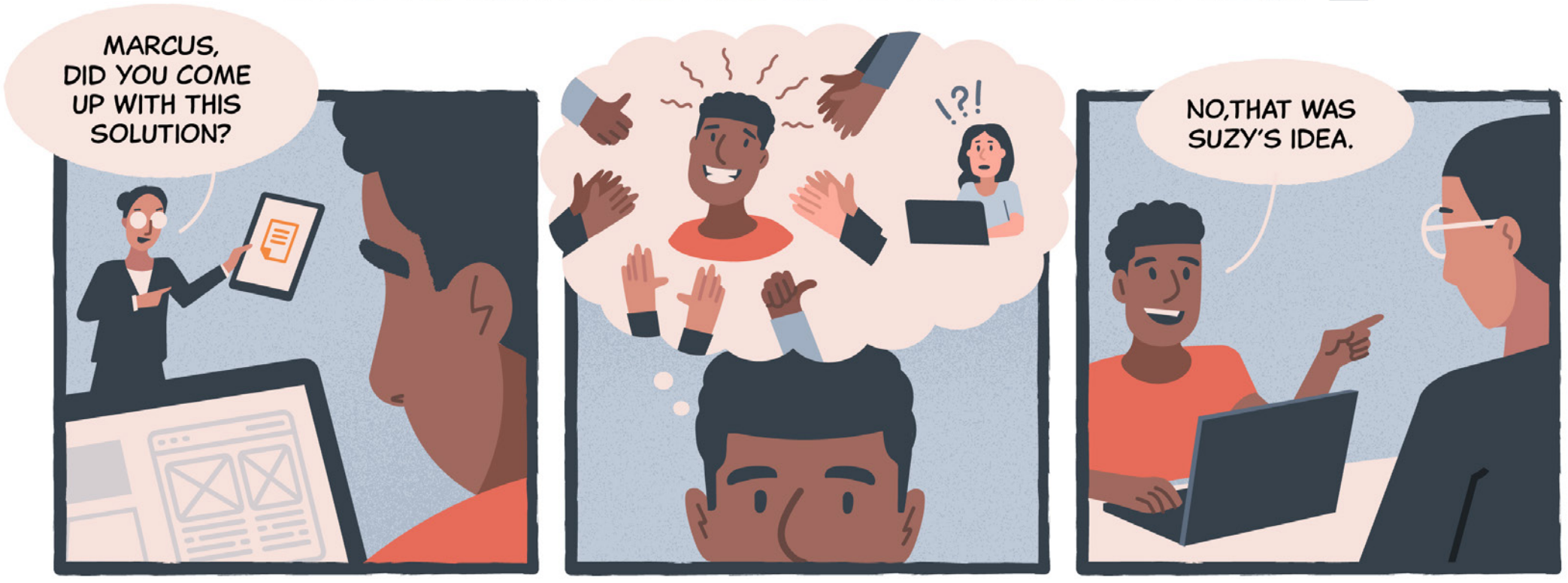

Consider a more directly relevant example. In an overburdened sector, the introduction of an automated system for making building sites available for development would vastly expedite housing delivery. The implementation of this AI system would thus produce a consequence that could be beneficial to the public, protecting public good.[6]

Yet, it may, among other things, simultaneously do damage to the value of Connect. This value safeguards interpersonal dialogue, meaningful human connection, and social cohesion. The implementation of the AI system would eliminate time intensive consultation processes that contribute to interpersonal communication, trust building, and social bonding between council staff and residents.[7]

How then could one go about weighing the value of improving public welfare against the value of respecting the integrity of interpersonal relations?

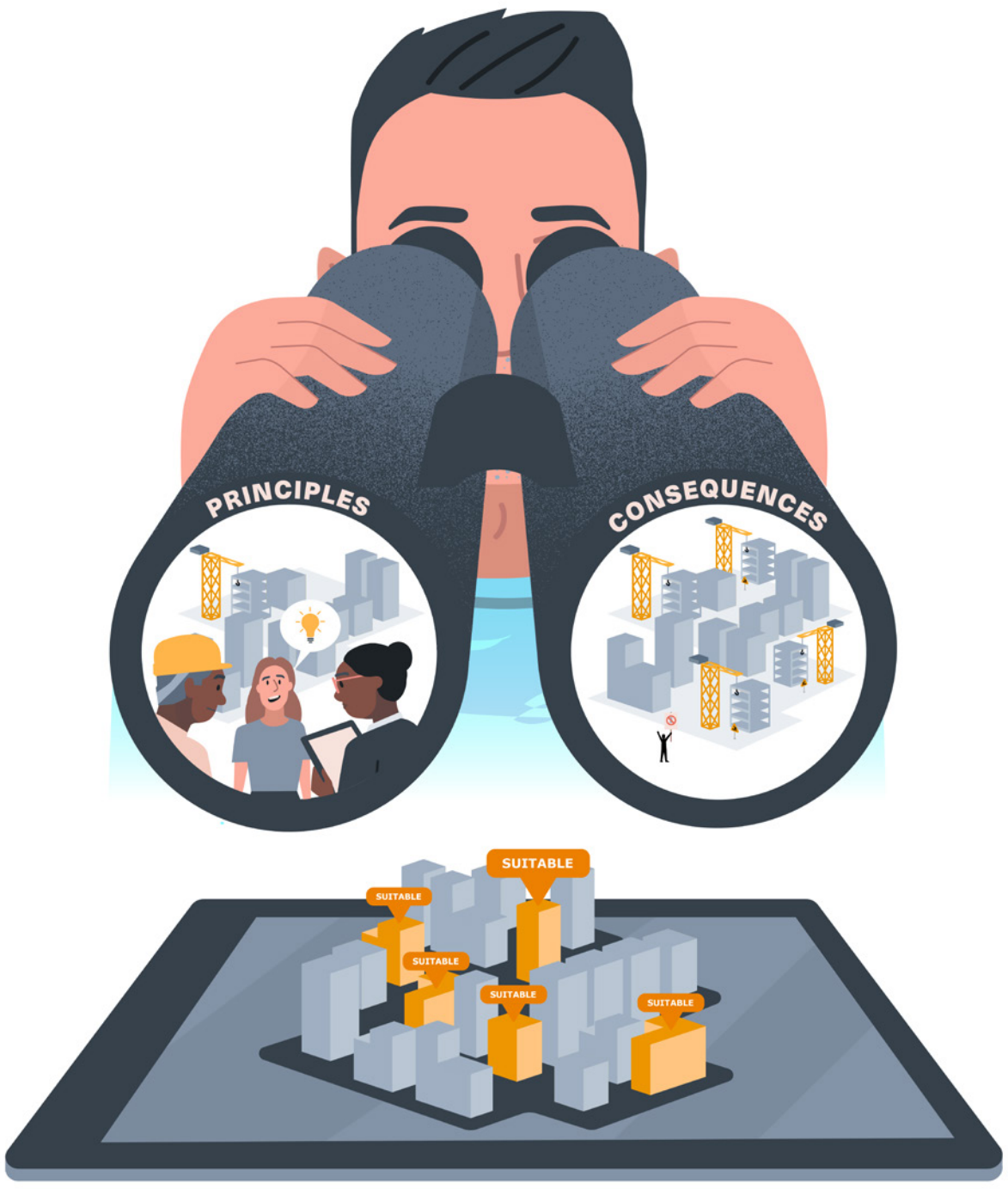

One way would be to place each side of this comparison under the rubric of either consequences or principles, and then measure them up against each other accordingly. From one perspective, the publicly beneficial consequences of improving service delivery might outweigh the publicly harmful consequences of impairing social cohesion. From another perspective, such a trade-off would be unacceptable, because the principle of respecting the integrity of social cohesion trumps any solidarity-harming but publicly beneficial consequences whatsoever.

Getting clear on the consequences and the principles involved in a specific case of conflicting values will allow you to get a better picture of the practical and moral stakes at play in a particular project. It will also help you gain a sharper idea of the proportionality of using of an AI technology to achieve a desired outcome given both its potential ethical impacts and the social needs to which it is responding. When drawn upon for guidance, consequentialism and deontology can provide you with a procedural scale upon which to place, measure, and weigh conflicting values. They are practical tools that can be used to enable meaningful deliberation.

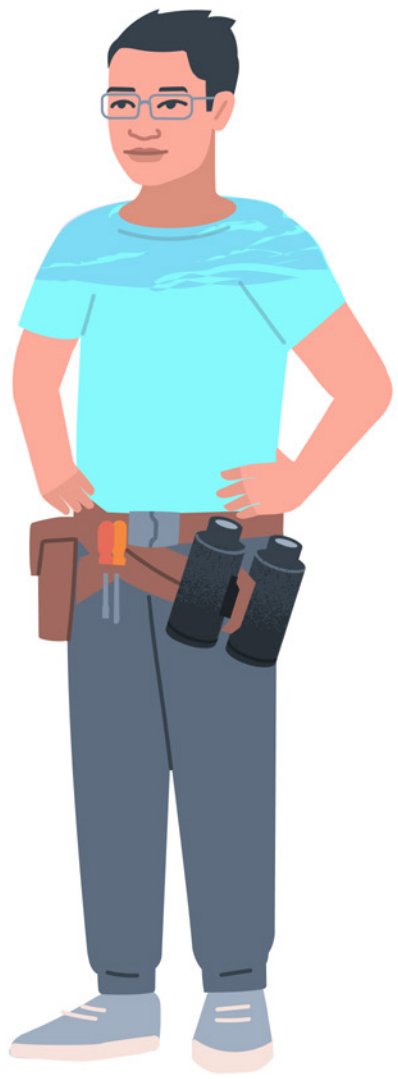

# Ensuring Meaningful and Inclusive Deliberation

The most general approach to ensuring meaningful and inclusive deliberation is to encourage respectful, sincere, and well-informed dialogue. Through this approach, reasons offered from all affected voices can be heard and considered. Deliberations that have been inclusive, open, and impartial tend to generate better and more sound conclusions. Approaching the adjudication of conflicting values in this manner will likely improve mutual understanding of the rationales and perspectives which inform those values.[8] The importance of cultivating a culture of innovation, which encourages respectful, open, non-coercive, and accountable communication, must be stressed. The success of the modern sciences has been built on the dynamic foundations of inclusive, rational, and democratic communication. This is perhaps evidence enough to support the validity of this emphasis.

The rational exchange and assessment of ideas and beliefs plays a central role in meaningful dialogue about balancing values. The validity of the claims we make in conversations about values is bounded by practices of giving, and asking for, reasons. A claim about values that is justified is one that convinces by the unforced strength of the better or more compelling argument. Rational justification and persuasive reason-giving are, in fact, central elements of legitimate and consensus-oriented moral decision-making. And, along the same lines, claims made about moral value or properties need to be carefully evaluated in terms of their inferential strengths and weaknesses.

### Meaningful Dialogue and Grice's Maxims[9]

In 1975, the British philosopher of language, Paul Grice, formulated what he called the "cooperative principle" to capture the assumptions that 'interacting people hold when engaging in meaningful conversation'. He broke these assumptions down into four maxims:

**Maxim of Quantity**
Make your contribution as informative as is required, providing the necessary information but no more.

**Maxim of Quality**
Be truthful, and do not provide information that is misleading, false or that is not evidence-based.

**Maxim of Relation**
Be relevant.

**Maxim of Manner**
Be clear. Avoid ambiguity. Avoid obscurity of expression. Be brief. Be orderly.

There is another way to understand the importance of meaningful dialogue in balancing values. This alternative focuses on the enabling conditions of meaningful deliberation. It focuses on how an inclusive and open exchange of reasons about balancing or prioritising values can be made possible without imposing substanive views about the values themselves. Here, an emphasis on rational communication in deliberations looks to secure a justified and equitable process of exchanging and evaluating reasons. It starts with the question: what are the preconditions and assumptions of meaningful communication that allow people, who are exchanging views on their values and beliefs, to come rationally acceptable moral judgments and reason-based agreement?

To answer this question, moral thinkers over the past century have endeavoured to reconstruct the practical assumptions behind, and presuppositions of, rational communication (a summary of the most essential of such assumptions and presuppositions is provided below).[10] [11] [12] [13] [14] [15] [16] [17] [18] [19] [20] Creating a reflective and practicable awareness of these assumptions and presuppositions among members of your team can play a crucial role in creating an innovation environment that is optimally conducive to meaningful and inclusive deliberation:

## Preconditions of Meaningful Deliberation

### Impartiality

Interlocutors engaging in meaningful deliberation must consider the interests of all those who are affected by their actions equally. Thinking impartially involves taking on the view of others to try to put oneself in their place.

### Non-Coercion

Meaningful deliberation must be free from any sort of implicit or explicit coercion, force, or restriction that would prevent the open and unconstrained exchange of reasons.

### Sincerity

Meaningful deliberation must be free from any sort of deception or duplicity that would prevent the authentic exchange of reasons. Interlocutors must mean what they say.

### Consistency and Coherence

Arguments and positions offered in meaningful deliberation must be clear, free from contradictions, and hold together collectively in an understable way.

### Mutual Respect and Egalitarian Reciprocity

All interlocutors must be treated with respect and given equal opportunity to contribute to the conversation. All voices are worthy of equal consideration in processes of exchanging reasons.

### Inclusiveness and Publicity

Anyone whose interests are affected by an issue and who could make a contribution to better understanding it must not be excluded from participating in deliberation. All relevant voices must be heard and all relevant information considered.

# Addressing and Mitigating Power Dynamics that May Obstruct Meaningful and Inclusive Deliberation

The stewardship of meaningful and inclusive dialogue is critical to safeguarding the collective weighing up of values. However, there is an important potential barrier to meaningful deliberation that challenges its feasibility and must be addressed. As guiding assumptions of rational communication, norms like sincerity, impartiality, non-coercion, and inclusiveness may strike some as overly idealistic. In the real world, discussions are rarely fully inclusive, informed, and free of assertive manipulation, coercion, and deception. Rather, deliberation and dialogue are often steered by, and crafted to protect, the interests of the dominant.[21] Likewise, differential power relationships (for instance, divergent educational backgrounds that derive from differential socioeconomic privileges) create power imbalances that fundamentally challenge the conditions of reciprocity and equal footing that are needed for justified and equitable communication.[22] [23] [24]

Your project team should confront these obstacles to meaningful deliberation head-on through a power-aware approach to facilitating collaborative reflection, dialogue, and engagement. By kindling an awareness of, and sensitivity to, the differential relationships of power that can suppress the full participation of disadvantaged or marginalised voices, you can better encourage an inclusive, open, and equal opportunity conversation between participants.[25] [26] Clear-headed explorations of power dynamics between civil servants, scientists, citizens, domain experts, and policymakers can assist you in avoiding the kind of deficiencies of representation and empowerment that risk reinforcing existing power structures and inequalities.[27] [28] This may involve active mitigation measures like the provision of training, upskilling, and technical resources to those who have lacked access to them. Above all, the norms of meaningful deliberation make you aware of the possible distortions of communication (i.e. a lack of egalitarian reciprocity, non-coercion, sincerity, etc.) that must be tackled and rectified for the hurdles of power disparities to be scaled.[29]

# Sustainability Throughout the AI Lifecycle

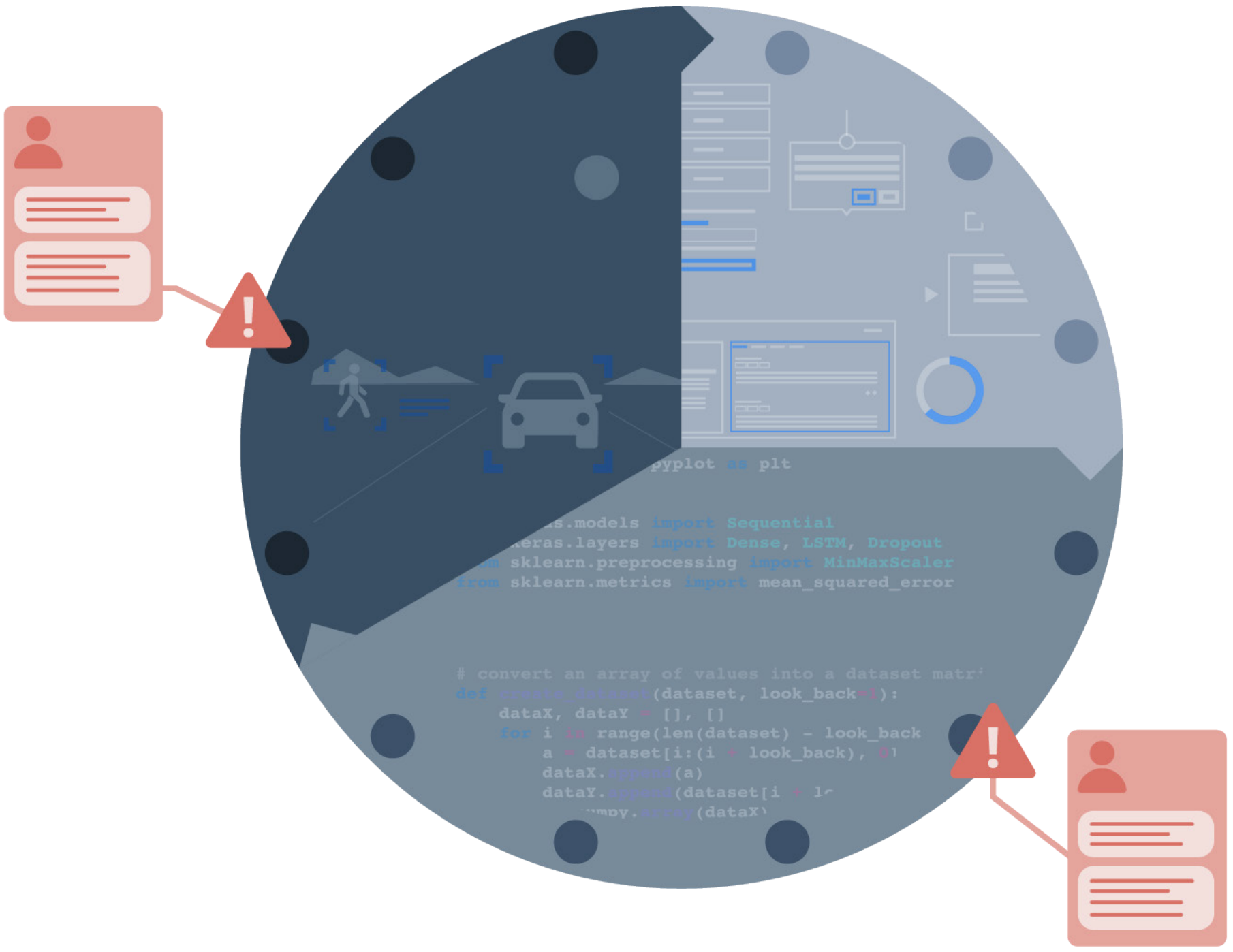

## The Need for Responsiveness Across the AI Lifecycle

In its general usage, the word 'sustainability' refers to the maintenance of, and care for, an object or endeavour over time. In the AI innovation context, this implies that building sustainability into your project is not a "one-off" affair. Carrying out a SIA at the beginning of an AI innovation project is a critical step. However, it is only a first step in a much longer, end-to-end process of responsive evaluation and re-assessment.

> **!** SIAs must pay continuous attention both to the dynamic and changing character of AI production and implementation lifecycles, and to the shifting conditions of the real-world environments in which AI models are embedded.

There are two sets of factors that necessitate this demand for responsiveness in sustainable AI innovation:

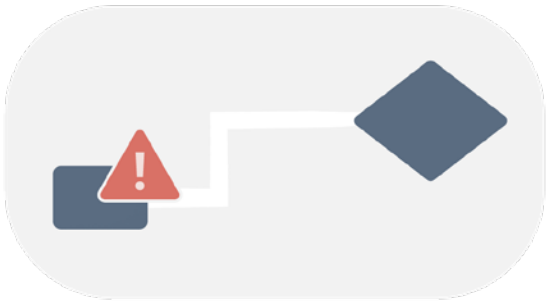

### 1. Production and Implementation Factors

Choices made at any point along the design, development, and deployment workflow may impact prior decisions and assessments. This leads to a need for re-assessment, reconsideration, and amendment. For instance, design and development choices could be made that were not anticipated in the initial impact assessment. Such choices might include adjusting the target variable, choosing a more complex algorithm, or grouping variables in ways that may impact specific groups. Changes may influence how an AI system performs or how it impacts affected individuals and groups. Processes of AI model design, development, and deployment are also iterative and frequently bi-directional. This often results in the need for revision and update. For these reasons, sustainable AI design, development, and use must remain agile, attentive to change, and at-the-ready to move back and forth across the decision-making pipeline as downstream actions affect upstream choices and evaluations.

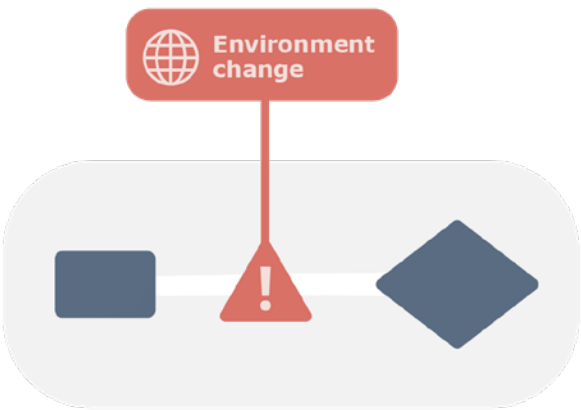

### 2. Environmental Factors

During the time in which the system is in production or use, changes in project-relevant social, regulatory, policy, or legal environments may occur. These changes may have a bearing on how well the model works and on how the deployment of the system impacts affected individuals and groups. Likewise, domain-level reforms, policy changes, or changes in data recording methods may take place in the population of concern. This could affect whether the data used to train the model accurately portrays phenomena, populations, or related factors in an accurate manner. In the same vein, cultural or behavioural shifts may occur within affected populations that alter the underlying data distribution. This can hamper the performance of a model, which has been trained on data collected prior to such shifts. All of these alterations of environmental conditions can have a significant effect on how an AI system performs. They can also have a significant effect on the way it impacts affected individual and communities.

# Example in Focus: Challenges to AI Sustainability in Children's Social Care

In the context of children's social care (CSC), the performance and impact of AI models over time can be influenced by amendments to underlying laws or procedures (i.e. changes to legal thresholds or definitions), population shifts, and alterations in social work practices and protocols.[30] Where new reforms or changes to law or service delivery procedures have taken place, the performance of predictive risk models, whose fit to the data distribution is based on prior/outdated social and legal structures, could potentially even undermine such reforms. For instance, a shift in the policy that determines the procedural steps involved in taking a child into care may work on a different set of assumptions about, or definition of, risk than that which was programmed into a predictive risk model designed before such a policy change took place. This could lead to the model underestimating or overestimating risk in a manner that is at cross-purposes with the policy reform and that harms children and families.

Changes and improvements in CSC services can likewise adversely affect the predictive qualities of models trained on data from the past. For example, whereas a parent's placement in foster care as a child and a child's prior contact with child services have both been found to be predictive of child abuse or neglect, the predictive power of such input features rests on the assumption (backed by statistics) that foster care and child services interventions have been ineffective in the past. Successful reforms that improve the effectiveness of these services would diminish the predictive power of these variables, so a model that retains the inferences from the prior data will end up identifying risks inaccurately and could inequitably impact affected decision recipient(s).

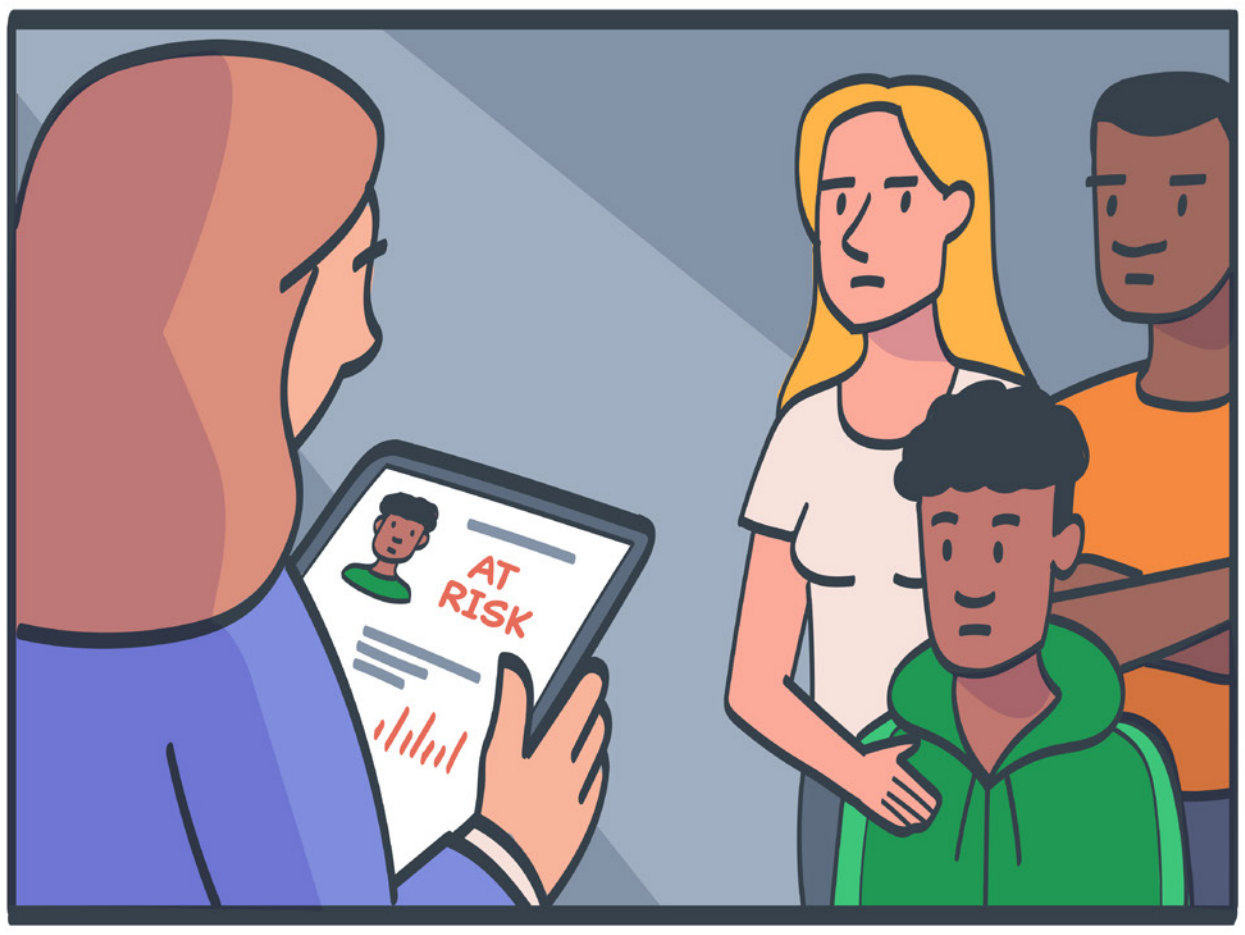

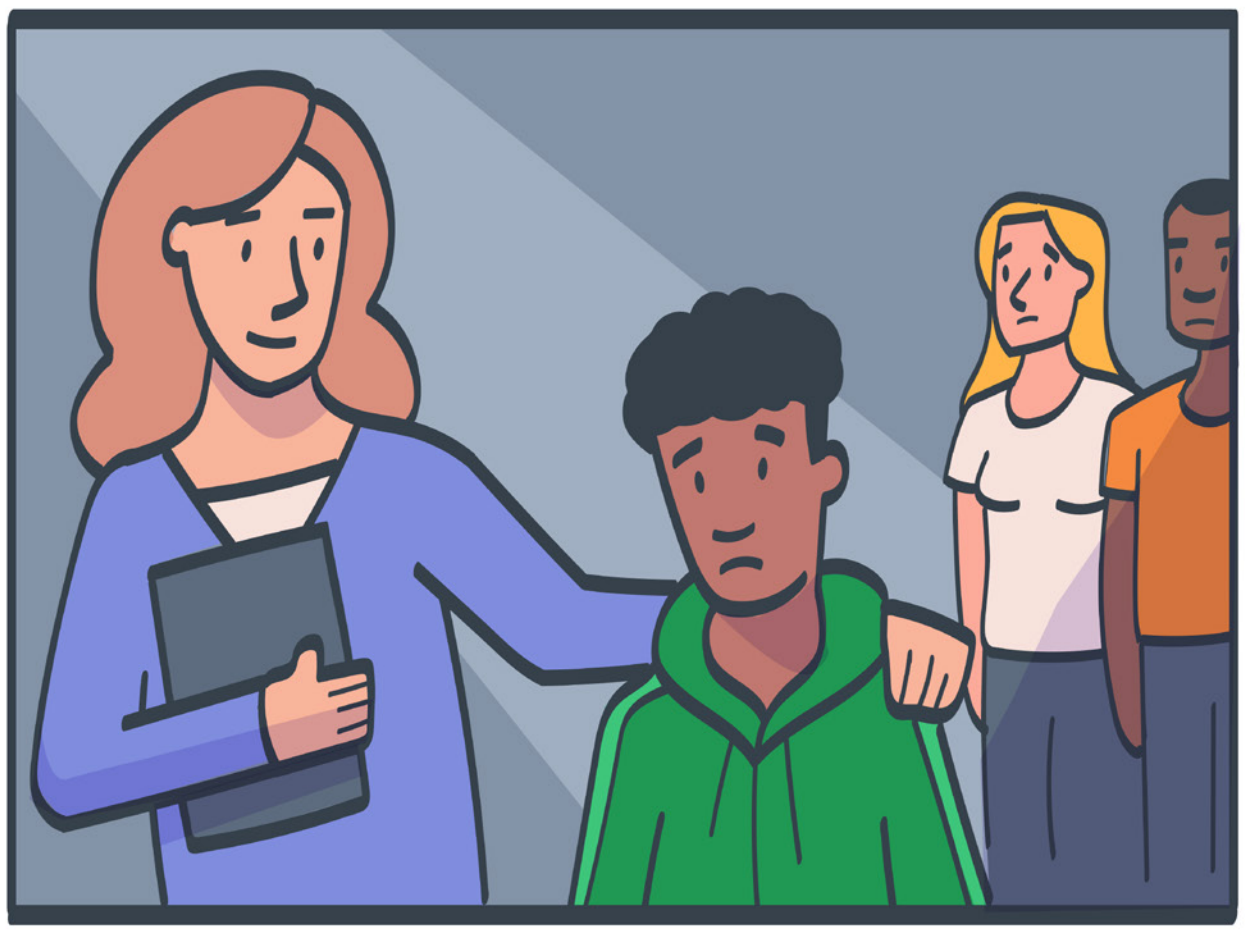

Similarly, the thresholds for access to CSC, as applied by different local authorities, are often adapted to the changing demand for services and to resource availability.[31] They are also influenced by changes in the central government's policy strategies. These are often external to, and independent of, the policy strategies of the local authorities providing CSC services. Furthermore, thresholds for access are frequently influenced by the outcome of the Office for Standards in Education inspections. All these factors are subject to unpredictable and frequent transformation. AI models used in CSC, especially ones deployed for risk assessments, should be continually monitored, re-assessed for potential stakeholder impacts, and updated whenever policies, procedures, and practices are changed.

The effectiveness of your project team's ability to bring AI sustainability into practice will largely hinge on the governance actions and procedures you set up to ensure that the AI innovation workflow is sufficiently responsive to changing production, implementation, and environmental factors. These procedures and mechanisms should involve both the public-facing, engagement dimension of your project and internal processes of reassessment, updating, monitoring, and deprovisioning.

## Proportional Governance of Engagement Goals and Methods

When significant changes occur in the production, implementation, and environmental factors over the course of the AI innovation lifecycle, your team will have to re-evaluate its engagement objectives and methods. This will ensure that affected stakeholders are appropriately consulted and involved. This means that impacted stakeholder groups are re-engaged to provide input on the relevant changes. It may also mean that your team chooses different engagement methods to align the level of engagement with the scale of impact that the changes may generate. For details on assessing appropriate levels of participation, please consult the Assessing Stakeholder Engagement Needs section of the AI Sustainability in Practice Part One workbook.

## Deployment Phase Re-Assessment and Other Necessary Monitoring, Updating, and Deprovisioning Activities

A pre-implementation re-assessment of your initial impact assessment is included as a second part of the SIA. This re-assessment directs you and your team to revisit your initial SIA to confirm that the trained or completed AI system is still in line with the evaluations and conclusions of your original assessment. It is at this juncture of the workflow that any changes occurring in production or environmental factors surrounding the project should first be identified, discussed, and addressed. Beyond this, a reasonable timeframe for monitoring, re-assessment, and updating (once the model is in operation) should be established that accords with the specific use-context and domain in which the system will operate as well as the scale of its potential impacts. Likewise, a plan should be generated, setting out details of the timeframes and procedures for re-assessment.

The re-training of the model may be planned around these timeframes to help maintain a high-level of performance. This type of updating can use the original model as a starting point, in order to retune the model's parameters or, where appropriate, to drop certain features that are no longer predictive. However, there is also the option of entirely deprovisioning (i.e. stopping use of) the model and system if performance simply drops too low to be addressed by mere re-training. Deprovisioning may not always mean simply removing a system. An existing, but retired, project may serve as a foundational input or constraint into the planning stages of a new project—starting the cycle once more.

At all events, Deployment Phase re-assessment and other necessary monitoring, updating, and deprovisioning activities should be determined by:

- the specific use-context of the systems;
- changes in production and environmental factors that may influence the system's performance; and
- changes in the scale or scope of system impacts.

AI Sustainability in Practice Part Two: Sustainability Throughout the AI Workflow

# Activities

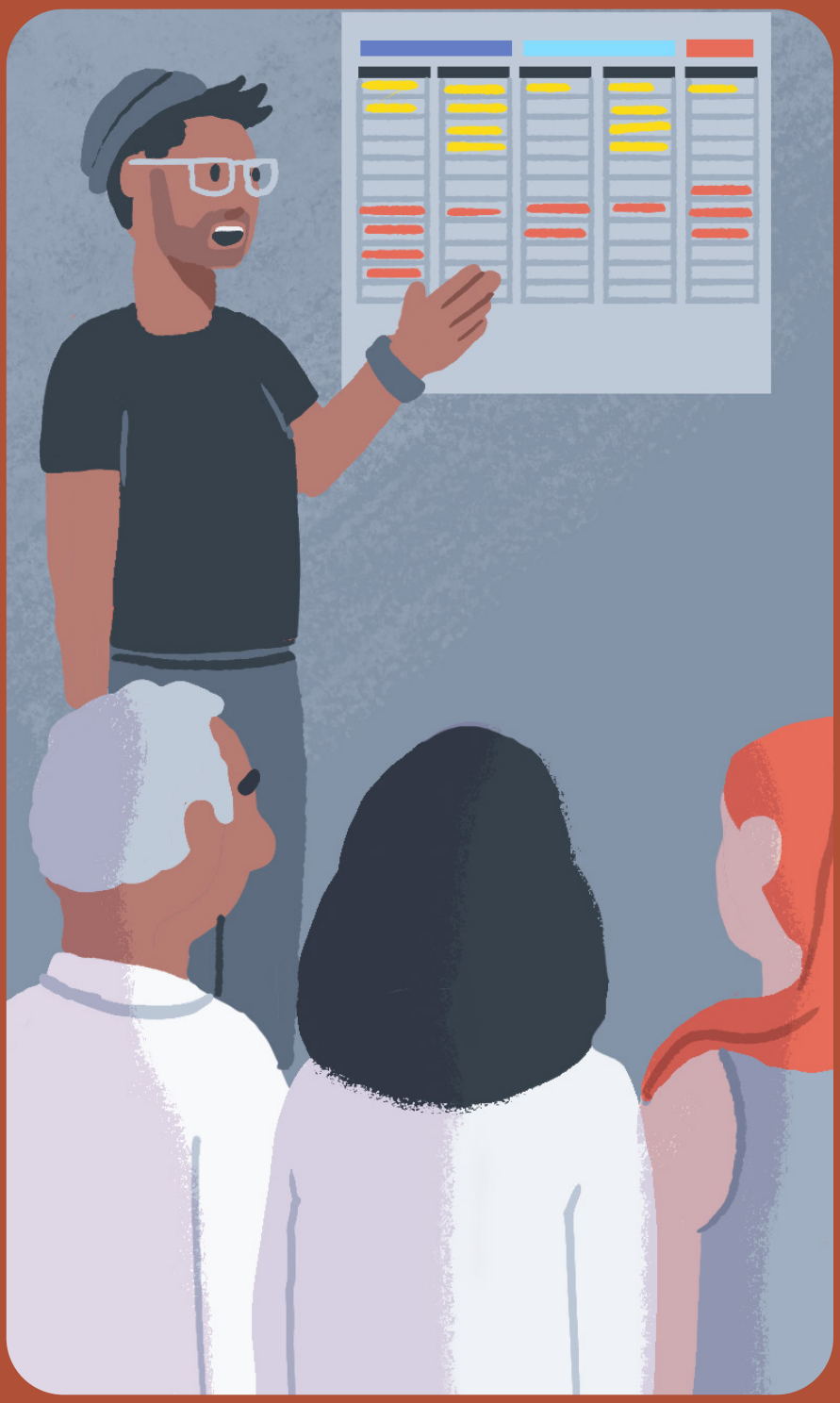

# Activities Overview

In the previous sections of this workbook, we have presented an introduction to the core concepts of AI Sustainability. In this section we provide concrete tools for applying these concepts in practice. Activities related to AI Sustainability in Practice Part Two will help participants conduct and respond to SIAs throughout the design and development of AI systems. Your team will continue engaging with the interactive case study presented in the previous workshop, playing the role of a local authority developing an AI model aimed to identify suitable building sites for housing development. Your team will plan stakeholder engagement activities, schedule impact assessments, and determine how to incorporate results from these engagements and assessments. **These activities are to be conducted following the completion of activities** from AI Sustainability in Practice Part One. Although new participants may join this session, outputs from the previous workshop board are necessary materials for the delivery of this workshop.

We offer a collaborative workshop format for team learning and discussion about the concepts and activities presented in the workbook. To run this workshop with your team, you will need to access the resources provided in the link below. This includes a Miro board with case studies and activities to work through.

Workshop resources for **AI Sustainability in Practice Part Two:**
turing.ac.uk/aieg-3-activities

### A Note on Activity Case Studies

Case studies within the Activities sections of the AI Ethics and Governance in Practice workbook series offer only basic information to guide reflective and deliberative activities. If activity participants find that they do not have sufficient information to address an issue that arises during deliberation, they should try to come up with something reasonable that fits the context of their case study.

**Note for Facilitators**

In this section, you will find the participant and facilitator instructions required for delivering activities corresponding to this workbook. Where appropriate, we have included Considerations to help you navigate some of the more challenging activities.

Activities presented in this workbook can be combined to put together a capacity-building workshop or serve as stand-alone resources. Each activity corresponds to a section within the Key Concepts in this workbook. Some activities have pre-requisites, which are detailed on the following page.

## Stakeholder Impact Assessment (Design Phase)

Practise answering key questions within SIAs.

**Corresponding Sections**

- → Introduction to Sustainability: Stakeholder Impact Assessments (page 10)
- → A Closer Look at Stakeholder Impact Assessments (page 14)
- → Stakeholder Impact Assessment Template (Design Phase) (page 15)

**Pre-Requisites**

- ↗ Key Concepts: AI Sustainability in Practice (Part One)
- ↗ Activity: Establishing an Engagement Objective from AI Sustainability in Practice (Part One)

## Balancing Values

Practise weighing tensions between values when assessing the ethical permissibility of AI projects by considering consequence-based and values-based approaches and engaging in deliberation.

**Corresponding Sections**

- → Skills for Conducting Stakeholder Impact Assessments (page 21)

**Pre-Requisites**

- ↗ Key Concepts: AI Sustainability in Practice (Part One)

## Revisiting Engagement Method

Practise undertaking practical considerations of resources, capacities, timeframes, and logistics as well as stakeholder needs to establish an engagement method for the following SIA.

**Corresponding Sections**

- → Sustainability Throughout The AI Lifecycle (page 28)
- → Determining Stakeholder Engagement Methods for Stakeholder Impact Assessments from AI Sustainability in Practice (Part One)

**Pre-Requisites**

- ↗ Key Concepts: AI Sustainability in Practice (Part One)
- ↗ Activity: Stakeholder Impact Assessment (Design Phase) (page 45)

## Stakeholder Impact Assessment (Deployment Phase)

Practise using SIAs to formulate proportional monitoring activities for the development and deployment of AI models.

**Corresponding Sections**

- → Introduction to Sustainability: Stakeholder Impact Assessments (page 10)
- → A Closer Look at Stakeholder Impact Assessments (page 14)
- → Stakeholder Impact Assessment Template (Deployment Phase) (page 20)

**Pre-Requisites**

- ↗ Key Concepts: AI Sustainability in Practice (Part One)

# Interactive Case Study Recap: AI in Urban Planning

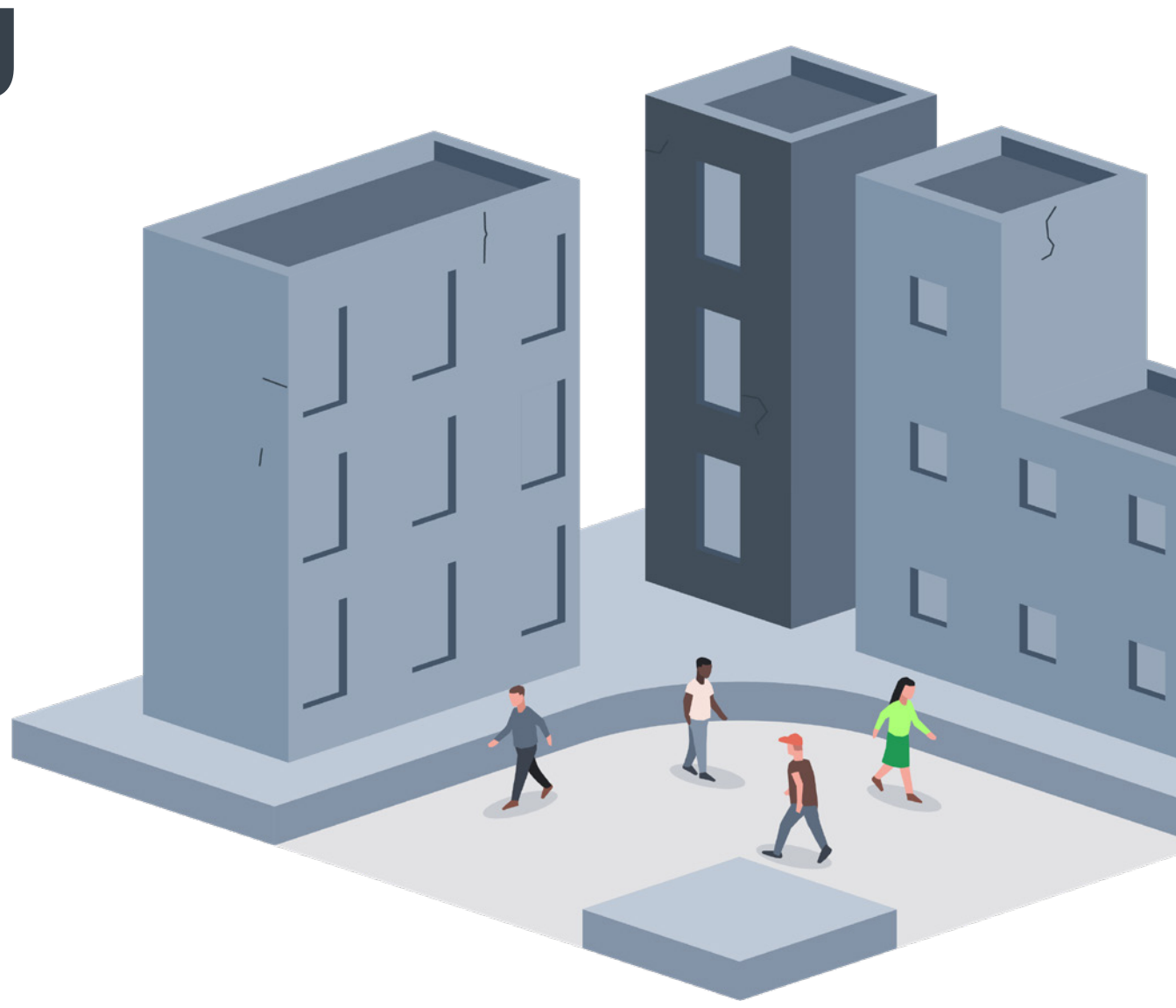

Your team is a local planning authority within a borough facing a housing crisis. The local poverty rate is higher than the national average and residents complain of sub-optimal living conditions.

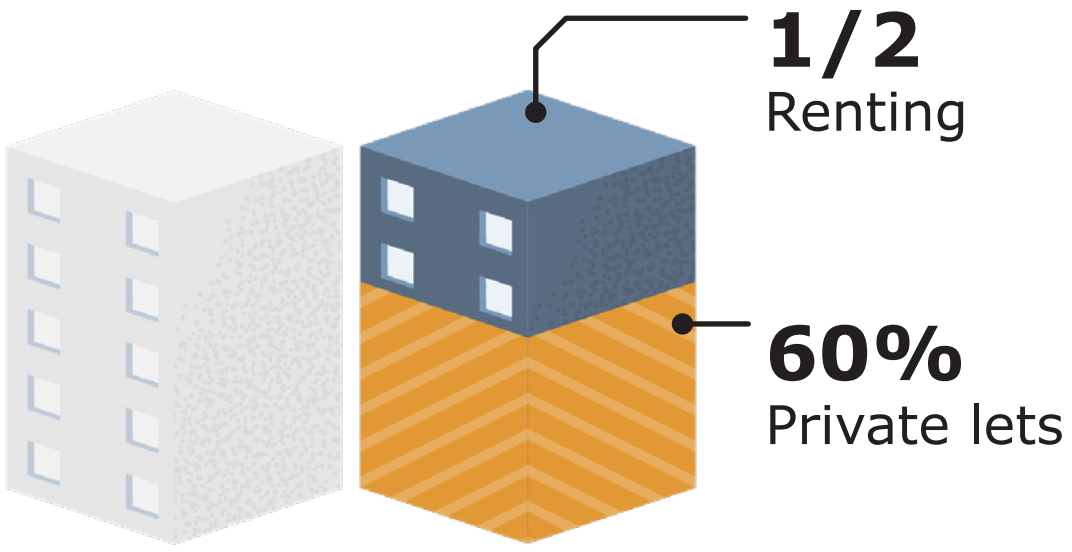

Around half of your residents are renters, 60% of whom live in private lets. The private letting sector is becoming increasingly unaffordable.

# 25%

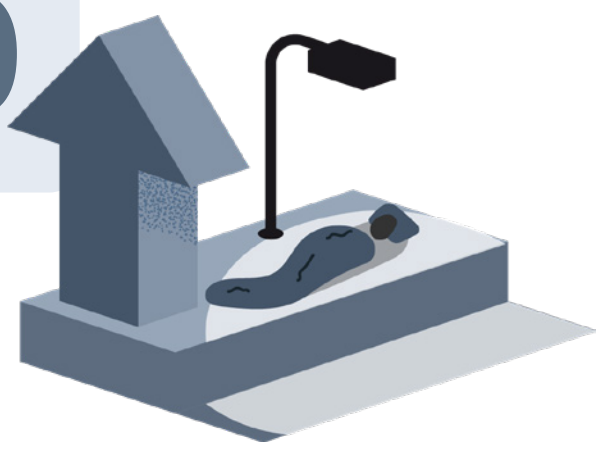

# 53%

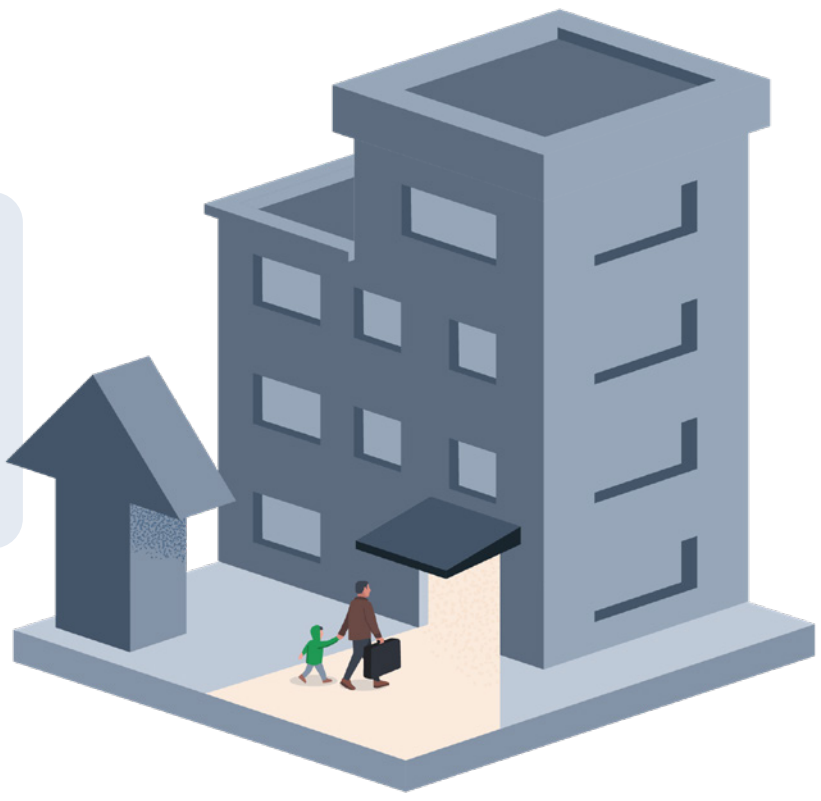

The number of homeless applications has **risen by 25% in the past three years.**

**The number of households in temporary accommodation has risen by 53%,** with an unprecedented number of applications submitted since 2020.

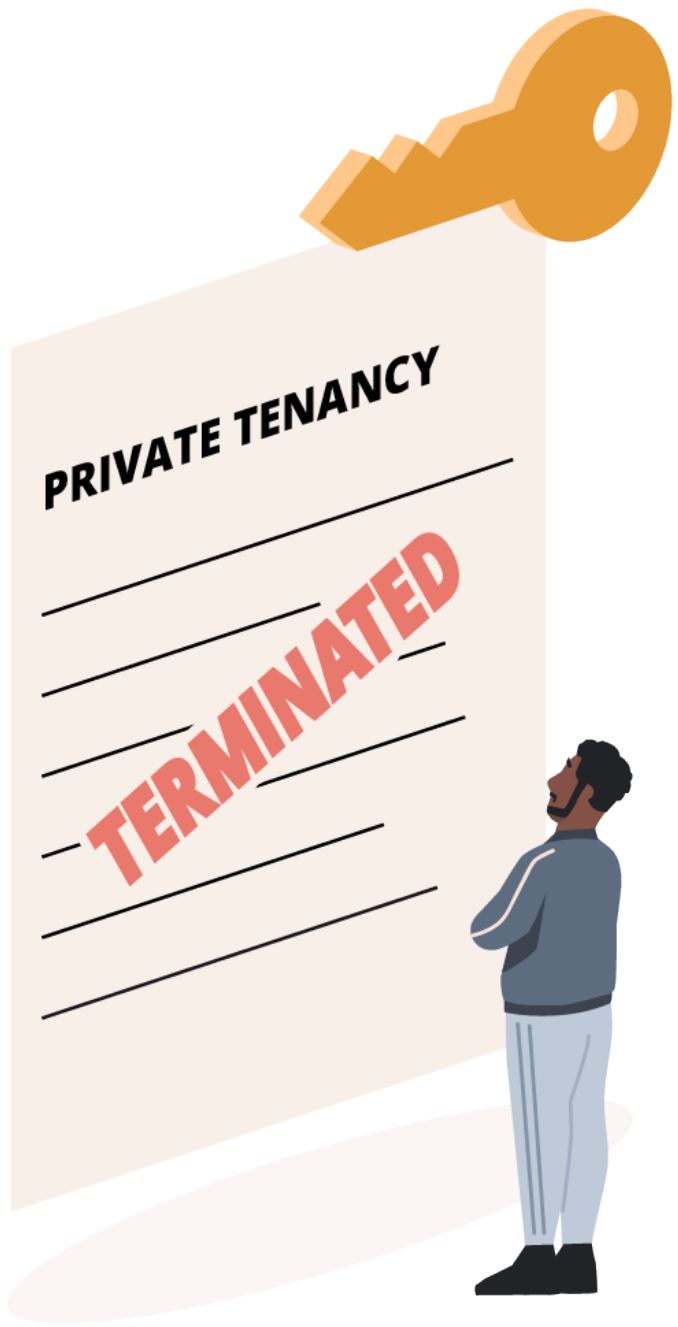

A recent council investigation found that **terminated private tenancy leases** are the **single greatest cause of homelessness in the borough.**

**10,000**
New homes

**50%**
Affordable homes

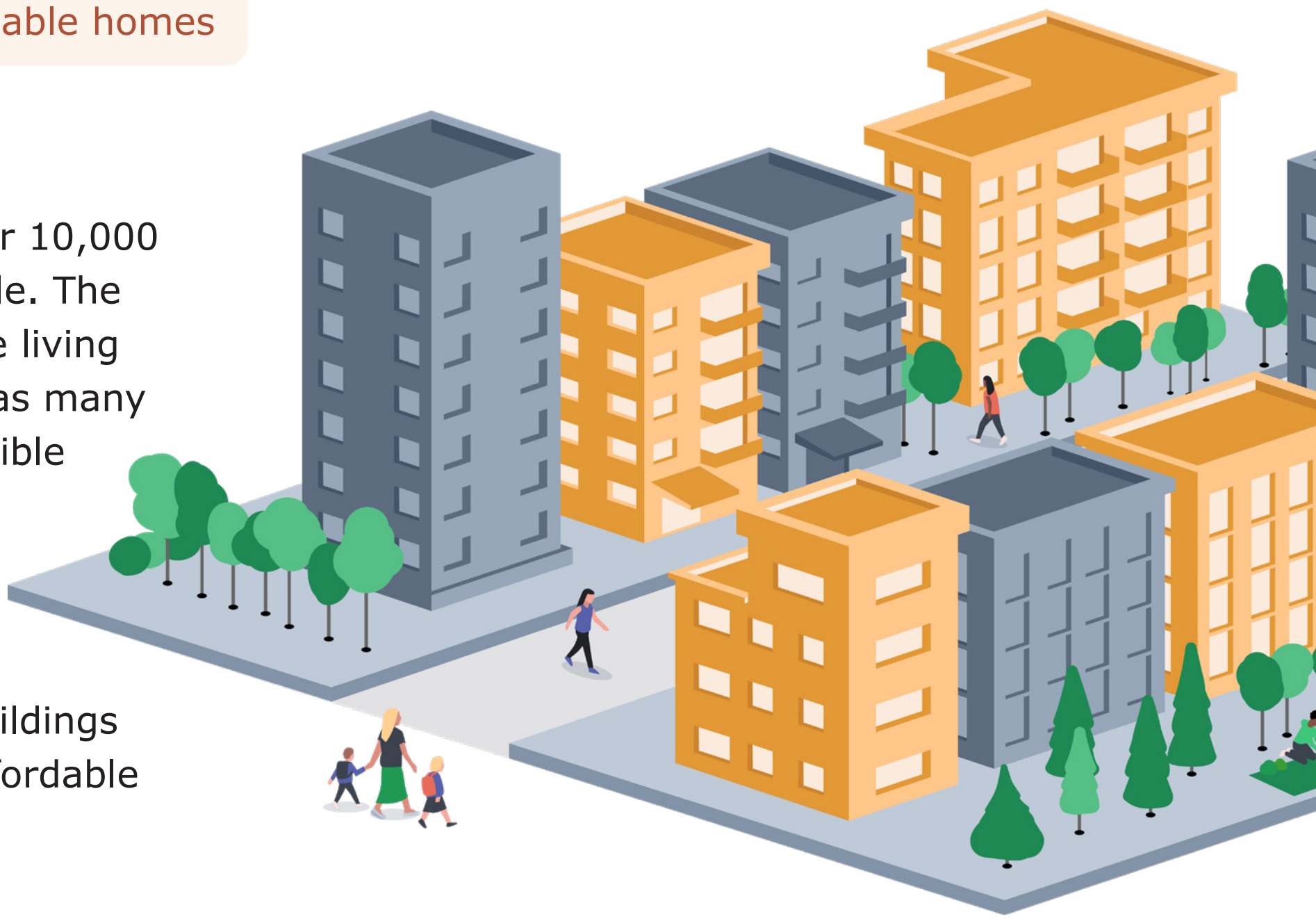

Your council has established a 10-year housing plan set out to deliver 10,000 homes, 50% of which will be affordable. The objective of this plan is to improve the living standards of residents by developing as many high-quality affordable homes as possible over the next ten years.

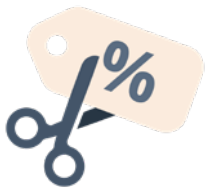

The council has offered to subsidise new residential buildings that deliver at least 50% affordable housing.

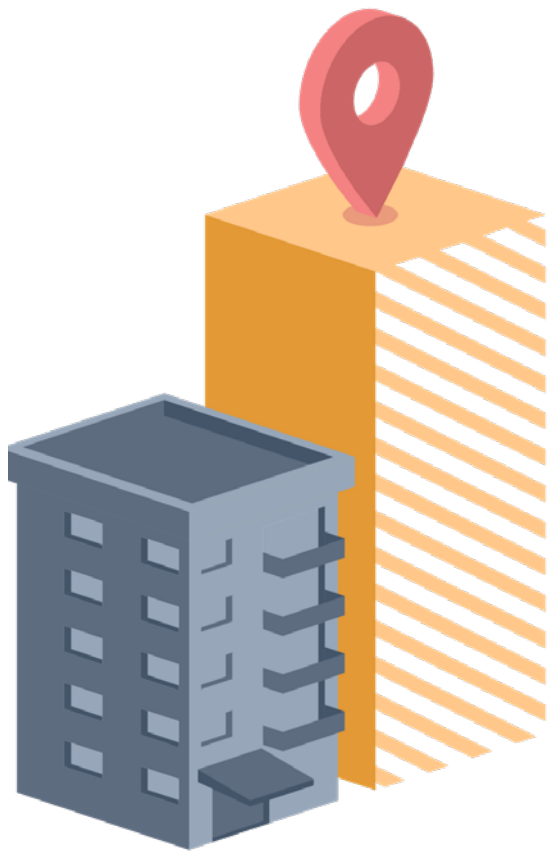

To support housing developments, your team will need to **expand your list of sites permitted for planning applications.** Achieving your target would mean **doubling** the number of local homes. Your team will need to review a much higher volume of planning applications, which may not be obtainable through your current process.

# Model Proposal

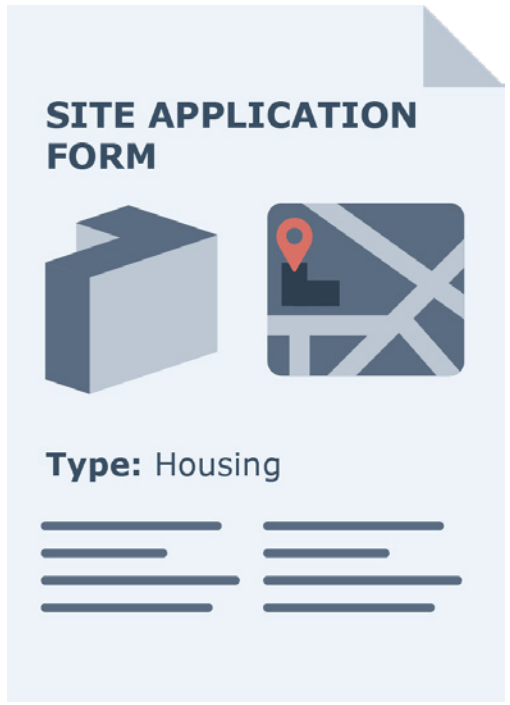

## Current Method

Your current method for allocating new development sites can take up to ten months to complete and considers **a limited number of sites proposed by developers, landowners, and estate agents.** These sites are manually reviewed by your team to ensure they meet policy standards (i.e. sites' ability to provide basic amenities) and are suitable for development in practice.

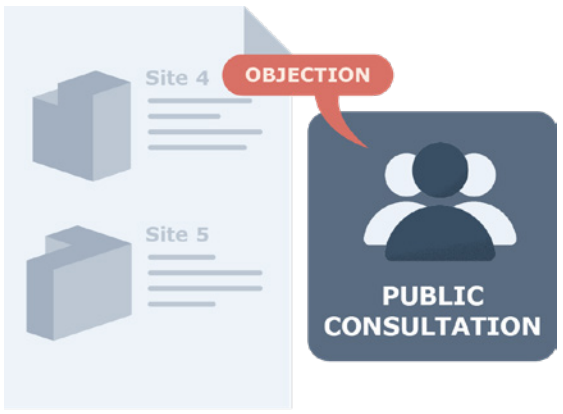

**Sites that pass your review process are taken forward for a public consultation.** This gives residents the opportunity to object to certain sites being open for planning applications. Your team considers public input to help determine which site proposals are accepted.

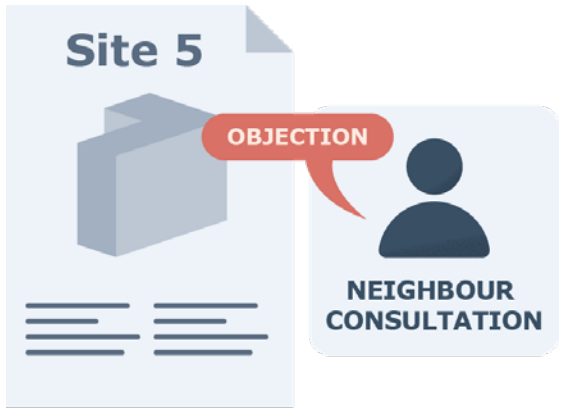

**Accepted sites are made available for planning applications.** Applications are detailed development proposals demanding in-depth review. **Your team manually reviews individual applications in a process that includes a second tier of consultations with neighbours of specific sites.**

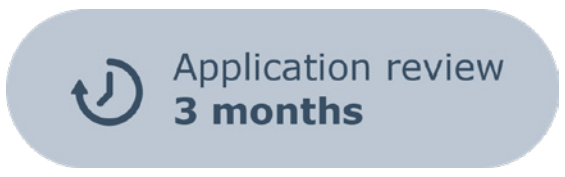

Granting planning permissions can take up to three months per application.

## Proposed Method

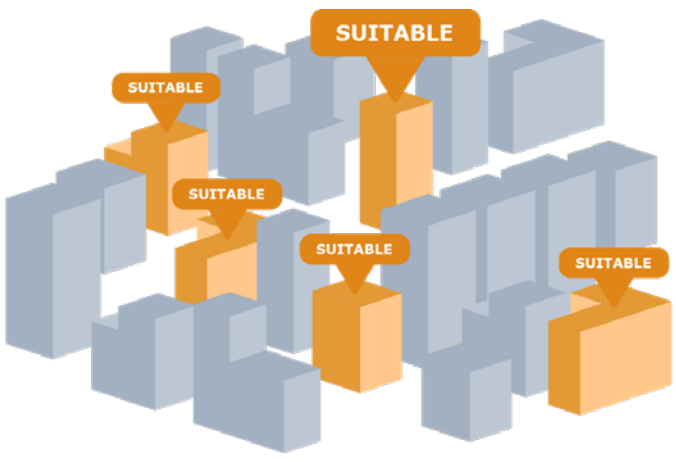

Your council has suggested you automate this process by using an machine learning (ML) model to automatically review every site in the local area, classifying them as suitable or unsuitable for housing development.[32] This approach would allow your team to scale-up the number of sites considered for development. Whereas your current method captures a number of submitted proposals, **the model would capture all local sites.** This model would consider sites that are outside the reach of your current method, such as council owned buildings that could be repurposed, and private parcels that could accept purchase offers.

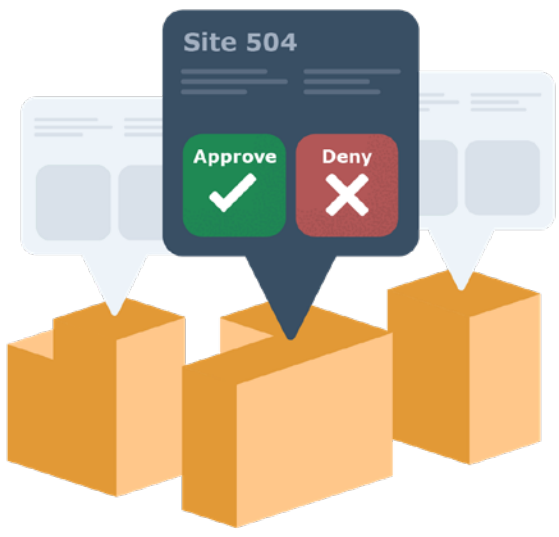

Sites categorised as suitable would be reviewed by your team. **Those that pass this review process would be brought forward for a three month public consultation** which your team would consider when accepting a final list of sites for development. **Accepted sites would be made public in a digital map and approved for development, forgoing the additional three-month application review process.**

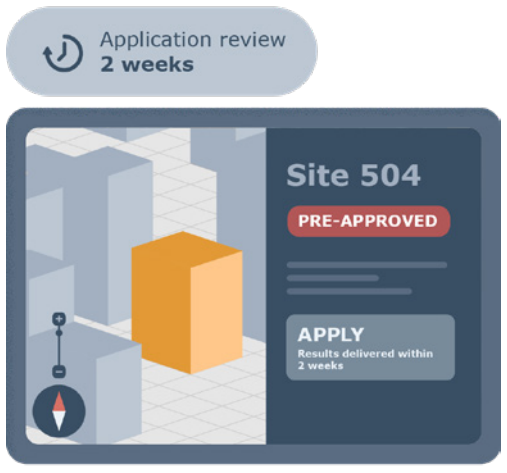

**The proposed method would remove neighbour consultations from the application review process.** By removing time-consuming steps, your team would be able to verify applications’ compliance with building design standards and grant approvals or request adjustments within two working weeks.

Your team conducted a Stakeholder Analysis and advised your council on how to engage stakeholders throughout the Design and Development process. The council has reviewed your advice and has decided to open your assessment to the public.

### If the team advised on Partnering or Empowering stakeholders...

They have approved your engagement objective, as they deemed that this model would have significant social impacts. Your team is now to (partner with or empower) stakeholders when conducting SIAs for this model.

### If the team advised on Informing or Consulting with stakeholders...

They deemed that this model would have significant social impacts and have decided that your team should partner with stakeholders as an engagement objective. Your team is now to engage with stakeholder when conducting SIAs for this model.

# Stakeholder Profiles

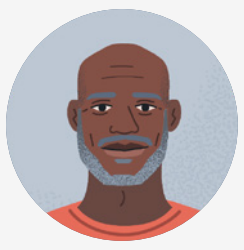

## George

60 | He/him | Impacted Stakeholder

### Profile

George is a 60-year-old black British man. He is a member of the Local Small Business Association and owns a popular restaurant at one of the local high streets.

### Goals and Aspirations

Locals love George’s restaurant. However, because of a shortage of supplies, he has had to temporarily close the shop twice in the past year. Having opened back up, he is hoping to increase sales especially given that the rent for his commercial space might increase when his contract is up.

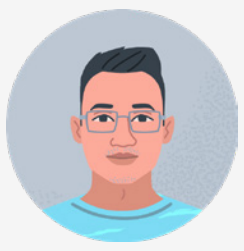

## Alex

35 | He/him | Project Team Member

### Profile

Alex is a 35-year-old Chinese man who lives in another borough. He is the Planning Authority Lead and has been working for the council for the past six years. Alex traditionally leads the site searching process for council developments, and will be involved in the Housing Delivery Plan.

### Goals and Aspirations

Alex is committed to delivering projects that directly involve the local community throughout the process. Criticism of previous uses of ML for urban planning worry him, and he wants to ensure that the housing plan values the knowledge of his team and the local community.

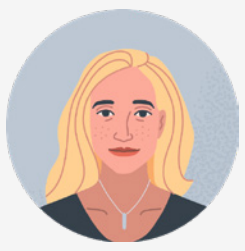

## Hayley

40 | She/her | Impacted Stakeholder

### Profile

Hayley is a 40-year-old white British woman on the housing register. She lives in an overcrowded flat with her family of five, she is waiting for a bigger home ideally in proximity to affordable childcare and a specialist school for her son who has autism spectrum disorder (ASD).

### Goals and Aspirations

Hayley and her husband’s living situation has been overcrowded since the birth of their third child a year ago, and greatly worsened when they started working from home. The lack of space has been extremely challenging for their eldest son in particular, with ASD. Their current flat is in close proximity to their eldest son’s specialist school, and affordable childcare for their two youngest. They are hoping to move to a bigger home that provides this level of access as soon as possible.

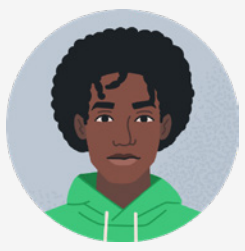

## Ali

17 | He/him | Impacted Stakeholder

### Profile

Ali is a mixed-race 17-year-old local. He moved to the UK from Jamaica with his family when he was two and has lived locally since. His family rents a house near a community garden he helps run.

### Goals and Aspirations

Ali spends his leisure time at a local community garden. The garden is small but run by a committed group of community members. It sits within a greater green space where young people like to gather. Ali is concerned with the high levels of development in the area. Not only are the construction noises overwhelming, but he has noticed an increasing number of green spaces being used for development over the past years. He is hoping this space remains run by the community.

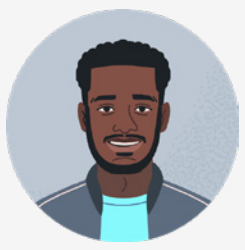

## Terry

27 | He/him | Impacted Stakeholder

### Profile

Terry is a 27-year-old black British man. He was born and raised locally and works at a local corner store owned by a family friend.

### Goals and Aspirations

Terry grew up in a local council estate where his neighbours helped raise him, as his mother worked two jobs. After the estate was demolished for replacement by mixed income homes, residents were relocated. In order to stay close to his job, Terry has since rented a room in a private letting. He aspires to open his own shop but is struggling to make enough money to get by. In recent years he has noticed more buildings being repurposed for mixed income housing and rent prices going up. Terry isn’t certain he will be able to afford staying in the neighbourhood for much longer.

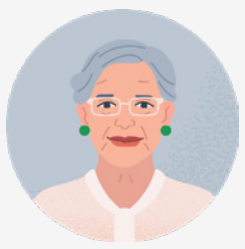

## Katherine

73 | She/her | Impacted Stakeholder

### Profile

Katherine is a 73-year-old white British woman. She is a regular at the local library, leisure centre, and church. She currently lives with her daughter and grandchildren but has recently been placed on a priority waiting list within the housing register in order to move into council housing that supports her mobility needs.

### Goals and Aspirations

Katherine has a very social lifestyle which she loves and intends to maintain. She wants to make sure her new home is close to leisure facilities and public transport in order to be able to see her friends and family regularly.

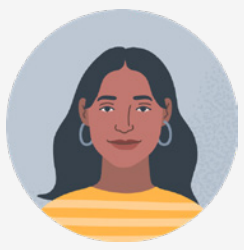

## Mia

28 | She/her | **Project Team Member**

### Profile

Mia is a 28-year-old British Indian woman and data scientist who rents an apartment in another borough. She has been a council employee for two years, and has experience using a variety of ML techniques.

### Goals and Aspirations

Mia is passionate about applying ML techniques to improve the quality of life of residents. Having recently finished her work utilising predictive analytics to identify families in need of support, Mia is eager to continue applying ML across council services, and is excited about the prospect of supporting your housing delivery plan.

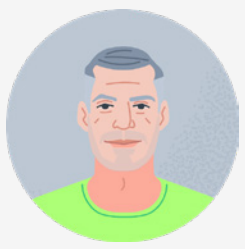

## Nick

55 | He/him | **Impacted Stakeholder**

### Profile

Nick is a 55-year-old white British man and electrician who has recently been placed in emergency accommodation after losing his job and after his private tenancy agreement wasn’t renewed.

### Goals and Aspirations

Although Nick is receiving immediate support, he is finding it difficult to cope with the difficult situation he finds himself in. He is a transgender man and is afraid of facing harassment at the emergency shelters located in neighbourhoods he is unfamiliar with. His goal is to find employment again and move into a home where he feels safe as soon as possible.

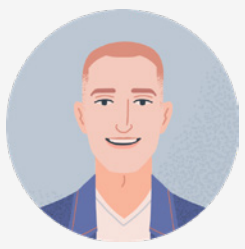

## Jamie

31 | He/him | **Impacted Stakeholder**

### Profile

Jamie is a 31-year-old white British man and graphic designer for a creative agency.

### Goals and Aspirations

Jamie moved into the area three years ago and has been renting privately since. He has fallen in love with the neighbourhood and is a regular at the variety of coffee shops and restaurants that have opened up in the past couple of years. He and his husband are looking to buy a home locally.

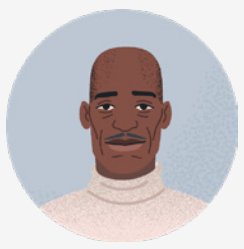

## Tom

59 | He/him | **Impacted Stakeholder**

### Profile

Tom is a 59-year-old Black French real estate owner. He inherited a property portfolio that includes a variety of local commercial properties, which he has been managing for around 10 years.

### Goals and Aspirations

Tom’s business has experienced a lack of consistency in rental payments in the last two years. Not all business were able to pay and some went bankrupt and terminated their leases. Tom has stopped extending some leases and is looking to sell a portion of his property.

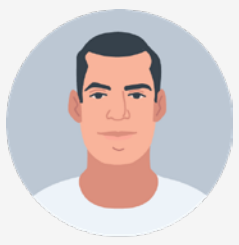

## Michael

31 | He/him | **Project Team Member**

### Profile

Michael is a 31-year-old white British man and the product manager for the proposed project.

### Goals and Aspirations

Michael has wanted to make a career move towards working in AI projects for a while, and has proposed this project as a way to support the housing delivery plan.

40 mins | Participant Instructions

# Stakeholder Impact Assessment (Design Phase)

## Objective

Practise answering key questions within SIAs.

## Role Play

In this activity, your team will conduct a Design Phase SIA. Your group will be assigned stakeholder profiles in order to consider a variety of perspectives that may be present in stakeholder engagements.

## Team Instructions

1. This activity will start with your facilitator reading out the activity context. They will split the team into groups, each with assigned personas.

2. Once groups have been assigned, take a few minutes to individually read over the **Project Proposal**. Team members are to consider the note on case studies at the beginning of the activities section of this workbook, imagining how stakeholders might relate to the content.

3. Once team members have read over the **Project Proposal**, the team will have some minutes to answer the questions on your assigned section of the **Stakeholder Impact Assessment (Design Phase)**. Consider how each persona might respond differently to questions.

4. A group member is to volunteer to write answers on the board and report back to the team.

5. You will then reconvene as a team, having volunteer note-takers share each group's answers to the questions and discussing the answers.

6. Having discussed as a team, individually use sticky notes to write answers to the questions under the **Sector-Specific and Use Case-Specific Questions** section.

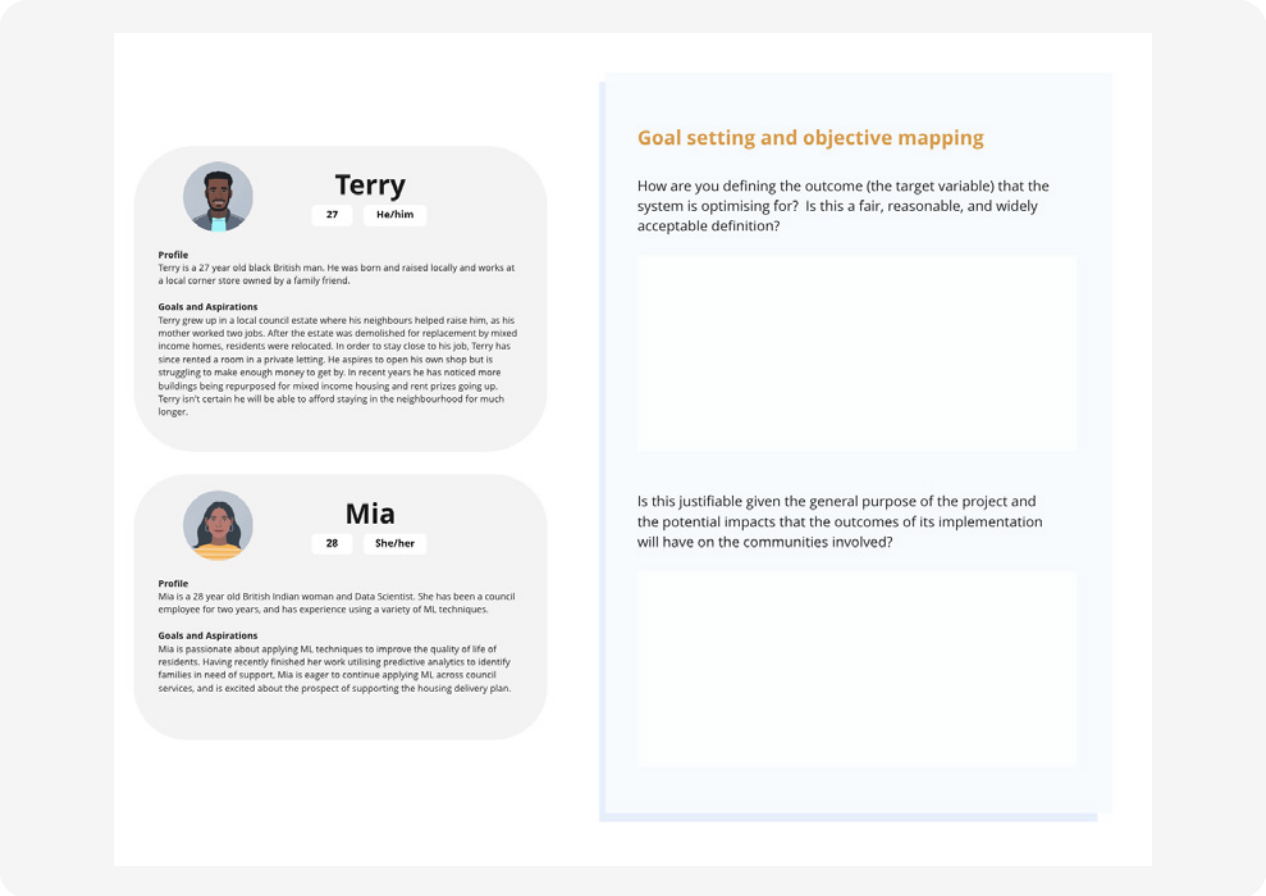

**Stakeholder Impact Assessment (Design Phase)**

# Project Proposal

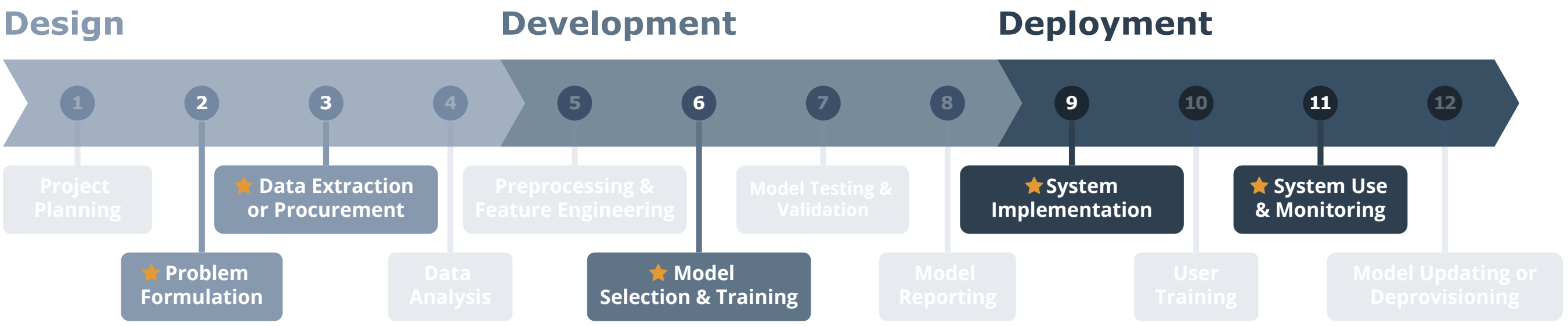

★ Stages in focus

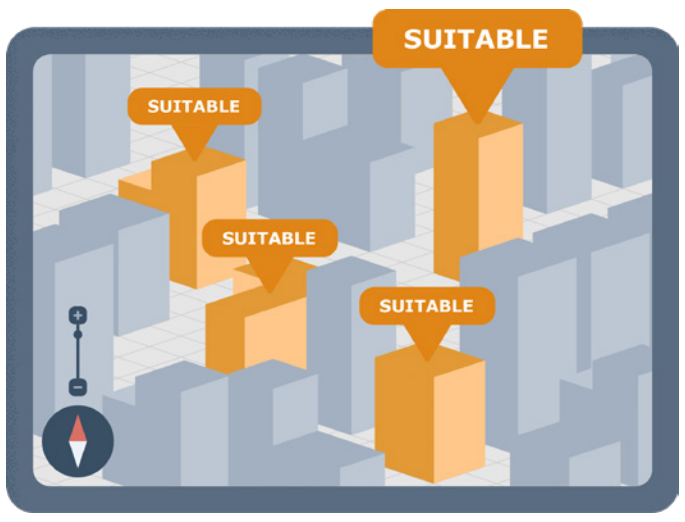

## ★ Problem Formulation

The proposed system is a visual interface that classifies sites as suitable or unsuitable. The target variable of suitability would indicate that sites classified as suitable would be reviewed by your team. Those that pass the review process would be brought forward for public consultation. Your team would consider public input to help determine which suitable sites are made public and pre-approved for planning applications for developments that include at least 50% affordable housing.

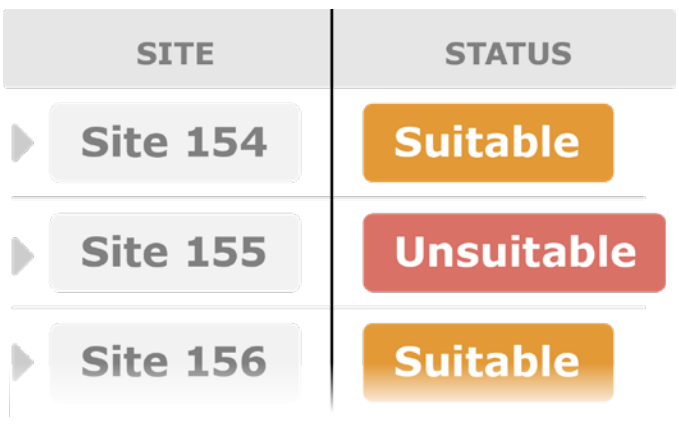

## ★ Data Extraction

The model would use pre-labelled data from sites currently allocated as accepted for residential use (suitable sites) and an equal amount of non-allocated sites (unsuitable sites).

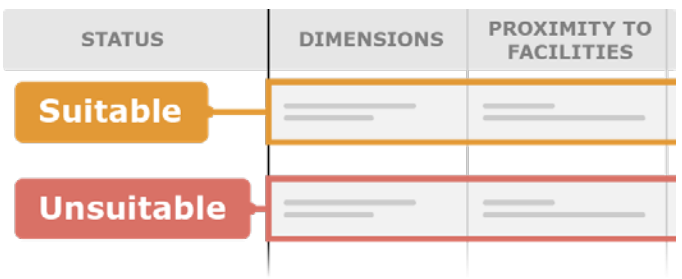

## ★ Model Selection & Training

This system would use a Random Forest Classification Model (a supervised ML model, described on page 49). The model is detailed in the second part of this proposal.

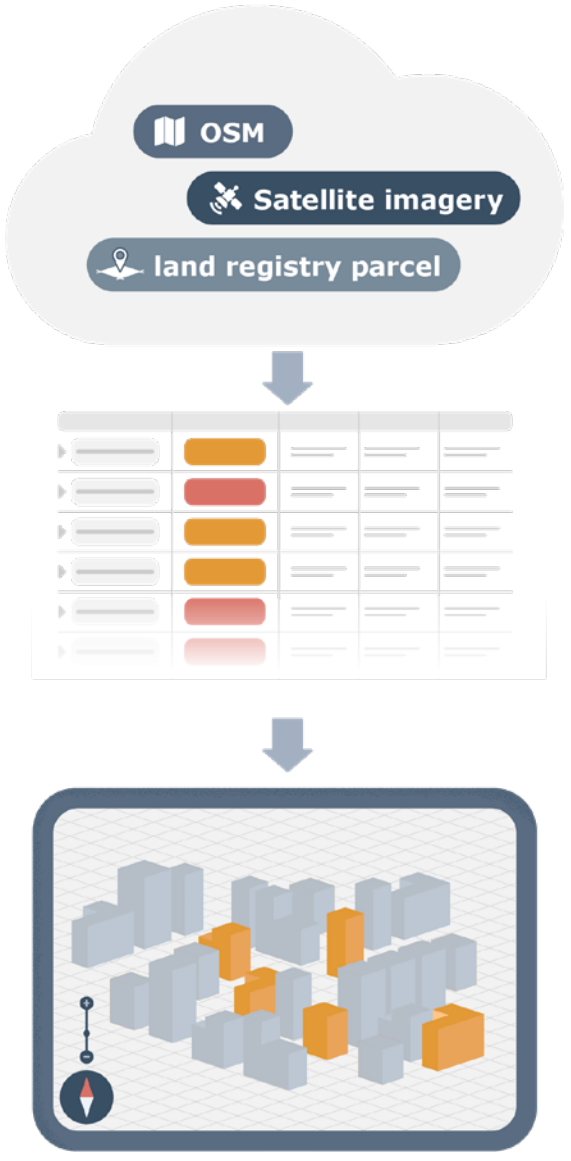

## System Implementation

When deploying the model, your team would import data representing the entire region, gathered from Open Street Maps (OSM), satellite imagery databases, land registry parcel databases, and your council's urban databases. This data would be pre-processed by the model, organising it into features that match the training data. The pre-trained model would then classify sites and illustrate outputs in a digital map highlighting suitable sites and providing key information for each.

---

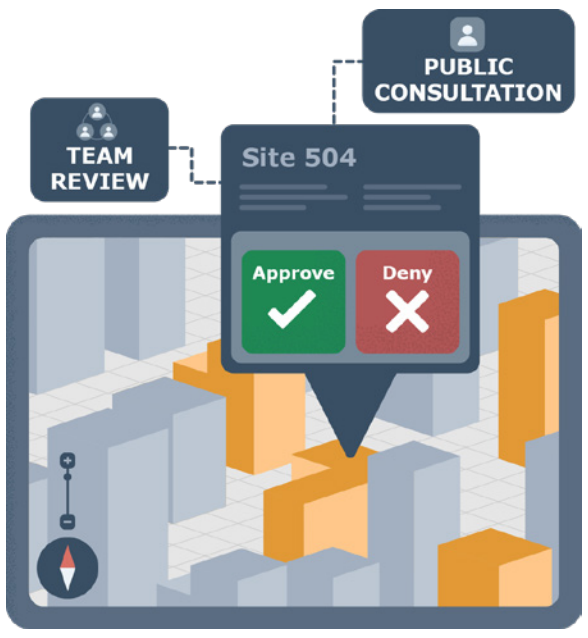

## Site Validation

Your team would review suitable sites and adjust your list as deemed appropriate based on local policy, landowners' interest in development, and a public consultation. **This process would take no longer than three months** since suitable sites will reflect features of currently accepted sites, and key information for validating sites will be found in a centralised web interface.

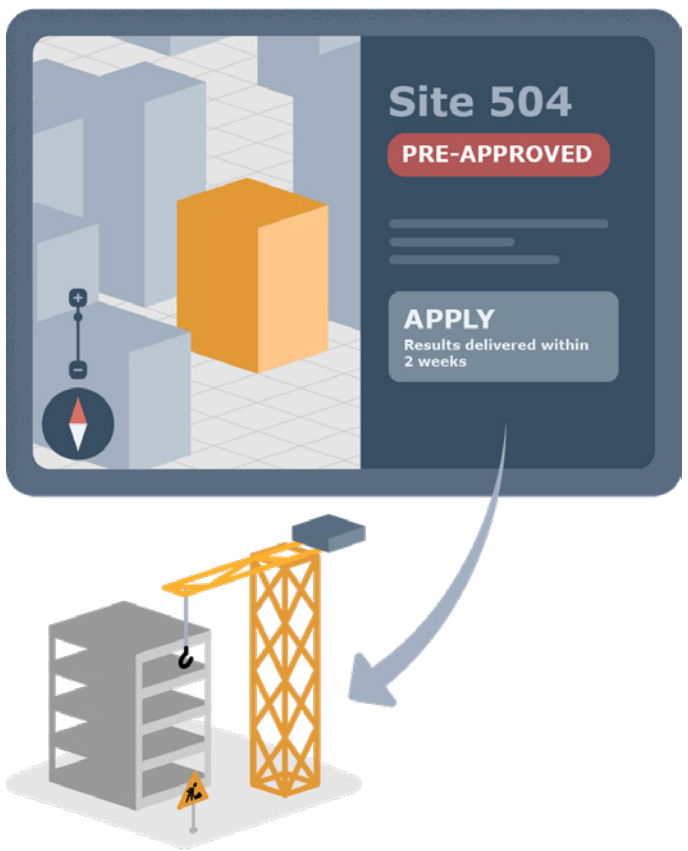

### Outcome

Accepted sites will be made publicly available in the council website. Planning applications for these sites will be deemed pre-approved, **waiving time-consuming elements of planning application reviews, such as consultations with neighbours**. Your work reviewing applications would be reduced to verifying compliance with building standards (i.e. compliance with accessibility, health and safety) and requesting any necessary adjustments. Application results are to be delivered within two working weeks or deemed approved in the event of no response. Your public map will be automatically updated as permissions are granted, reflecting availability.

# A Closer Look at the Model

The model is a Random Forest Classifier trained to map patterns between features (individual characteristics or properties within the data) and suitability (target variable), using a pre-labelled database of 1300 local sites.[33]

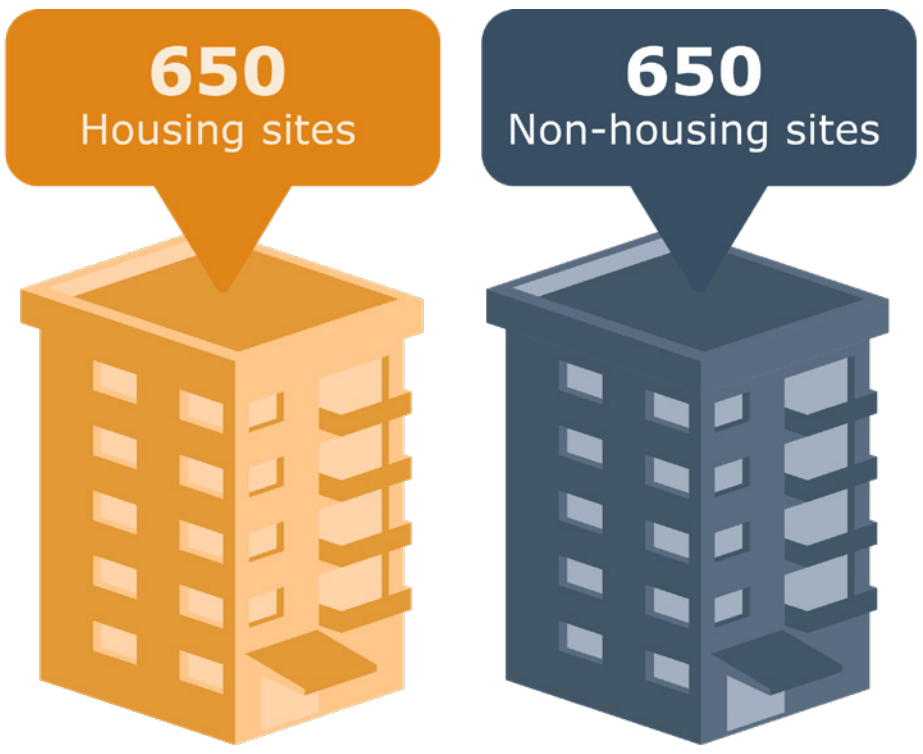

This database would be split in a 70/30 ratio, where 70% of the data (1000 evenly split sites) would be used to train the model, while 30% (300 evenly split sites) would be concealed from the model during training, and later used for testing.

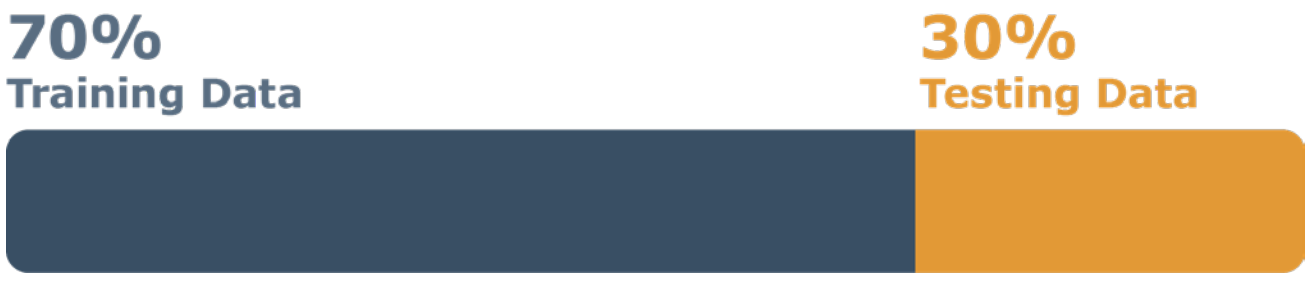

Each site in the dataset is represented by features used in the Random Forest to determine whether the site is suitable:

- Area/dimensions (m2)
- Location (coordinates) (within local area)
- Address (zip code)
- Road access and presence of parking area
- Access to energy, utilities, water, waste management
- Policy area
- Restricted location (listed buildings, conservation areas, safety hazard areas)
- Building materials
- Market value (£)
- Current use
- Current ownership
- Proximity to leisure, recreation, entertainment, green spaces (km)

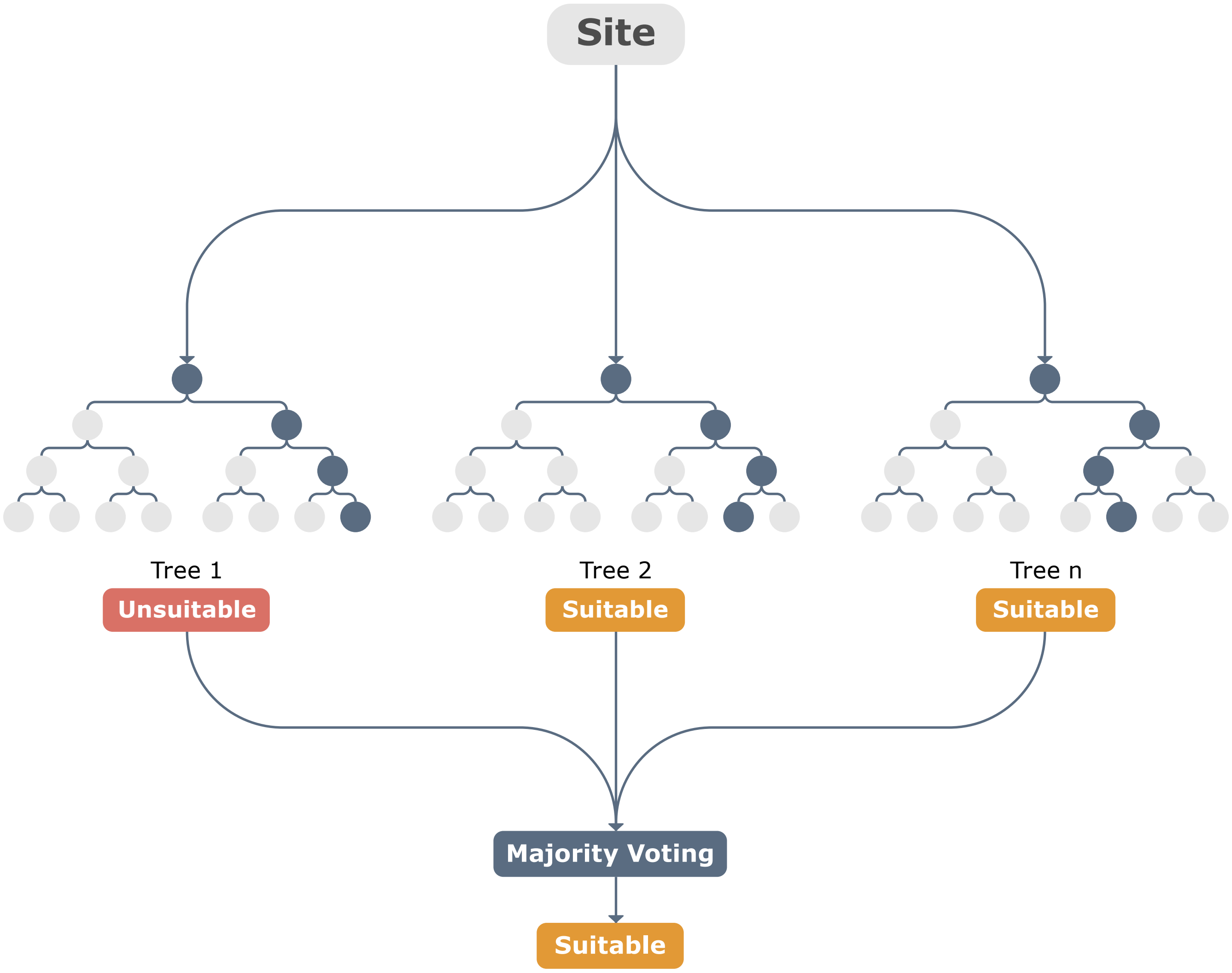

Random Forest models determine classifications based on the majority vote of a large number of individual decision trees (flow charts analysing features that lead to a classification).

40 mins | Facilitator Instructions

# Stakeholder Impact Assessment (Design Phase)

1. Read the following statement to the team:

> Our team has conducted an iteration of the Stakeholder Engagement Process and advised our council on an engagement objective for this project. The council has determined we will conduct our Design Phase SIA by **(partnering with or empowering, depending on your team's engagement objective, determined in Part One of this workshop)** stakeholders and engaging them in a citizens' panel. They have provided a model proposal containing further details about the model and its intended placement within our team.

2. Give the team some minutes to read the instructions for this activity. When they finish reading the instructions, ask them if they have any questions.

3. Split team members into groups, each assigned a section within the **SIA (Design Phase)** template on the board. Each group will answer two key questions within each section.

   **If your team has chosen** Partner **as an objective,** assign the profiles marked with either Impacted Stakeholder or Project Team Member.

   **If your team has chosen** Empower **as an objective,** only assign the profiles marked with Project Team Member.

**Group 1** (Goal Setting and Objective Mapping):

- Terry — Impacted Stakeholder
- Mia — Project Team Member

**Group 2** (Horizon Setting and the Decision to Design):

- Hayley — Impacted Stakeholder
- George — Impacted Stakeholder
- Tom — Impacted Stakeholder

**Group 3** (Possible Impacts on the Individual):

- Alex — Project Team Member
- Katherine — Impacted Stakeholder

**Group 4** (Possible Impacts on Society and Interpersonal Relationships):

- Jamie — Impacted Stakeholder
- Ali — Impacted Stakeholder
- Nick — Impacted Stakeholder
- Michael — Project Team Member

4. Let the team know they have some minutes to individually read over the Project Proposal and then to go through the questions together.

5. Facilitators and co-facilitators should check in with each group, using the Considerations section of this activity to help answer any questions.

6. When the time is up, ask the team to reconvene.

7. Give volunteer note-takers a few minutes to share their teams' answers to the questions in their section.

8. After all questions are shared, lead a discussion about the answers. Ask participants if there is anything they would like to adjust, take away, or add to the answers.
   - **Co-facilitator:** adjust any necessary answers to the questions based on the team discussion.

9. When the group discussion is up, ask the team to use sticky notes to individually write answers to the questions under the **Sector-Specific and Use Case-Specific Questions** section of the board, placing their answers on the board.

## Facilitator Considerations

## Key Discussion Points for Potential Harms

When facilitating the discussion on potential harms, it may be useful to refer back to the 'Origins of the SUM Values: Drawing principles from real-world harms' section of AI Sustainability in Practice Part One. In particular, the mapping of risks that emerge from the use of AI/ML technologies to the ethical concerns underwriting responsible AI/ML research and innovation provides a helpful starting point for examining potential negative impacts:[34]

| Risks that Emerge From the Use of AI/ML Technologies | Ethical Concerns Underwriting Responsible AI/ML Research and Innovation |
| --- | --- |
| 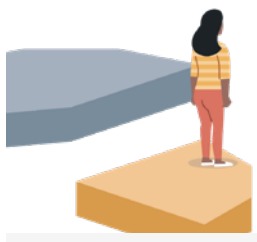 Loss of autonomy, interpersonal connection, and empathy | 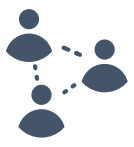 Human agency and social interaction |

For instance: residents may not be sufficiently consulted about a development project that results from the AI model's use and may thus lose a sense of agency and autonomy; meaningful community participation in decision-making about local affairs may be circumvented by the use of the model, harming social solidarity and interpersonal connection;[35] the significance of the professional judgment of city planners and local officials may be diminished through this form of automation, thereby harming their agency and decision-making authority; certain citizens may be displaced or severely inconvenienced as the result of the automated classification of suitability without having a say in the way suitability is being defined by the system, thereby harming their sense of autonomy and agency.[36]

| Risks that Emerge From the Use of AI/ML Technologies | Ethical Concerns Underwriting Responsible AI/ML Research and Innovation |
|---|---|
| 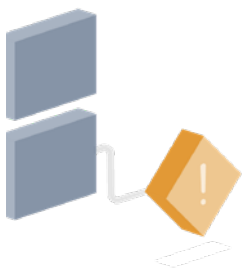 Poor quality outcomes | 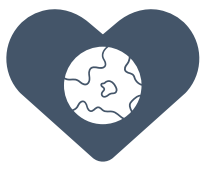 Wellbeing of each and all |

For instance, poor quality data (i.e. city records or property and land use information that contain human errors), gaps in measurement (poor/inconsistent recording of geographic proximity to essential services and amenities), or out-of-date information (i.e. dated/obsolete information about current property use, essential services and amenities, or access to utilities), can lead to outputs that inaccurately indicate a site's suitability for development and that end up harming the wellbeing of future residents, who then have to inhabit inhospitable or deprived living environments.[37] [38] [39] [40]

| Risks that Emerge From the Use of AI/ML Technologies | Ethical Concerns Underwriting Responsible AI/ML Research and Innovation |
|---|---|
| 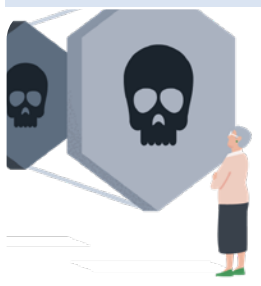 Bias, injustice, inequality, and discrimination | 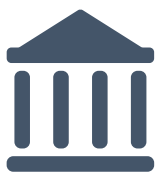 Social justice, equity, public interest, and the common good |

For instance, where past determinations of suitability contained discriminatory or biased patterns (as in cases where features like location operated as a proxy for socioeconomic status or race or, where deficient access to essential services and amenities disqualified deprived populations of development opportunities), the model trained on data containing such patterns could replicate or augment those discriminatory harms and injustices.[41] [42] [43] Moreover, trends to toward gentrification and the displacement of local, and potentially vulnerable, social groups in urban development processes my influence AI project design decisions and overall project planning choices. Considerations of these potential ecosystem-level inequities should factor into impact assessment.

45 mins | Participant Instructions

# Balancing Values

## Objective

Practise balancing and navigating tensions between values when assessing the ethical permissibility of AI projects. Learn to employ consequences-based and principles-based approaches when engaging in deliberation.

## Activity Context

When answering questions about the possible impacts of the proposed system, members of your team may have noticed that, at times, different values come into tension with each other. Decisions that are made around how to balance these tensions can both influence the direction that AI projects take, and shape their outcomes.

In this activity, your team will consider three values that can come into conflict when considerations are undertaken about the ethical permissibility of using a model to automate the site selection process. Your team is to consider these tensions from both consequences-based and principles-based approaches, establishing a plan for how you will balance these values. This plan will be used both to inform your recommendation to the council on whether to develop the system, and to specify any amendments you would need to the Project Proposal in order to consider it ethically permissible.

Your group will be assigned a pair of conflicting values in this activity. The goal is for your group to keep in mind the identities and circumstances of the stakeholder profiles in order to consider how they might evaluate tensions discussed in this activity.

## Team Instructions

1. Split into groups, each assigned a pair of conflicting values. Take a look at your group's section of the **Conflicting Values Diagram**.

2. Fill out the **Consequence-Based Approaches**, **Principle-Based Approaches**, and **Balancing Plan** sections, answering the prompts listed.

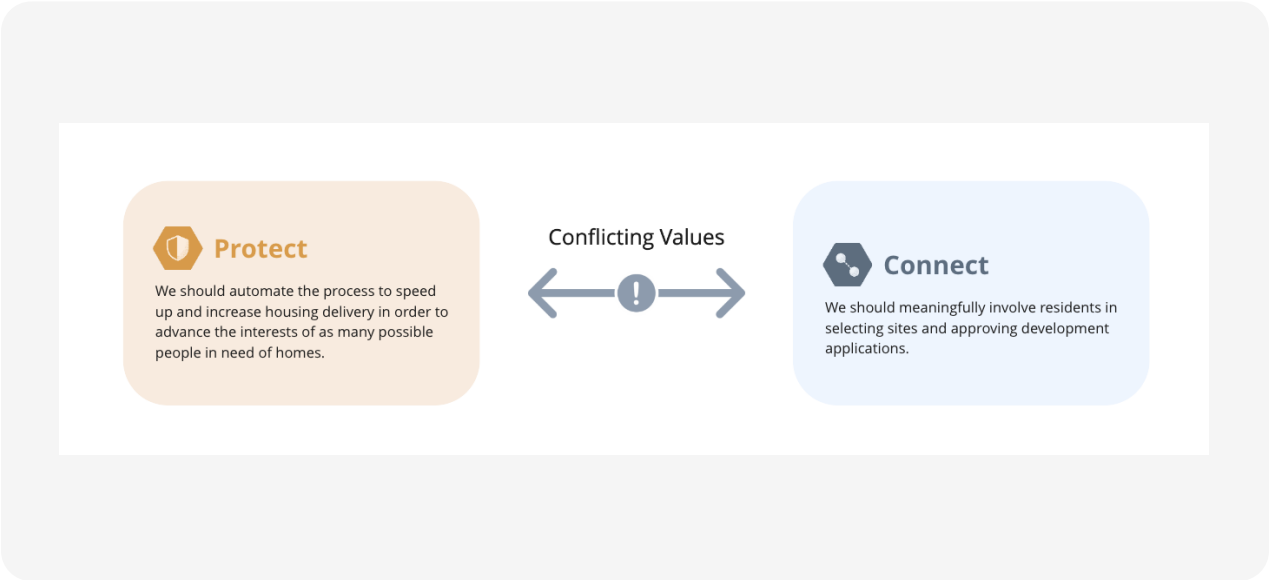

**Conflicting Values Diagram**

3. When the time is up, reconvene as a team, having your group's volunteer note-taker share your conflicting values and plans with the team.

4. Engage in a team discussion about the extent to which your plans address anticipated concerns or questions, and how you might adjust them to better address these. Discuss any promising aspects of developing this system, (considering your plans).

   - Your co-facilitator will write your answers on sticky notes, placing them under the **Potential Project Benefits** section.

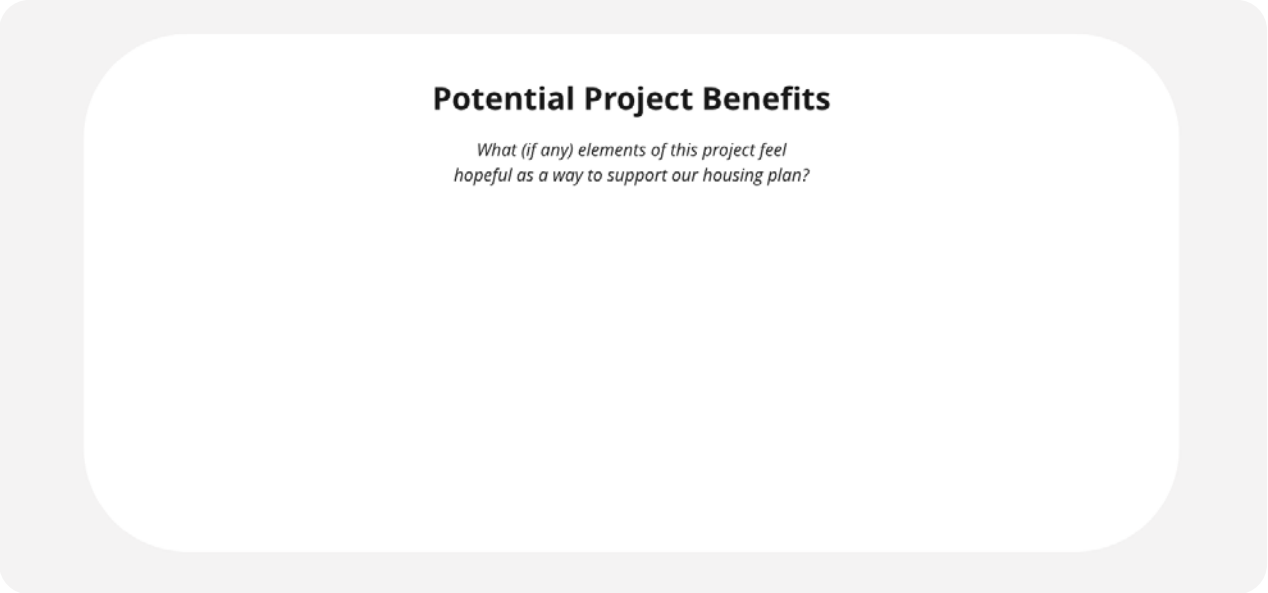

**Potential Project Benefits**

5. Individually take a look at the **Potential Project Benefits** and the **Sector-Specific and Use Case-Specific Questions** in the Stakeholder Impact Assessment (Design Phase) activity, then vote on the question:

   - Should our team develop this system to automate our site selection process?

6. Choose the notes within these sections which best explain your decision, marking them with a dot.

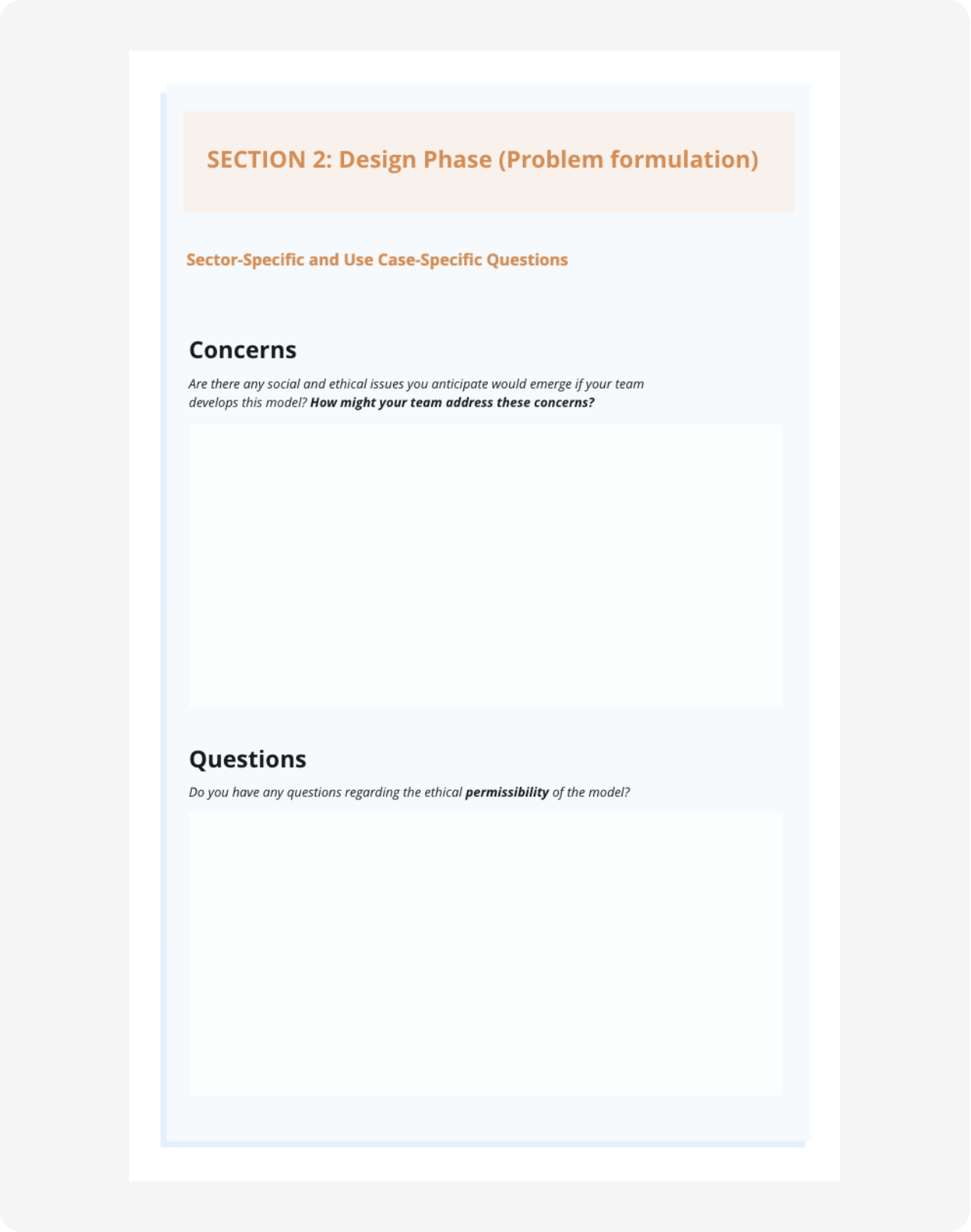

**Sector-Specific and Use Case-Specific Questions** in the Stakeholder Impact Assessment (Design Phase)

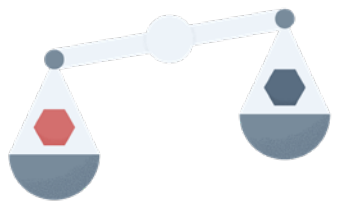

45 mins | Facilitator Instructions

# Balancing Values

1. Give the team a moment to read over the activity instructions, answering any questions.

2. Next, split the team into groups, each assigned a pair of conflicting values. Let the team know that they will have some minutes to answer the questions and come up with a plan.

3. Facilitators and co-facilitators are to touch base with each of the groups in this time, using the Considerations section of this activity to support.

4. When time is up, ask the team to reconvene. Have each group's volunteer note-taker take a few minutes to share their conflicting values and plans with the team.

5. Next, lead a team discussion about the extent to which these plans address any anticipated questions or concerns in the Design Phase SIA, and how you might adjust the plans to address these.
   - **Co-facilitator:** Write these answers on sticky notes, placing them under the **Potential Project Benefits** section on the board.

6. Give team members a moment individually take a look at the **Potential Project Benefits** and **the Sector-Specific and Use Case-Specific Questions** from the Stakeholder Impact Assessment (Design Phase) activity, then vote on the question:
   - Should our team develop this system to automate our site selection process?

7. When the team has voted, let them know that this decision, alongside the plans developed in this activity, will be used to draft your recommendation to the council.

## Facilitator Considerations — Approaches Within the Model Proposal

When facilitating discussion on the role that principle-based and consequence-based approaches to ethical deliberation play in the model proposal, you may want to stress that some tensions might be better addressed by drawing one approach and other tensions by drawing on the other. In other words, at times and in certain cases, consequences and outcomes matter more for resolving tensions (say, for instance, when the benefits of an action far outweigh the costs of abiding by a principle). At other times and in certain other cases, abiding by principles may matter more (for instance, when, regardless of the consequences, adhering to a principle is crucial for upholding a cherished right or maintaining individual integrity).

## Facilitator Considerations — Balancing Plan

In stewarding discussions of balancing plans, you might want to point workshop participants to the 'SUM Values in Focus: Respect, Connect, Care, and Protect' section of the AI Sustainability in Practice Part One workbook. The details of the ethical values contained in this section provide a good launching pad for in-depth dialogue about balancing tensions between them. 'The Values Map' section of the same workbook may also be helpful in this activity. On the following page are some useful guiding questions to assist workshop participants in resolving value tensions and in weighing up principles against the benefits of using the model to speed up delivery of development sites.

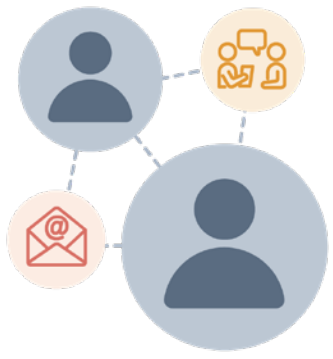

65 mins | Participant Instructions

# Revisiting Engagement Method

### Objective

Undertake practical considerations of resources, capacities, timeframes, and logistics as well as stakeholder needs to establish an engagement method for the following SIA.

### Activity Context

Your team has conducted the Design Phase SIA and shared it with your council along with your recommendation regarding the development of the proposed project. The council has chosen to move forward with the project but incorporated the following amendments to the project proposal:

#### Amendments

- The target variable of suitability will now indicate that sites categorised as suitable will, once passing your team's review and public consultation, be made publicly available for developers to submit planning applications.
- Your team will review individual applications in a process that will include consulting with neighbours of specific sites. Your team will use the model's web interface to assess centralised information about each site, streamlining the process while enabling human oversight.
- The model will be deployed in a small-scale area and expanded in increments, subject to SIAs at each phase.

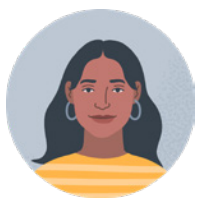

With the help of Mia, your team's Data Scientist, your team has designed and developed the model. You are now in the Model Reporting step within the Development Phase of the lifecycle and are finishing up your Development Phase SIA, which you are conducting through a Citizens' Jury. You are finishing up your Design to Development Phase SIA and need to schedule proportional Development Phase assessments and engagement activities.

## Full Team Instructions (Part One)

1. Take a moment to individually look over the activity instructions and **Development Phase SIA**, the **Model Performance Metrics Report**, and the updates on the **Project Lifecycle** section on the board.

2. Your team will be split into two groups. Go to your relevant team's instructions below to continue the activity.

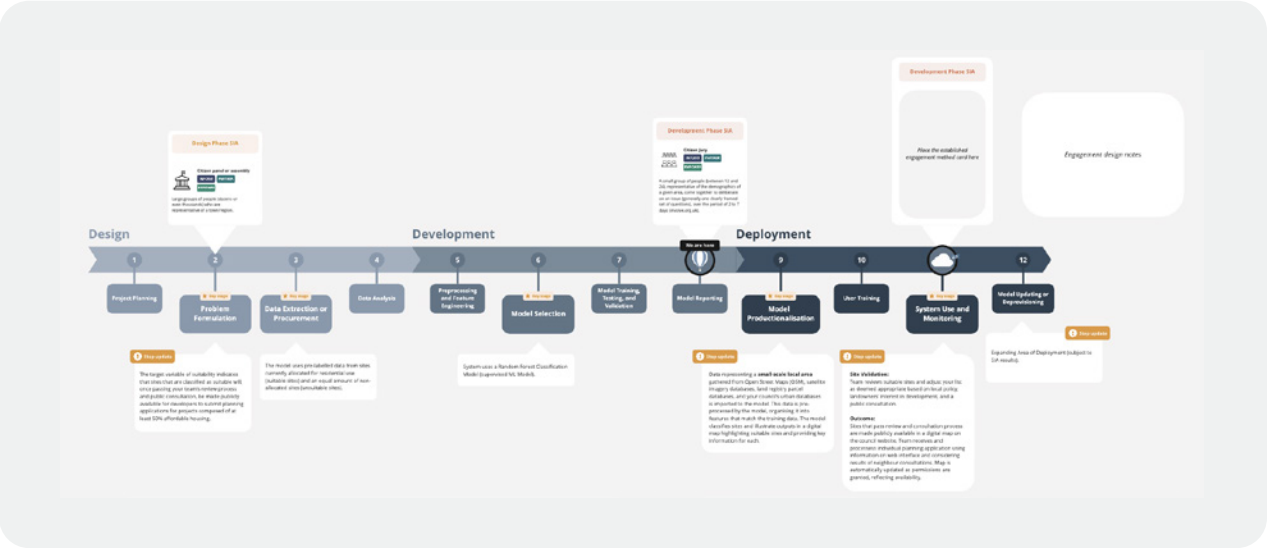

**Project Lifecycle**

## Group 1 Instructions

1. Take a moment to individually look over the **Engagement Method Cards** as well as the **Overview of Engagement Resources and Constraints**. You will consider your established engagement objectives along with practical considerations of resources, capacities, timeframes, and logistics to determine which engagement method to put forward for the next SIA.

2. Consider the following questions:
   - Which methods meet your established engagement objective?
   - What resources are available for conducting engagements?
   - What, if any, practical considerations regarding resources, capacities, timeframes, and logistics may pose constraints when selecting engagement methods?
   - What engagement methods may be most feasible within these constraints?
   - Considering the above questions, which engagement method would you establish for the next SIA?

3. Your group note-taker is to write out answers within the Notes section of this activity.

4. Once instructed to by your facilitator, jump to the **Full Team Instructions**.

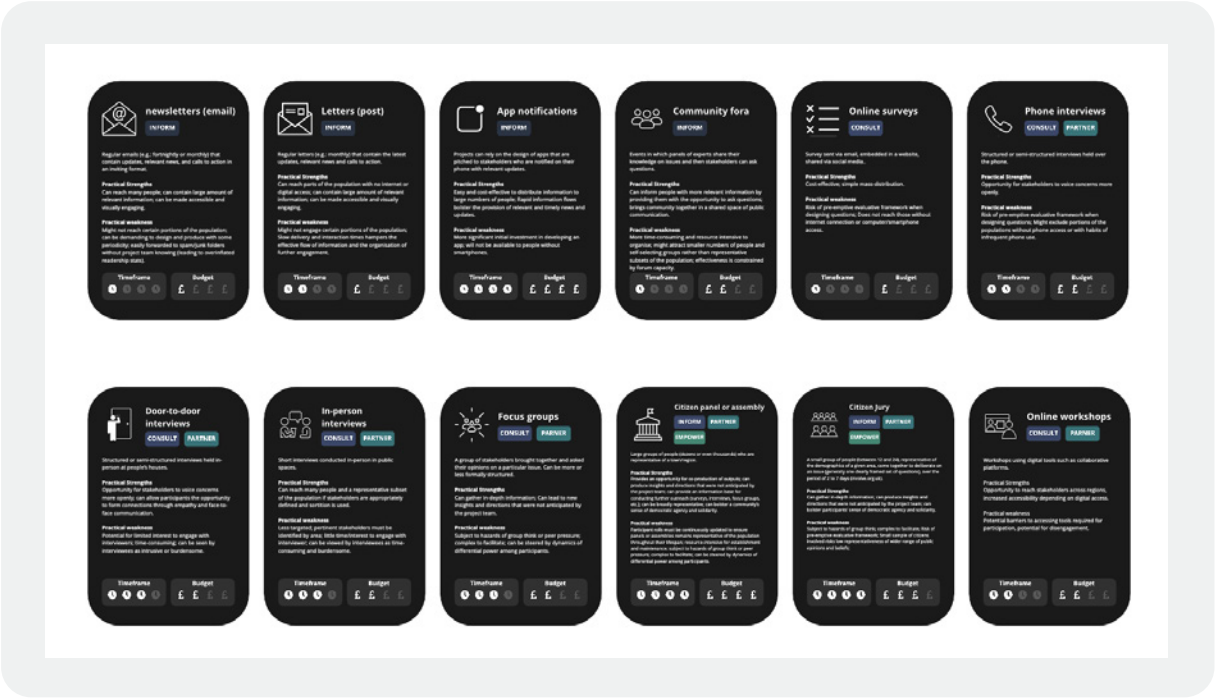

**Engagement Method Cards**

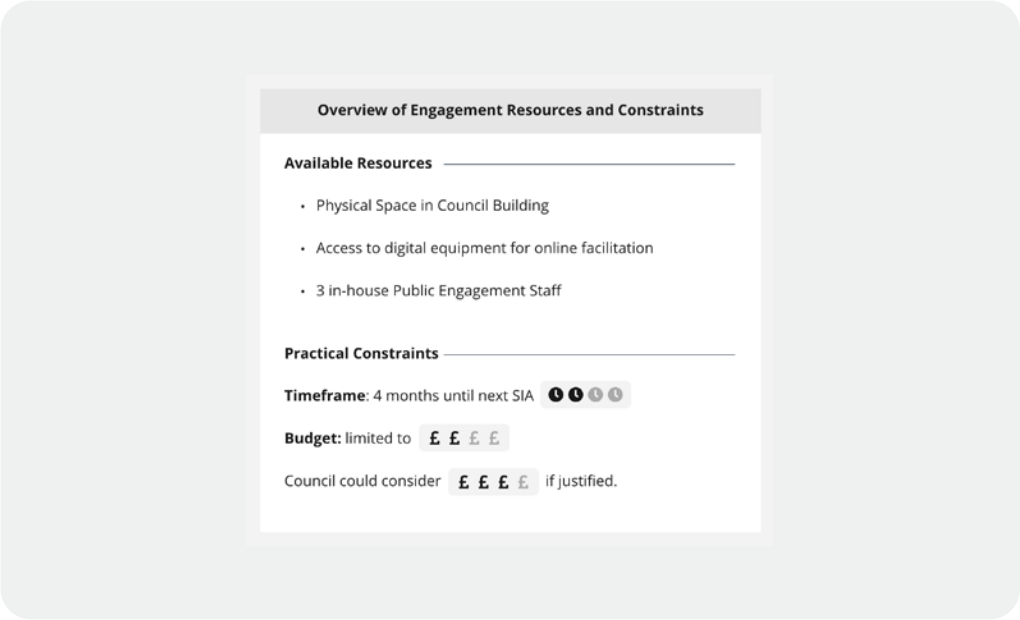

**Overview of Engagement Resources and Constraints**

## Group 2 Instructions

1. In your group, discuss the results of the Development Phase SIA. Consider the questions:

   - How might the model update harm stakeholders?
   - To what extent do the model's performance metrics safeguard salient stakeholders against poor quality outcomes?
   - To what extent do the updates to the project plan (including changes to the target variable and the incremental deployment of the model) mitigate potential risks posed within this SIA?

2. Next, take a moment to individually look over the **Engagement Method Cards**. You will consider SIA results along with stakeholder needs to put forward an engagement method for the next SIA.

3. Consider the following questions:

   - What accessibility requirements might stakeholders have?
   - Will online or in-person methods (or a combination of both) be most appropriate to engage salient stakeholders?
   - Which methods meet your established engagement objective?
   - Considering the above questions, what engagement method would you establish for the next SIA?

4. Your group note-taker is to write out answers within the **Notes** section of this activity.

5. Once instructed to by your facilitator, jump to the **Full Team Instructions**.

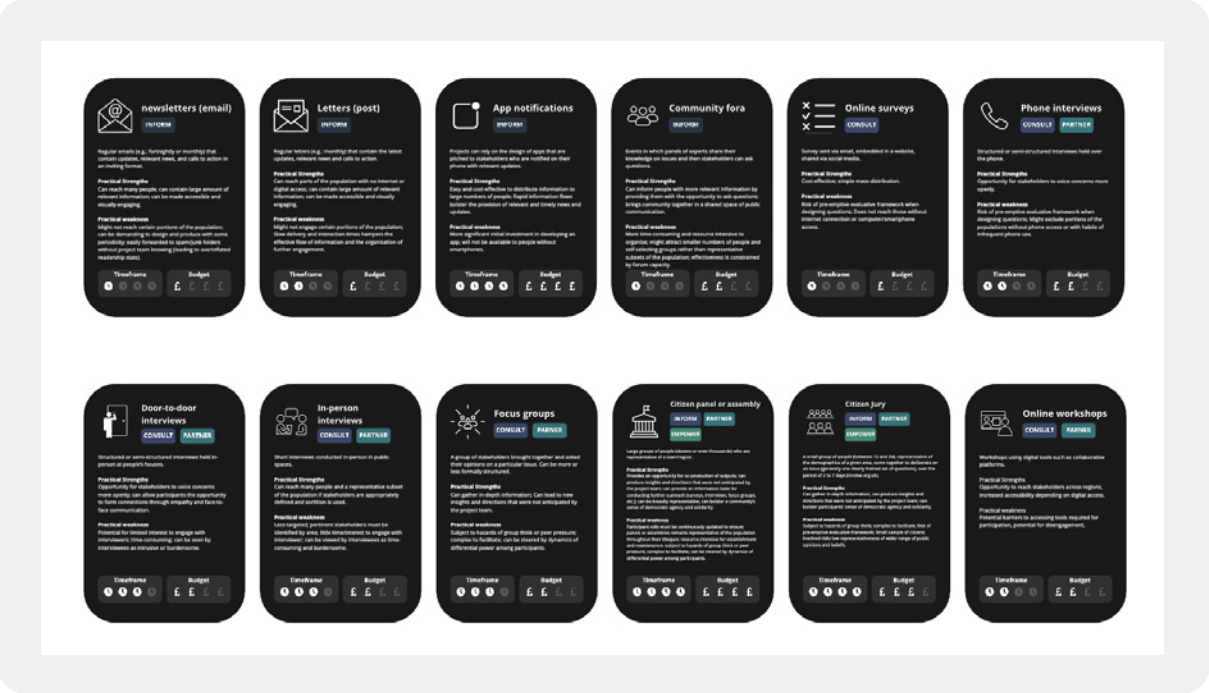

**Engagement Method Cards**

## Full Team Instructions (Part Two)

1. Reconvene as a group, having group note-takers present their chosen methods and each group's reasoning.

2. Have a group discussion about what method might be best suited to balance practical constrains with stakeholder needs.

3. Vote on an engagement method.

4. Next, have a group deliberation about the design of the engagement method, considering:

    - How might the team make sure that this chosen method accommodates different types of stakeholders?
    - How might the team ensure that the PS report used to conduct the SIA is accessible to stakeholders?
    - How might the team ensure that engagement method feeds useful information to your SIAs?

5. Consider what feedback mechanisms will be in place.

6. Your co-facilitator will place the established **Engagement Method Card** on the appropriate section of the **Project Lifecycle** on the board, and outline engagement details within the card.

## SECTION 2: Development Phase

Model Reporting

After reviewing the results of your initial SIA, answer the following questions:

- Are the trained model's actual objective, design, and testing results still in line with the evaluations and conclusions contained in your original assessment? If not, how does your assessment now differ?

**RESULTS**

The model has been adjusted to address concerns with the original assessment:

1. Concerns with the meaning attributed to the target variable of suitability:
    - The target variable of suitability will now indicate that sites categorised as suitable will, once passing your team's review and public consultation, be made publicly available for developers to submit planning applications for projects composed of at least 50% affordable housing.
2. Concerns with the degree of monitoring in the Deployment Phase:
    - The Deployment Phase of this project has been adjusted for the model to be deployed in increments, expansion being subject to SIAs.

Testing results meet acceptable performance metrics.

- Have any other areas of concern arisen with regard to possibly harmful social or ethical impacts as you have moved from the Development to the Deployment Phase of this project?

**RESULTS**

The model has been updated to account for a new planning policy that enables commercial buildings to be repurposed for housing development. The model now considers commercial buildings as potentially suitable sites. This update has enabled the model to identify suitable sites accurately under current local policy.

During model development, our team determined that the model was not generating enough suitable sites due to a feature indicating the percentage of green or public space within sites. The model was only classifying sites with a low percentage of green or public spaces as suitable. Our team removed this feature in order for the model to generate a greater number of sites, irrespective of the percentage of public or green space within these.

As these changes were not accounted for in the Design Phase SIA, our team will closely monitor potential issues and feedback.

**Revisiting Stakeholder Analysis and Positionality**

Considering the current state of the project as reflected in the Development Phase SIA:

- Stakeholder groups identified in the PS report accurately reflect current stakeholders in the project.
- There are new potential impacts posed by updates to this model, such as the potential harm to local businesses that may be caused by the model categorising commercial sites as suitable for housing development, and our team promoting them for housing development.
- Considering these new potential impacts, local business owners are to be considered increasingly salient stakeholders at this phase in the project.
- Given that there have been no changes to the team's composition or identified stakeholders, the positionality reflection conducted in the Design Phase remains relevant at this phase in the project.

**Revisiting Engagement Objective**

The engagement objective of partnering with stakeholders is considered proportional to the new potential impacts outlined in the SIA results and the team positionality reflection.

## Model Performance Report

The following metrics summarise the model's performance when predicting suitability within unseen testing data. These metrics were reported with a 95% confidence interval and error bars.

97%

**Precision**
Number of true positives divided by the number of all sites classified as suitable (true positives and false positives)

97%

**Accuracy**
Number of correct classifications divided by total number of classifications made

95%

**Recall**
Number of true positives divided by the number of all actual suitable sites (true positives and false negatives)

**True Positive:**
Model correctly classifies a site as suitable.

**False Positive:**
Model incorrectly classifies site as suitable.

**True Negative:**
Model correctly classifies site as unsuitable.

**False Negative:**
Model incorrectly classifies a site as unsuitable.

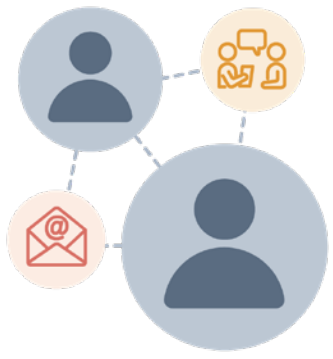

65 mins | Facilitator Instructions

# Revisiting Engagement Method

1. Give the team some minutes to individually look over instructions for this activity, as well as the **Development Phase SIA** and the **Model Performance Metrics Report**.

2. When time is up, ask the team if they have any questions.

3. Next, split the team into groups, asking for a volunteer note-taker from each group. Note-takers will be responsible for reporting back group findings to the team.
   - **Group 1** will be responsible for discussing practical considerations of resources, capacities, timeframes, and logistics that may pose constraints when selecting engagement methods.
   - **Group 2** will discuss stakeholder needs.

4. Give each group some minutes to decide on an engagement method by discussing the questions in their group instructions.
   - Facilitators should join and support one group, and co-facilitators another.

5. When time is up, ask the team to reconvene, giving each group's note-taker a few minutes to share their team's chosen engagement method, as well as their reasoning behind this decision.

6. After the two groups have shared, lead a team discussion about what method might be better suited to balance practical constrains with stakeholder needs.

7. After the team discussion, ask the team to vote on an engagement method.
   - **Co-facilitator:** place the established **Engagement Method Card** on the 'System Use and Monitoring' step within the Project Lifecycle on the board.

8. Next, lead a group deliberation about the design of the engagement, considering the questions:
   - How might the team make sure that this chosen method accommodates different types of stakeholders?
   - How might the team ensure that the PS Report used to conduct the SIA is accessible to stakeholders?
   - How might the team ensure that method feeds useful information to your SIA?

9. Consider what feedback mechanisms will be in place.
   - **Co-facilitator:** write engagement details on sticky notes within the **Engagement Details** section on the board.

**Facilitator Considerations** | **Engagement Methods**

In facilitating discussion of the re-evaluation and re-crafting of engagement methods and of setting timeframes for Development Phase re-assessment, you should stress how proportional monitoring acts on the need for responsiveness across sustainable AI lifecycles. You might want to emphasise the points made in the relevant passages from AI Sustainability in Practice Part One:

> “ Stakeholder analyses may be carried out in a variety of ways that involve more-or-less stakeholder involvement. This spectrum of options ranges from analysis carried out exclusively by a project team without active community engagement, to analysis built around the inclusion of community-led participation and co-design from the earliest stages of stakeholder identification. The degree of stakeholder involvement will vary from project to project based upon a preliminary assessment of the potential risks and hazards of the model or tool under consideration.
>
> AI Sustainability in Practice Part One

35 mins | Participant Instructions

# Stakeholder Impact Assessment (Deployment Phase)

### Objective

Practise using SIAs to formulate proportional monitoring activities for the development and deployment of AI models.

### Activity Context

Your team has deployed the system within an initial deployment area, and you are due to conduct your first Deployment Phase SIA.

## Team Instructions

1. In this activity, your team will be split into three groups. Each group will be assigned samples of stakeholder feedback that represent greater stakeholder reactions to the deployment of the model.

2. Each group will have an assigned note-taker who is to record team discussions on the group's section on the board and report back to the team, considering:

   - What was the feedback sample and how, if at all, was it connected to production, implementation, or environmental factors?
   - What harmful impacts were raised by this sample?
   - Did your team decide updating or deprovisioning was a better option? What informed this choice?

### Part One: Production, Implementation, and Environmental Factors

3. In your groups, discuss how your assigned feedback samples may be connected to changes in production and implementation factors, or to environmental factors.

   - Revisit the section *Case Study: Challenges to AI sustainability in AI for Urban Planning* of the workbook for support in this discussion.

### Part Two: SIA Question

4. Next, turn to the Deployment Phase SIA question assigned to your group on your section of the board, and have a group discussion to come up with an answer to the question. Consider any harmful impacts that have arisen from the deployment of this system.

## Part Three: Updating or Deprovisioning

5. Having assessed impacts, your team is to discuss whether updating the model may serve to mitigate possible harmful impacts, and/or to amplify beneficial ones. Consider the possibilities identified on your group's **Updating or Deprovisioning** section.

6. As a group, decide on whether you believe updating the model would be a feasible solution for addressing the harmful impacts of this model, or if deprovisioning the model is a better option.

7. Your volunteer note-taker is to indicate your decision on your group's **Updating or Deprovisioning** section.

8. Reconvene as a group, having volunteer note-takers share each group's decision and discussing each decision as a team.

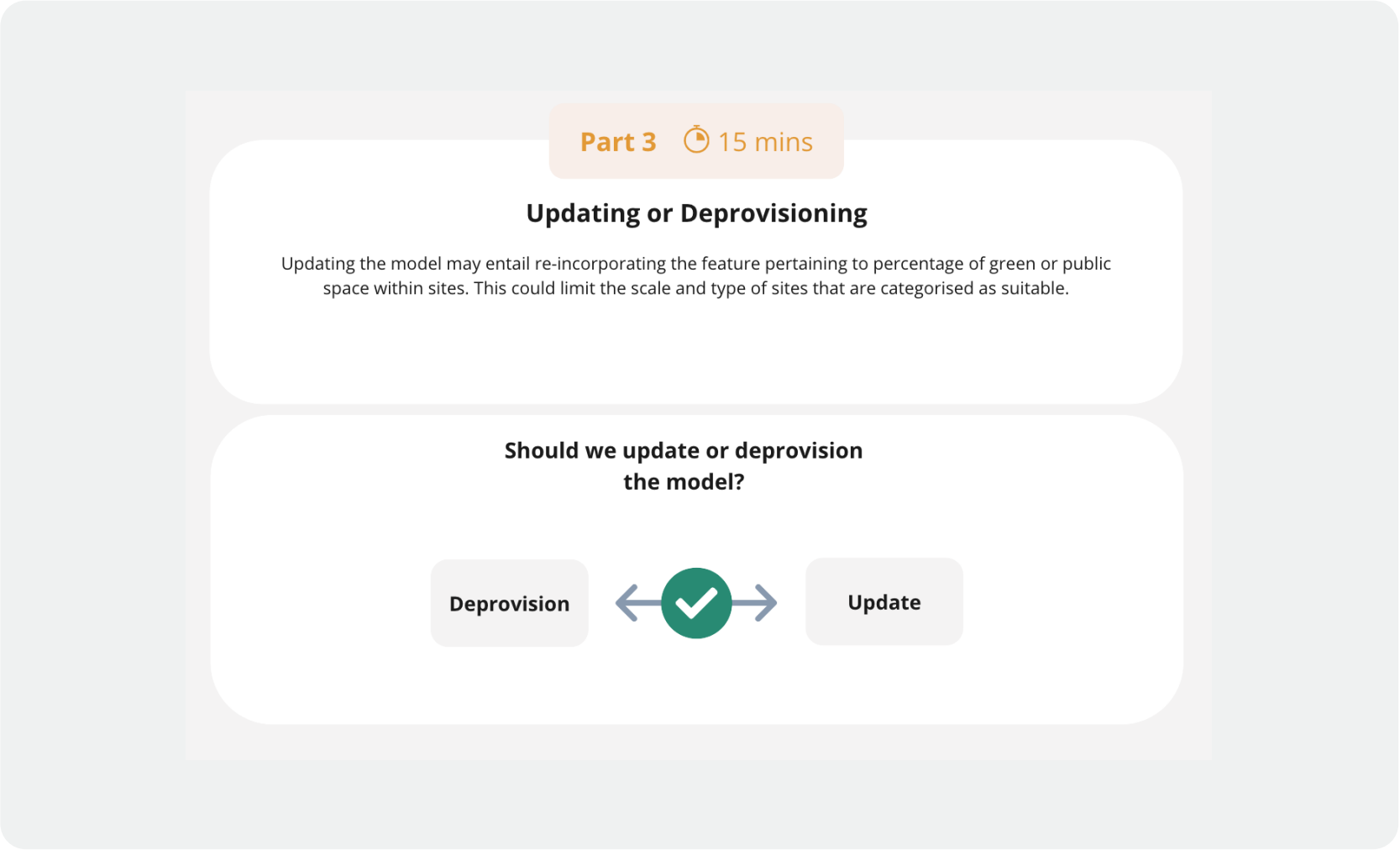

**Updating or Deprovisioning**

35 mins | Facilitator Instructions

# Stakeholder Impact Assessment (Deployment Phase)

1. In this activity, your team will be split into groups. Each group will be assigned samples of stakeholder feedback that represent wider stakeholder reactions to the deployment of the model, and a relevant Deployment Phase SIA question.

2. Harmful impacts raised by each feedback sample are connected to a production, implementation,or environmental factor. Each group will deliberate on what production, implementation, or environmental change their assigned feedback sample may be connected to:

   - Groups are to answer their assigned question, identifying harmful impacts.

   - Lastly, groups will decide if updating the model would be a feasible solution for addressing harmful impacts, or if deprovisioning the model is a better option.

3. Give the team a moment to individually read over the activity instructions, answering any questions.

4. Next, split the team into groups, asking for a volunteer note-taker per group that will report back to the team.

5. Give the team enough minutes to conduct this activity. Inform the team of the maximum allocated minutes for each part of this activity:

   - Part One: Production, implementation, and environmental factors - **10 minutes**

   - Part Two: SIA question - **10 minutes**

   - Part Three: Updating or deprovisioning - **15 minutes**

6. Facilitators and co-facilitators are to touch base with each group and provide support, using the Considerations section of this activity.

7. When time is up, ask the team to reconvene.

8. Give note-takers a few minutes to report back to the team. After each volunteer shares, give the team a few minutes to discuss the decision.

9. Once all decisions have been shared and discussed, consider the overall group view on updating or deprovisioning the model. Based on the overall group view, choose the corresponding scenario from the following section to read out to the group. These scenarios represent the outcomes of updating or deprovisioning the model.

## Facilitator Considerations — Updating

Model has been updated to include current datasets, and an updating protocol has been set up to ensure data remains timely and relevant. The feature indicating the percentage of green or public spaces within sites was re-integrated into the model to ensure it doesn't select sites that have a significant amount of green space, and that less sites are consequently selected. Model updates that would result in outputs that are at odds with planning policy were not permitted, but the council has taken note of stakeholders' feedback for further consultation on the policy itself.

Local development has continued to grow at a pace that meets the target in our 10-year housing plan, and the model is utilising datasets that accurately reflect real-world access to essential services, safeguarding the quality of outputs. Residents have responded positively to the shift to decrease the scale of development and limit the percentage of green and public spaces being used, although rising prices continue to be an area of concern.

Our team does, however, continue to receive negative feedback regarding the impact of the model on local businesses. Residents critique the time-consuming nature of updating local policy compared to the speed at which commercial sites are being repurposed for housing. We also continue to receive negative feedback regarding what constitutes the definition of affordable housing.

## Facilitator Considerations — Deprovisioning

Model deployment has been stopped while there is public consultation regarding:

1. the percentage of affordable housing that is deemed appropriate by local residents;
2. the definition of "affordable"; and
3. the constraints that are to be put on what types of commercial sites can be repurposed for housing.

The outputs of this consultation will serve to define objectives of a new project, for which components of the current project (i.e. re-validated datasets, model) will serve as a foundation. Residents are happy to be involved in defining outputs once these points are democratically addressed. The new project is likely to provide outputs that reflect residents' self-articulated interests.

The pace of local housing development has, however, temporarily returned to the growth rate it had prior to the deployment of this project. This is challenging our team's ability to meet the targets in our 10-year housing plan.

## Group 1 Activity Considerations and Answers

### Feedback Samples

- Quote from Alex, Planning Authority Lead, highlighting increase in housing development in the deployment area:

  > *The model has attracted much more development in a short time period. Our team has been able to review planning applications faster while considering residents' input. Using the model has been useful, but we will need to review all available feedback prior to assessing next steps.*

- Quote from Terry, Local Resident, highlighting harmful impacts of the model, including high rates of development pricing-out of local residents:

  > *We are seeing development left right and centre, bringing people from outside the area who can actually afford to rent or buy. It doesn't seem like the council is interested in those of us who have always been here. There are new shops none of us can afford, public spaces turned private, rent prices going up. Your plan is helping change our neighbourhood for the worse. I myself need affordable housing, but this plan is kicking us out.*

- Quote from Ali, Local Resident, highlighting harmful impacts of the model including green spaces being built over:

  > *A planning application has been submitted for a development to be built over our community garden. We won't let this happen. The council needs to protect the green spaces that make this neighbourhood a community.*

### Relevant Production, Implementation, or Environmental Factor

- **Production Factor:** The meaning attributed to target variable of suitability includes promotion of sites for developments of at least 50% housing.

### Deployment Phase SIA Question and Potential Answers

- *Q: How does the content of the existing SIA compare with the real-world impacts of the AI system as measured by available evidence of performance, monitoring data, and input from implementers and the public?*
- *A:* The deployment of the model seems both to confirm many of the concerns expressed in the original SIA and to uncover new harms that were previously unanticipated. Concerns about affordability and the inequitable impacts of gentrification and displacement have been validated in light of the rapid pace of development and the influx of new residents. Concerns about the diminishment of public and green spaces have been confirmed by spreading privatisation and the filing of new planning applications, though there is disagreement among residents about the costs and benefits of streamlined planning.

### Considerations for Updating or Deprovisioning

- There are likely significant constraints to changing meaning attributed to the target variable, namely significant stakeholder consultation and council approval.

## Group 2 Activity Considerations and Answers

### Feedback Samples

- Quote from Katherine, Local Resident, highlighting model categorising sites without access to essential services as suitable:

> *It was such a relief to hear I was one of the first people offered a disabled-adapted home in these new houses, I can't even walk up the flight of stairs in my current flat! It is a shame that the only leisure within two kilometres of the building was closed last year. I took the house because I simply cannot stay here, but I don't know what I will do without my exercise routine. This something that needs to be thought about.*

### Relevant Production, Implementation, or Environmental Factor

- **Environmental Factor:** Change in real-world proximity to service provision (used as features in the model) have diminished the predictive power of these features. The model has retained the inferences from outdated data and is inaccurately classifying sites as suitable.

### Deployment Phase SIA Question and Potential Answers

- *Q: Have the maintenance processes for your AI model adequately taken into account the possibility of distributional shifts in the underlying population? Has the model been properly re-tuned and re-trained to reflect changes in the environment?*
- *A:* Katherine's feedback indicates that the model has not been adequately updated to keep pace with the relevant distributional shift (i.e. that the closing of the leisure centre has changed certain people's access to essential services). This suggests that more frequent model updating may be necessary.

### Considerations for Updating or Deprovisioning

- Extracting or producing up-to-date datasets that accurately reflect the state of essential services in the deployment area, and establishing monitoring protocols that ensure the data used by the model remains current are likely feasible avenues to addressing this issue.

## Group 3 Activity Considerations and Answers

### Feedback Samples

- Quote from Mia, Project Data Scientist, highlighting the model's ability to identify sites that meet requirements set out in current planning policy:

  > *Having tested, validated, and verified the system, our team was happy to see the model perform with strong performance and safety metrics. Having incorporated new features, our model is also up to date with local policy.*

- Quote from George, local business owner, highlighting harmful impacts of promoting commercial sites for residential repurposing, namely, it's correlation with local businesses not receiving rental contract renewals:

  > *Your model is closing down long-standing local businesses. More and more property owners are refusing to renovate our contracts. By publishing commercial buildings, you have attracted purchase offers that small business owners simply cannot match.*

### Relevant Production, Implementation, or Environmental Factor

- **Change in Environmental Factors:** Change in policy standards regulating the repurposing of sites was incorporated into a model update.

### Deployment Phase SIA Question and Potential Answers

- *Q: Have any unintended harmful consequences ensued in the wake of the deployment of the system?*
- *A:* Incentives for building owners to sell to property developers, who are converting commercial buildings to residential properties, are driving local businesses out of their spaces. Though the pace of local development is allowing for the local authority to meet its targets, negative impacts on local businesses have been an unintended harmful consequence of this success.

### Considerations for Updating or Deprovisioning

- Updating the model for it to not categorise commercial sites as suitable would entail significant stakeholder consultation and is likely to raise tensions as the system's categorisations would be at odds with policy.

# Endnotes

1 Information Commissioner's Office. (2021). *Guide to the General Data Protection Regulation (GDPR).* https://ico.org.uk/for-organisations/guide-to-data-protection/

2 Harrington, C., Erete, S., & Piper, A. M. (2019). Deconstructing Community-Based Collaborative Design: Towards More Equitable Participatory Design Engagements. *Proceedings of the ACM on Human-Computer Interaction, 3*(CSCW), 1-25. https://doi.org/10.1145/3359318

3 OECD. (2005). *Evaluating Public Participation in Policy Making.* https://doi.org/10.1787/9789264008960-en

4 OECD. (2013). *Government at a Glance 2013.* https://doi.org/10.1787/gov_glance-2013-en

5 Dawkins, C. E. (2014). The principle of good faith: Toward substantive stakeholder engagement. *Journal of Business Ethics,* 121, 283-295. https://doi.org/10.1007/s10551-013-1697-z

6 Catapult. (2019, May 11). *Building a 21st century digital planning system: A quick start guide.* UKRI Innovate UK. https://cp.catapult.org.uk/news/building-a-21st-century-digital-planning-system-a-quick-start-guide/

7 Leslie, D. (2019). Understanding artificial intelligence ethics and safety: *A guide for the responsible design and implementation of AI systems in the public sector.* The Alan Turing Institute. https://doi.org/10.5281/zenodo.3240529

8 Hartz-Karp, J. (2005). A case study in deliberative democracy: Dialogue with the city. *Journal of Public Deliberation, 1*(1), 1-15. https://doi.org/10.16997/jdd.27

9 Grice, H. P. (1975). Logic and Conversation. In P. Cole, & J. L. Morgan. (Eds.), *Syntax and Semantics, Volume 3, Speech Acts* (pp. 41-58). Academic Press.

10 Arendt, H. (1958). *The human condition.* University of Chicago Press.

11 Bachrach, P., & Baratz, M. (1962). Two faces of power. *American Political Science Review, 57*(4), 947–952. https://doi.org/10.2307/1952796

12 Bohman, J. (2000). *Public deliberation: Pluralism, complexity, and democracy.* MIT press.

13 Gutmann, A., & Thompson, D. (1996). *Democracy and disagreement.* Harvard University Press.

14 Habermas, J. (1984). *The theory of communicative action I: Reason and the rationalization of society.* Beacon Press.

15 Hindess, B. (1996). *Discourses of power: From Hobbes to Foucault.* Blackwell Publishers

16 Cohen, J. (1989). Deliberation and democratic legitimacy. In A. Hamlin, & P. Pettit (Eds.), *The good polity: normative analysis of the state* (pp. 17–34). Basil Blackwell.

17 Manin, B. (1987). *On legitimacy and political deliberation. Political Theory, 15*(3), 338–368.

18 McLeod, J. M., Scheufele, D. A., Moy, P., Horowitz, E. M., Holbert, R. L., Zhang, W., Zubric, S., & Zubric, J. (1999). Understanding Deliberation: The effects of discussion networks on participation in a public forum. *Communication Research, 26*(6), 743–774. https://doi.org/10.1177/009365099026006005

19 Przeworski, A. (1998). Deliberation and ideological domination. In J. Elster (Ed.), *Deliberative Democracy* (pp. 140–160). Cambridge University Press.

20 Landwehr, C. (2014). Facilitating deliberation: The role of impartial intermediaries in deliberative mini-publics. In Grönlund, K., Bächtiger, A., & Maija Setälä (Eds.), *Deliberative mini-publics: Involving citizens in the democratic process* (pp. 77-92). ECPR Press.

21 See for instance Nagoda, S., & Nightingale, A. J. (2017). Participation and power in climate change adaptation policies: Vulnerability in food security programs in Nepal. *World Development,* 100, 85-93. https://doi.org/10.1016/j.worlddev.2017.07.022

22 Lupia, A., & Norton, A. (2017). Inequality is always in the room: Language & power in deliberative democracy. *Daedalus, 146*(3), 64-76.https://doi.org/10.1162/DAED_a_00447

23 Mendelberg, T., Karpowitz, C. F., & Oliphant, J. B. (2014). Gender Inequality in Deliberation: Unpacking the Black Box of Interaction. *Perspectives on Politics, 12*(1), 18–44. http://doi.org/10.1017/S1537592713003691

24 Mendelberg, T., & Oleske, J. (2000). Race and public deliberation. *Political Communication, 17*(2), 169-191. https://doi.org/10.1080/105846000198468

25 Parsons, M., Fisher, K., & Nalau, J. (2016). Alternative approaches to co-design: insights from indigenous/academic research collaborations. *Current Opinion in Environmental Sustainability,* 20, 99-105. https://doi.org/10.1016/j.cosust.2016.07.001

26 Lupia, A., & Norton, A. (2017). *Inequality is always in the room: Language & power in deliberative democracy. Daedalus, 146*(3), 64-76.https://doi.org/10.1162/DAED_a_00447

27 Tschakert, P., Das, P. J., Pradhan, N. S., Machado, M., Lamadrid, A., Buragohain, M., & Hazarika, M. A. (2016). Micropolitics in collective learning spaces for adaptive decision-making. *Global Environmental Change,* 40, 182-194. https://doi.org/10.1016/j.gloenvcha.2016.07.004

28 Garcia, A., Tschakert, P., Karikari, N. A., Mariwah, S., & Bosompem, M. (2021). Emancipatory spaces: Opportunities for (re) negotiating gendered subjectivities and enhancing adaptive capacities. *Geoforum, 119,* 190-205. https://doi.org/10.1016/j.geoforum.2020.09.018

29 Leslie, D., Katell, M., Aitken, M., Singh, J., Briggs, M., Powell, R., Rincon, C., Perini, A. M., & Jayadeva, S. (2022). *Data Justice in Practice: A Guide for Policymakers.* The Alan Turing Institute in collaboration with The Global Partnership on AI. https://doi.org/10.5281/zenodo.6429475

30 Glaberson, S. K. (2019). Coding over the cracks: Predictive analytics and child protection. *Fordham Urban Law Journal, 46*(2), 307-363. https://ir.lawnet.fordham.edu/ulj/vol46/iss2/3

31 The Office for Standards in Education, Children's Services and Skills (Ofsted). (2023). *Inspecting local authority children's services.* https://www.gov.uk/government/publications/inspecting-local-authority-childrens-services-from-2018/inspecting-local-authority-childrens-services#inspection-methodology

32 Sideris, N., Bardis, G., Voulodimos, A., Miaoulis, G., & Ghazanfarpour, D. (2019). Using Random Forests on Real-World City Data for Urban Planning in a Visual Semantic Decision Support System. *Sensors, 19*(10), 2266. https://doi.org/10.3390/s19102266

33 Sideris, N., Bardis, G., Voulodimos, A., Miaoulis, G., & Ghazanfarpour, D. (2019). Using Random Forests on Real-World City Data for Urban Planning in a Visual Semantic Decision Support System. *Sensors (Basel, Switzerland), 19*(10), 2266. https://doi.org/10.3390/s19102266

34 Learn more about risks in urban planning through Koseki, S., Jameson, S., Farnadi, G., Rolnick, D., Régis, C., Denis. J., Leal, A., de Bezenac, C., Occhini,, G., Lefebvre, H., Gallego-Posada, J., Chehbouni, K., Molamohammadi, M., Sefala, R., Salganik, R., Yahaya, S., & Téhinian, S. (2022). *AI and Cities Risks, Applications and Governance.* UN-Habitat. https://unhabitat.org/sites/default/files/2022/10/artificial_intelligence_and_cities_risks_applications_and_governance.pdf

35 Ryan, R. M., & Deci, E. L. (2017). *Self-determination theory: Basic psychological needs in motivation, development, and wellness.* Guilford Publications.

36 World Economic Forum (2019). *AI Governance: A Holistic Approach to Implement Ethics into AI [White Paper].* https://www.weforum.org/whitepapers/ai-governance-a-holistic-approach-to-implement-ethics-into-ai/

37 Kilkenny, M. F., & Robinson, K. M. (2018). Data quality: "Garbage in–garbage out". *Health Information Management Journal, 47*(3), 103-105. https://journals.sagepub.com/doi/pdf/10.1177/1833358318774357

38 Babbage, C. (1864). *Passages from the life of a philosopher.* Longman, Green, Longman, Roberts, and Green.

39 Mellin, W. (1957). Work with new electronic 'brains' opens field for army math experts. *The Hammond Times,* 10, 66.

40 Suresh, H., & Guttag, J. V. (2019). A framework for understanding unintended consequences of machine learning. arXiv preprint *arXiv:1901.10002, 2*(8). https://doi.org/10.48550/arXiv.1901.10002

41 O'Neil, C. (2017). Weapons of math destruction: How big data increases inequality and threatens democracy. Crown.

42 Prince, A. E., & Schwarcz, D. (2020). Proxy discrimination in the age of artificial intelligence and big data. *Iowa Law Review, 105*(3), 1257-1318. https://ssrn.com/abstract=3347959

43 d'Alessandro, B., O'Neil, C., & LaGatta, T. (2017). Conscientious classification: A data scientist's guide to discrimination-aware classification. *Big data, 5*(2), 120-134. https://doi.org/10.1089/big.2016.0048

# Bibliography and Further Readings

## Stakeholder Impact Assessment

AI Now Institute. (2018). Algorithmic Impact Assessments: Toward Accountable Automation in Public Agencies. Retrieved from: https://ainowinstitute.org/publication/algorithmic-impact-assessments-toward-accountable-automation-in-public-agencies

Diakopoulos, N., Friedler, S., Arenas, M., Barocas, S., Hay, M., Howe, B., Jagadish, H. V., Unsworth, K., Sahuguet, A., Venkatasubramanian, S., Wilson, C., Yu, C., & Zevenbergen, B. (n.d.). Principles for accountable algorithms and a social impact statement for algorithms. Fairness, Accountability, and Transparency in Machine Learning. Retrieved from: http://www.fatml.org/resources/principles-for-accountable-algorithms

Karlin, M. (2018). A Canadian algorithmic impact assessment. Retrieved from https://medium.com/@supergovernance/a-canadian-algorithmic-impact-assessment-128a2b2e7f85

Karlin, M., & Corriveau, N. (2018). The government of Canada's algorithmic impact assessment: Take two. Retrieved from: https://medium.com/@supergovernance/a-canadian-algorithmic-impact-assessment-128a2b2e7f85

Reisman, D., Schultz, J., Crawford, K., & Whittaker, M. (2018). Algorithmic impact assessments: A practical framework for public agency accountability. AI Now institute. Retrieved from: https://ainowinstitute.org/publication/algorithmic-impact-assessments-report-2

Vallor, S. (2018) An ethical toolkit for engineering/design practice. Retrieved from: https://www.scu.edu/ethics-in-technology-practice/ethical-toolkit/

# AI in Urban Planning

Cabinet Office, & Geospatial Commission. (2021). Planning and Housing Landscape Review — Executive Summary. https://assets.publishing.service.gov.uk/government/uploads/system/uploads/attachment_data/file/965740/Planning_and_Housing_Landscape_Review.pdf

Geospatial Commission. (2019). Future Technologies Review. https://www.gov.uk/government /publications/future-technologies-review

Ministry of Housing, Communities & Local Government, & The Rt Hon Esther McVey MP. (2019, November 5). PropTech dragons form new expert property innovation council. GOV.UK. https://www.gov.uk/government/news/proptech-dragons-form-new-expert-property- innovation-council

Sideris, N., Bardis, G., Voulodimos, A., Miaoulis, G., & Ghazanfarpour, D. (2019). Using Random Forests on Real-World City Data for Urban Planning in a Visual Semantic Decision Support System. Sensors (Basel, Switzerland), 19(10), 2266. https://doi.org/10.3390/s19102266

The Open Data Institute. (2020, August 6). Case study: Unlocking data on brownfield sites. https://theodi.org/article/case-study-unlocking-data-on-brownfield-sites/

# Resources Informing Activities

Boal, A. (2002). Games for actors and non-actors. London: Routledge.

Coimbra, T. C., & Caroli, P. (2020). FunRetrospectives: Activities and Ideas for Making Agile Retrospectives More Engaging. Amazon Digital Services LLC - KDP Print US. https://books.google.co.uk/books?id=1KHMzQEACAAJ

To find out more about the AI Ethics and Governance in Practice Programme please visit:

turing.ac.uk/ai-ethics-governance

The Alan Turing Institute